\documentclass[10pt,a4paper,final,twoside,titlepage]{book}

\usepackage{ThesisStyle}
\usepackage{epigraph}

\title{Rotating Isospectral Drums}

\begin{document}

\frontmatter

\begin{titlepage}
\begin{center}
        Master Thesis\\
        \vspace{2cm}
        \textbf{\huge{ Rotating Isospectral Drums } }\\
        \vspace{2cm}
        A master thesis submitted to the\\
        \vspace{0.3cm}
        \Large{\textbf{Institut f\"ur Theoretische Physik}}\\
        \vspace{0.2cm}
        of the\\
        \vspace{0.2cm}
        \Large{\textbf{Eberhard-Karls Universit\"at T\"ubingen}}\\
        \vspace{0.5cm}
        \normalsize
        in partial fulfilment of the requirements for the degree of\\
        \vspace{0.3cm}
        \Large{\textbf{Master of Science}}\\
        \vspace{1.8cm}
        \normalsize
        presented by\\
        \vspace{0.5cm}
        \textbf{\Large{Anton Lebedev}}\\
        \vspace{1.8cm}

        born Apr 3$^{rd}$, 1989\\
        resident of T\"ubingen, Federal Republic of Germany\\

        \vspace{1.8cm}
        supervised by\\
        \vspace{0.5cm}
        Prof. Dr. N. Schopohl\\
        \vspace{0.5cm}
        2016
        \date{13.10.2016}
\end{center}
\thispagestyle{empty}
\end{titlepage}
\tableofcontents
\clearpage\thispagestyle{plain}
\section*{Introduction - Can one hear the shape of rotating drums?}
The thesis you currently hold in your hand is intended to provide an answer to a simple question:
\begin{center}\Large
Can one hear the shape of a rotating drum?
\end{center}

To be more precise I shall try to provide an answer to the question whether two isospectral domains - or drums - remain isospectral when rotated.
Therefore the behaviour of the eigenfrequencies of such domains has been investigated in the course of this work.
Prior to embarking on the lengthy quest in search of an answer a short overview of the subject is provided here. It shall serve as a motivation
and a guide for the first part of the present work - the theory of the subject.

\begin{figure}[tbp]
\centering
\subfloat[Isospectral domains 1]{
\resizebox{0.8\linewidth}{!}{
\begin{tikzpicture}
	\begin{axis}[plot box ratio=1 1]
	\addplot[mark=none,fill=yellow, fill opacity=0.8] file {Plots/Geometries/ID1_1_Coord.dat} --cycle;
	\coordinate (COM1_1) at (axis cs:0.4762,-0.6667);
	\fill[blue] (COM1_1) circle(1pt);
	\coordinate (origin) at (axis cs:0,0);
	\fill[red,fill opacity=0.5] (origin) circle(1.5pt);
	\end{axis}
\end{tikzpicture}
\begin{tikzpicture}
	\begin{axis}[plot box ratio=1 1]
	\addplot[mark=none,fill=yellow, fill opacity=0.8] file {Plots/Geometries/ID1_2_Coord.dat} --cycle;
	\coordinate (COM1_2) at (axis cs:0.33333,0.047619);
	\fill[blue] (COM1_2) circle(1pt);
	\coordinate (origin) at (axis cs:0,0);
	\fill[red,fill opacity=0.5] (origin) circle(1.5pt);
	\end{axis}
\end{tikzpicture}
}
\label{subfig:SimpleIDs}
}\\
\subfloat[Isospectral domains 2]{
\resizebox{0.8\linewidth}{!}{
\begin{tikzpicture}
	\begin{axis}[plot box ratio=1 1]
	\addplot[mark=none,fill=yellow, fill opacity=0.8] file {Plots/Geometries/ID2_2_Coord.dat} --cycle;
	\coordinate (COM2_2) at (axis cs:1.0095,-2.7048);
	\fill[blue] (COM2_2) circle(1pt);
	\coordinate (origin) at (axis cs:0,0);
	\fill[red,fill opacity=0.5] (origin) circle(1.5pt);
	\end{axis}
\end{tikzpicture}
\begin{tikzpicture}
	\begin{axis}[plot box ratio=1 1]
	\addplot[mark=none,fill=yellow, fill opacity=0.8] file {Plots/Geometries/ID2_1_Coord.dat} --cycle;
	\coordinate (COM2_1) at (axis cs:0.9905,-0.4381);
	\fill[blue] (COM2_1) circle(1pt);
	\coordinate (origin) at (axis cs:0,0);
	\fill[red,fill opacity=0.5] (origin) circle(1.5pt);
	\end{axis}
\end{tikzpicture}
}
\label{subfig:ElaborateIDs}
}
\caption[The isospectral domains studied in this work.]{Two pairs of isospectral domains treated in \cite{gordon_isospectral_1992}. These two-dimensional geometries constitute the
main subjects of study of the present work. The red dot signifies the origin of the coordinate system, the blue dot represents
the centre of mass of the figures. The importance of both shall become apparent in due course. }
\label{fig:Intro_IsospectralManifolds}
\end{figure}
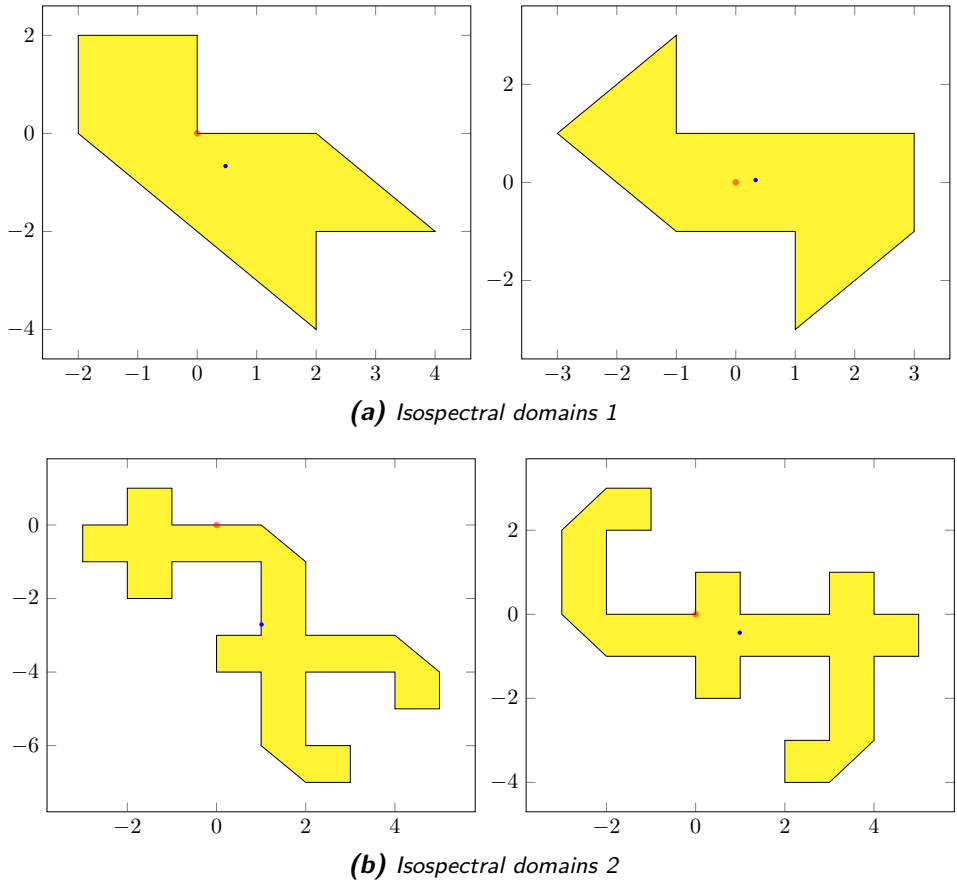

Isospectral domains are formally just domains which posses identical spectra. Whilst sets of domains which were obtained from each other by some 
sort of transformation (rotations, reflections and translations) are obviously isospectral they are not in the focus of the present work.
Here 2-dimensional surfaces have been investigated whose spectra are identical but whose shapes are different. Two such pairs are shown in \ref{fig:Intro_IsospectralManifolds}
and other examples may be found in \cite{buser_planar_2010}.

The origin of the work on isospectral manifolds is commonly attributed to Mark Kac. He asked, in \cite{kac_can_1966}, the question eponymous to
the study of isospectral domains: "Can one hear the shape of a drum?". The motivation for such a question has been provided by Hermann Weyl's work on
the connection of the spectrum of a domain and its geometric properties.
With Kac being a mathematician the physical question was immediately translated into a mathematical one. The latter being whether it is possible to
identify the shape of a domain if one were to know the complete spectrum of the Laplacian defined on it.
It is thus perhaps not surprising that a lot of work on the generalizations and specializations of the original question has been done by mathematicians,
as can be seen i.e. in \cite{sunada_riemanninan_????, gordon_isospectral_1992,buser_isospectral_1986} (and many more). Abstract proofs of isospectrality 
flourish in the domain and the astute reader will notice that no reference is made as to a way of measuring the spectrum.
A generalization of Kac' question was almost immediately answered by John Milnor \cite{kac_can_1966} in the negative for 16-dimensional torii.

The problem did not go unnoticed by physicists. \citeauthor{sridhar_experiments_1994} have determined some eigenfrequencies of the domains featured in \cite{gordon_isospectral_1992}
(and shown in \ref{subfig:SimpleIDs}) experimentally \cite{sridhar_experiments_1994}. In such a way they have shown that the domains are indeed isospectral.
The result was achieved by fabricating said domains and measuring their resonance frequencies as well as mapping the associated eigenmodes.

The standard reference for planar isospectral domains is \cite{gordon_isospectral_1992}. The proofs of isospectrality mainly rely either
on the theorem of Sunada \cite{sunada_riemanninan_????} or the transplantation method of Buser \cite{buser_isospectral_1986}. We will encounter both when 
surveying the mathematical literature on the subject in \ref{sec:Theo_IDsPriorArt} and deriving analytic results in \ref{ch:TheoAnalysis}.
Prior to immersing oneself into mathematics the definitions necessary for a basic understanding of the subject are provided in \ref{ch:MathematicalMinimum}.

The analysis of the mathematical literature is done in \ref{sec:Theo_IDsPriorArt}, where I attempt to determine whether the problem at hand has not already been
solved - perhaps in a convoluted way. In anticipation of the results I shall state here that to the best of my knowledge the central question of this thesis
has not yet been answered.

I shall make no attempt at being thorough in the mathematical explanations for that has been done \textit{in extenso} by others and it would 
distract from the main point of this work. The interested reader is referred to \cite{Tu2010,Marsden1987,Nakahara2015,hehl_foundations_2003}
for an introduction  to manifolds and differential forms as well as their use in classical mechanics and electrodynamics.

The mathematical introduction is followed by a presentation of the physical basis in \ref{ch:PhysicalBasis}. Therein classical electrodynamics
is formulated using differential forms. This beautiful formulation is then used in \ref{ch:Derivations} to obtain a wave equation with
additional terms for a uniformly rotating medium and a co-rotating observer.

Prior to engaging in a full-on numerical assault on the question stated above a brief theoretical analysis of the
equation obtained in \ref{ch:Derivations} is provided in \ref{ch:TheoAnalysis} by subjecting simple shapes
to uniform rotation and studying their properties.

Such analytic considerations are of limited use only and the main investigatory work has been performed numerically
and is presented in the second part of this work. Due to a multitude of methods for the solution of partial differential equations,
an overview is provided in \ref{ch:Exp_Introduction} along with the reasoning behind the choice of the finite element method (FEM).
An algebraic equivalent of the differential equation determined in \ref{ch:Derivations} is then derived in \ref{ch:Exp_AlgebraicFormulation}
using FEM. It is then used to obtain the results presented in \ref{ch:Exp_TheExperiment} and to answer the question posed at the outset of this
work.

The closing act is then made up of the appendix, where miscellaneous results and derivations are given, as well as a summary and a set of questions not yet
answered. The latter arose in the course of the research of the topic at hand and are still awaiting further treatment.

Last but not least it should be mentioned that the question whether the spectrum of the Laplace(-Beltrami) operator may be used to identify shapes has surfaced
in a different variation in the field of computer vision \cite{reuter_laplacebeltrami_2006}. There one is concerned with whether said spectrum may be used
to find similarities between shapes. 
\thispagestyle{plain}
\clearpage

\mainmatter
\part{Theory}

\chapter{Mathematical Minimum}\label{ch:MathematicalMinimum}
\epigraph{
The book of nature is written in the language of mathematics.
}{Galileo Galilei - \textit{The Assayer} }
\section*{Introduction}
As has been remarked in the introduction the subject of isospectral domains has been studied in extenso by mathematicians.
Therefore, to understand the state of the literature requires one to first acquire some familiarity with the language which is often employed.

An extensive treatment of all mathematical peculiarities is quite outside the scope of this work, nevertheless
we shall bite the bullet and introduce the formalism of differential forms and basics of differential geometry,
not least because it is my opinion that the formalism is practical \textit{and} beautiful and permits a better treatment of 
some aspects of electrodynamics than, for instance \cite{Post1962} or the tensor calculus referenced therein and 
presented by \cite{Schouten2011}. The language of differential forms shall prove to be more than just
a mathematical dialect in due course.

The ultimate goal of this chapter is to introduce the Laplace-Beltrami operator, which is central to the subject of this work.
This is done in the following section. In the last part of this chapter the Laplace-Beltrami operator is then considered in the
context of isospectral domains and the difference between mathematics and physics.

\section{Definitions}\label{sec:Theo_DGDefinitions}

\subsection{Bare Structure}\label{subsec:Theo_DG_BareStructure}
\subsubsection{Manifolds and Tangent Spaces}
\begin{defn}[Charts, Atlases and Manifolds]\label{def:Theo_ChartAtlasmanifold}
Let $\Mf$ be a topological manifold. Then it is locally euclidean if
\begin{equation}
\forall p\in\Mf\;\exists U(p)\subset\Mf\;:\;\varphi: U\rightarrow \RR^n \text{ is a homeomorphism}.
\end{equation}
Then $U$ is designated a local coordinate neighbourhood of $p$, $\varphi$ a local coordinate system and the pair 
$(U(p),\varphi(p))=(U(p), (x^1(p),\dots,x^n(p) )$ is called a (local) chart. The coordinate functions on $U(p)$ are $x^i:=r^i\circ\varphi$, where $r^i$
are the canonical coordinates on $\RR^n$.

A set $\mathcal{A}=\bigcup_i(U_i,\varphi_i)$ of charts which cover the manifold and for any two of which there exists a smooth mapping between the
charts is called an atlas of the manifold.

Finally, a locally euclidean topological manifold $\Mf$ equipped with an atlas is called a smooth manifold.
\end{defn}

The above definitions are collated from \cite{Tu2010} and are given for the sake of completeness. In what follows we shall deal mostly with
 $\Mf := \RR^{\lbrace 2,3,4\rbrace }$ as our underlying manifold. A general formulation facilitates the transfer of the methods presented in
\ref{ch:Exp_Introduction} to other geometries. The terms coordinate system and coordinate frame will used interchangeably.
 Next the definition of the actual spaces we shall be
dealing with is provided.

\begin{defn}[(Co)tangent spaces]\label{def:Theo_TangentCotangentSpaces}
Let $\Mf$ be a manifold and $\gamma: \RR \supset I \rightarrow \Mf$, with $0\in I$, be a smooth curve s.t. $\gamma(0)=p\in\Mf$. Then a 
tangent vector $v$ to $\Mf$ at $p$ is a mapping $C^\infty(\Mf)\rightarrow\RR$ given by 
\begin{multline}
v_pf = v_p(f) = \frac{d}{dt}\biggm\vert_{t=0}(f\circ\gamma)
 = \frac{d}{dt}\biggm\vert_{t=0}(\psi\circ h) = \frac{\del\psi}{\del r^i}\biggm\vert_{h(0)}\frac{dh^i}{dt}\biggm\vert_{t=0}\\
= \frac{\del(f\circ\varphi^{-1})}{\del r^i}\biggm\vert_{h(0)}\frac{d(r^i\circ\varphi\circ\gamma)}{dt}\biggm\vert_{t=0}
= \frac{\del(f\circ\varphi^{-1})}{\del r^i}((\varphi\circ\gamma)(0))\frac{d(r^i\circ\varphi\circ\gamma)}{dt}(0)\\
 =: \frac{\del f}{\del x^i}(\gamma(0))\frac{d(x^i\circ\gamma)}{dt}(0)
= \frac{\del f}{\del x^i}\biggm\vert_{p=\gamma(0)}\cdot \frac{d\gamma^i}{dt}\biggm\vert_{t=0}
\end{multline}
where $\del_{x^i}\vert_p f = \frac{\partial}{\partial r^i}\vert_{\varphi(p)}(f\circ\varphi^{-1})$.
The space spanned by all vectors tangent to $\Mf$ at $p$ is called the tangent space $T_p\Mf$. It's dual is the cotangent space $T_p^\ast\Mf$.
The natural bases of both are
\begin{equation}
\CalB_{T_p\Mf}:=\bigl\{ \del_{x^i}\vert_p\bigr\}_{i=1}^n,\qquad \CalB_{T_p^*\Mf}^\ast :=\bigl\{ dx^i\vert_p\bigr\}_{i=1}^n\; .
\end{equation}
With elements of $\CalB$ being denoted as (contravariant) vectors and elements of $\CalB^\ast$ as co(variant)vectors or coordinate 1-forms.
\end{defn}
The basis $\CalB$ may be obtained from the tangent vector by abstracting away the function $f$. The tangent space is thus just a vector space
with a 'strange' basis.
The definition above mixes the definitions of a geometric and algebraic tangent vectors, that is - derivations, as presented in \cite{Janich2005}.
This definition of the tangent vector at $p$ is identical to the directional derivative of $f$ in the direction of the
velocity vector of $\gamma$. Furthermore, for completeness' sake note that the set of all tangent spaces, along with their base point $p$, to a manifold is
designated the tangent bundle $T\Mf$.

Though the above definitions may seem abstract at first they shall not remain so for long, as they shall be put to good use in due course.
It is important to note that the action of a coordinate 1-form, at an arbitrary but fixed $p\in\Mf$ on $v\in T_p\Mf$, is similar to an orthogonal projection
of $v$ onto the coordinate axis associated with $\del_{x^i}$, but without
a concept of an inner product or metric (which shall be defined below in \ref{def:Theo_MetricTensor}).
The similarity is such that an application of $dx^i\vert_p$ to $v$ will yield the coefficient associated with $\del_{x^i}$ (this follows trivially from the definition of the dual space).
We shall encounter this peculiarity when dealing with so-called adapted coframes (c.f. \ref{subsec:Theo_Phys_Time}).

Henceforth, whenever two manifolds $\Mf,\Nf$ will be necessary we shall assume $\dim\Mf = m,\dim\Nf = n$.
\begin{defn}[Tangent Map]\label{def:Theo_TangentMap}
Let $\Mf,\Nf$ be manifolds as defined above and let $\Phi:\Mf\rightarrow\Nf$. Then $\Phi$ is a diffeomorphism of manifolds
if it is smooth w.r.t. all charts $(U,\varphi)=(U,(x^1,\dots,x^m))$ of $\Mf$ and $(V,\psi)=(V,(y^1,\dots,y^n))$ of $\Nf$.

Then $\Phi$ induces a map for all points of $\Mf$, the so called tangent map $d\Phi:\T_p\Mf\rightarrow T_{\Phi(p)}\mathcal{N}$.
Using charts we then have
\begin{equation}
\left[d\Phi\right]_{\CalB_{T_p\Mf},\mathcal{C}_{T_{\Phi(p)}\Nf}} = \left[ \frac{\partial y^i}{\partial x^j}\right]^{i=1,\dots, n}_{j=1,\dots,m}\;,
\end{equation}
that is, a Jacobi matrix of $d\Phi$.
\end{defn}
Although coordinate transformations are performed on the same manifold, they can be considered w.l.o.g. as maps between two copies of the same manifold,
which facilitates the visualization of the transformations.
\subsubsection{Tensors, Fields and Differential Forms}
\begin{defn}[Tensor]\label{def:Theo_Tensor}
A multilinear map 
\begin{align*}
T:\underbrace{T_p^\ast\Mf \times\dots\times T_p^\ast\Mf}_{s-\text{times}} \times \underbrace{T_p\Mf \times\dots\times T_p\Mf}_{l-\text{times}} &\rightarrow \KK\\
\left(T\vert v_1^\ast,\dots,v_s^\ast,v_1,\dots,v_l\right) &= T(v_1^\ast,\dots,v_s^\ast,v_1,\dots,v_l)
\end{align*}
is called a \textit{tensor} of contravariant rank $s$ and covariant rank $l$. The space of such tensors is
\[ T\indices{^s_{l,(p)}}(\Mf) := \underbrace{T_p\Mf \times\dots\times T_p\Mf}_{s-\text{times}}\times \underbrace{T_p^\ast\Mf \times\dots\times T_p^\ast\Mf}_{l-\text{times}} \].
\end{defn}

Given a basis $\CalB$ of the tangent space to $\Mf$ as well as its associated cobasis a tensor  $T\in T\indices{^s_{l,p}}(\Mf)$ is
specified unambiguously by its components $T\indices{^{i_1,\dots,i_s}_{j_1,\dots,j_l} }$ as
\begin{equation}
T = T\indices{^{i_1,\dots,i_s}_{j_1,\dots,j_l} }
\partial_{x^{i_1}}\otimes\dots\otimes\partial_{x^{i_s}}\otimes dx^{j_1}\otimes\dots\otimes dx^{j_l}\quad .\label{eq:Theo_TensorWRTaBasis}
\end{equation} 
What physicists refer to as 'tensor' is in fact this same set of coefficients (c.f. \cite{Schouten2011}). In this work we will
take care to always give the full representation of a tensor because the explicit basis shall prove quite useful.

A tensor field of contravariant rank $s$ and covariant rank $l$ is a function which assigns to each point of the manifold $\Mf$ a
tensor in $T^s_{l,(p)}\Mf$. Henceforth we shall generally use only tensor fields. The space of all such fields on the manifold is denoted
$\Tf{s}{k}{\Mf}$. It is obvious that $\Tf{1}{0}{\Mf}$ is the space of all vector fields on $\Mf$.

\begin{defn}[Differential $k$-Form]\label{def:Theo_DifferentialForm}
A function which, at every point $p\in\Mf$ provides an alternating $(0,k)$ tensor in $T^0_{k,p}(\Mf)$
is called a differential form of degree $k$, or simply an $k$-form.
Given a cobasis it is expressed by 
\begin{equation}
\omega_p(v_1,\dots,v_k)  := \omega_{i_1,\dots,i_k}(p) \begin{vmatrix}
dx^{i_1}(v_1) & \cdots & dx^{i_1}(v_k)\\
\vdots & \ddots & \vdots \\
dx^{i_n}(v_1) & \cdots & dx^{i_n}(v_k)
\end{vmatrix}\;.
\end{equation}
A function $f:\Mf\rightarrow\RR$ may be defined as a $0$-form.
The space of all $k$-forms on $\Mf$ with smooth coefficients is denoted $\Omega^k(\Mf)$ and is of dimension $\binom{m}{k}$.
\end{defn}
\begin{notation}
Let $\omega\in\Omega^k(\Mf)$. Often times one indicates the degree of the differential form as a superscript, thus $\omega^k$.
Furthermore, if there is no danger of confusing the manifolds a form is defined on we shall omit the explicit mention of the manifold.
\end{notation}

\begin{defn}[Exterior Product]\label{def:Theo_WedgeProduct}
The exterior product, or wedge-product, of forms is a generalization of the vector-product of classical analysis. 
It is a mapping $\wedge : \Omega^p(\Mf)\times\Omega^s(\Mf) \rightarrow \Omega^{p+s}(\Mf)$ with the following properties
\begin{subequations}\label{eq:Theo_WedgeProperties}
\begin{align}
(\omega^p+\omega^s)\wedge\omega^l &= \omega^p\wedge\omega^l+\omega^s\wedge\omega^l\\
(\omega^p\wedge\omega^s)\wedge\omega^l &= \omega^p\wedge(\omega^s\wedge\omega^l)\\
\omega^p\wedge\omega^s &= (-1)^{s\cdot p}\omega^s\wedge\omega^p
\end{align}
\end{subequations}
for $\omega^s\in\Omega^s,\omega^p\in\Omega^p,\omega^l\in\Omega^l$.
\end{defn}
It follows trivially (c.f. \cite[Ch. 15]{Zorich2006}) that any $k$-form on $\Mf$ may be expressed in a chart as 
\begin{equation}
\omega^k_p = \sum_{i_1\leq \dots\leq i_k} \omega_{i_1, \dots, i_k}(p) dx^{i_1}\wedge
\dots\wedge dx^{i_k}\quad .\label{eq:Theo_lFormFrom1Forms}
\end{equation}
The exterior product permits a simple construction of differential forms from $1$-forms. Furthermore it is
trivial to see that any differential form of degree higher than the dimension of the manifold it is defined on is identically zero.
As is evident from the above equation differential forms are basically integrals without an integral sign. They, or to be more precise
their densities (to be defined below) are the natural candidates for quantities to be integrated over manifolds.

\subsubsection{Pre-Metric Operators}

\begin{defn}[Interior Multiplication I]\label{def:Theo_IPVectorFieldAndDiffForm}
Let $\mathcal{X}\in\Tf{1}{0}{\Mf},t\in\Tf{0}{k}{\Mf}$. Then the interior multiplication of $\mathcal{X}$ with
$t$ is defined as
\begin{align*}
\iota : \Tf{1}{0}{\Mf}\times\Tf{0}{k}{\Mf} &\rightarrow \Tf{0}{k-1}{\Mf}\\
(\mathcal{X},t) &\mapsto \iota_\mathcal{X}(t) = t(\mathcal{X},\cdots)\quad .
\end{align*}
For a function $\iota_\mathcal{X}f = 0$.
\end{defn}
The above definition serves mainly as a shorthand notation in cases where a tensor is to be evaluated on a vector field.
The latter reduces the degree of the tensor by permanently occupying the position first argument.
Next we define the most important operator:
\begin{defn}[Exterior Derivative]\label{def:Theo_ExteriorDerivative}
Let $\omega^l\in\Omega^l(\Mf)$ be given by \eqref{eq:Theo_lFormFrom1Forms} w.r.t. a coordinate chart. 
The exterior derivative is defined as
\begin{align}
d:\Omega^{l} &\rightarrow \Omega^{l+1}\notag\\
\omega^l &\mapsto \omega^{l+1} = d\omega = \sum_{i_1\leq \dots\leq i_l} d\omega_{i_1,\dots, i_l}\wedge dx^{i_1}\wedge
\dots\wedge dx^{i_l}
\end{align}
where $d\omega_{i_1,\dots, i_l}$ is the common derivative of a function and hence a 1-form.
\end{defn}
\begin{rem}\label{rem:Theo_ExteriorDerivativeProperties}
Following properties hold for the exterior derivative
\begin{enumerate}
\item Distributivity w.r.t. addition: $d(\omega_1 + \omega_2) = d\omega_1 + d\omega_2\quad \omega_i\in\Omega^l(\Mf)$.
\item Product rule: $d(\omega^l\wedge \omega^s) = d\omega^l \wedge \omega^s + (-1)^{\deg\omega^l}\omega^l\wedge d\omega^s$.
\item Nilpotency: $d(d\omega) = 0\qquad \forall \omega\in\Omega^l(\Mf)$.
\end{enumerate}
\end{rem}
\subsubsection{Maps of Tensor-Fields}
\begin{defn}[Pullback]\label{def:Theo_Pullback}
Let $\Mf,\Nf,\Phi$ be as in \ref{def:Theo_TangentMap} and let $\nu\in\Tf{0}{k}{\Nf}$. Then the pull-back of $\nu$ by $\Phi$
is $\Phi^\ast\nu=:\omega\in\Tf{0}{k}{\Mf}$:
\begin{align*}
\Phi^\ast : \Tf{0}{k}{\Nf} & \rightarrow \Tf{0}{k}{\Mf}\\
\nu &\mapsto \omega = \Phi^\ast\nu\\
\intertext{such that}
\omega_p(v_1,\dots,v_k) = (\Phi^\ast\nu)_p(v_1,\dots,v_k) &:= \nu_\Phi(p)\left(d\Phi(v_1),\dots,d\Phi(v_k)\right)\quad ,
\end{align*}

where $v_i\in T^1_{0,(p)}\Mf$.
\end{defn}
A pull-back is thus a map in the direction opposite to $\Phi$ and to the 
\begin{defn}[Pushforward]\label{def:Theo_Pushfwd}
Let $\Mf,\Nf,\Phi$ be as above and $\mathcal{X}\in\Tf{s}{0}{\Mf}$. Then the push-forward of $\mathcal{X}$ by $\Phi$
is
\begin{equation}
\mathcal{Z} = \Phi_\ast(\mathcal{X}) = \Phi_\ast \sigma^{i_1,\dots,i_s} \del_{i_1}\otimes\cdots\otimes \del_{i_s}
= (\sigma^{i_1,\dots,i_s}\circ\Phi^{-1}) \left(d\Phi(\del_{i_1})\right)\otimes\cdots\otimes\left(d\Phi( \del_{i_s})\right)\ .
\end{equation}
A push-forward of a covariant tensor is defined as its pull-back by the inverse map.
\end{defn}
A push forward thus maps vector fields (or contravariant tensors) along $\Phi$ whilst the pull-back maps covariant tensors in the
opposite direction.
Note that the exterior derivative of \ref{def:Theo_ExteriorDerivative} behaves naturally under the pull-back operation, i.e.
\begin{equation*}
\Phi^\ast d\omega = d(\Phi^\ast\omega)\quad .
\end{equation*}
\subsubsection{Tensor Densities}
The orientation of a manifold is given by a set of compatible (under a change of coordinate charts) orientations of its tangent spaces, for
which the orientation is defined as the sign of the determinant of a matrix composed of the basis vectors. In a more palatable way it is the order of the
basis vectors.
\begin{defn}[Twisted Form]\label{def:Theo_TwistedForm}
Let $\Mf$ be an $n$-dimensional manifold and  $\omega\in\Omega^n(\Mf)$. Then a (tensor) density on $\Mf$ is defined by
\begin{equation}
\tilde{\omega}(v_1,\dots,v_n) := \begin{cases}
\omega(v_1,\dots,v_n) & (v_1,\dots,v_n) \text{ positively oriented}\\
\frac{1}{-1}\omega(v_1,\dots,v_n) & \text{else}\;.
\end{cases}
\end{equation}
Formulated w.r.t. coordinate charts $(U,\varphi),(V,\psi)$ with a coordinate change $h:=\psi\circ\varphi^{-1}$ on $U\cap V\neq \emptyset$
the above is equivalent to:
\begin{equation}
\omega_{i_1,\dots,i_n}(\psi(p))dy^{i_1}\wedge\cdots\wedge dy^{i_n} = 
\frac{1}{\sgn\det (J_h\vert_p)}\frac{\del y^{i_1}}{dx^{j_1}}\dots\frac{\del y^{i_n}}{dx^{j_n}} \omega_{i_1,\dots,i_n}(\phi(p)) dx^{j_1}\wedge\cdots\wedge dx^{j_n}\;,
\end{equation}
where $J_h$ is the Jacobi matrix of the coordinate change. A twisted form collects a sign of the Jacobian under the pull-back operation.
\end{defn}
The above is a modification of a definition given in \cite[Ch. 5]{Janich2005}. As remarked therein the difference between forms and densities is a
formal one, as long as the manifold is orientable and an orientation is chosen. Furthermore densities and not differential forms
are actually the prime candidates for integration on manifolds. The strength of densities (or twisted forms) is that, unlike common differential forms,
they can be integrated over non-orientable manifolds. 
A very nice explanation of forms and densities is given in \citetitle{Janich2005} by \citeauthor{Janich2005}, whereas \citeauthor{hehl_foundations_2003} provide concise, technical definitions and
an example with the M\"obius strip as a manifold. It remains to be added that in comparing the above definition with \cite{Schouten2011} one notices that we have only 
defined a density of weight $+1$, densities of different weights may be defined as delineated therein. For the purposes of this work the above definition will suffice.
The orientation as it is described above is called 'intrinsic orientation' due to being defined by quantities defined on the manifold only. An 'extrinsic' or 'exterior' orientation
of a manifold may be obtained by restricting a volume form of the ambient space onto the manifold. The 3-space equipped with a screw-sense induces an orientation (sense of rotation)
on a plane embedded therein. The M\"obius strip possesses neither an intrinsic, nor an exterior orientation.
\subsubsection{Integration}
Although the concept of integration is important we shall not dwell upon it for its role in this work is rather limited. 
The necessary notions are subsumed in the following
\begin{rem}
Let $\Mf$ be an $n$-dimensional manifold and $\omega\in\Omega^k(\Mf)$, then the integral of $\omega$ over $\Mf$
is denoted
$ \int_\Mf \omega$ 
and its explicit evaluation is performed using charts. With the exterior derivative the Stokes theorem
takes on the following form
\begin{equation}
\int_\Mf d\omega = \int_{\del\Mf} \omega\quad ,
\end{equation}
where $\del\Mf$ is the (topological) boundary of the manifold. The nilpotency of $d$, as remarked upon in rem. \ref{rem:Theo_ExteriorDerivativeProperties}, is equivalent to the
statement that the boundary of a boundary is an empty set.
Given a map as in \ref{def:Theo_TangentMap} and $\nu^l\in\Omega^l(\Nf)$ the integral behaves in the following way (assuming $\Phi$ is orientation-preserving)
\begin{equation}
\int_\Nf \nu^l = \int_{\Phi(\Mf)} \nu^l = \int_{\Mf} \Phi^\ast\nu^l = \int_{\Mf} \omega^l\quad .
\end{equation}
\end{rem}
\subsection{Metric Structure}\label{subsec:Theo_DG_MetricStructure}
\subsubsection{Metric Tensor and Duality}\label{subsubsec:MetricTensorAndDuality}
To be useful in practice the manifolds should be equipped with further structure. First and foremost being the metric tensor (physically: a 
measure for distances).
\begin{defn}[Metric Tensor]\label{def:Theo_MetricTensor}
Let $g\in \Tf{0}{2}{\Mf}$ be non-degenerate, i.e.,
\[ g(u,v) = 0\quad\forall u\in T_p\Mf \Rightarrow v=0. \]
Then $g$ is a \textit{pseudo-metric} tensor if
it is symmetric \[g(u,v) = g(v,u)\quad \forall u,v\in T_p\Mf\;.\]
If, additionally, it is positive definite, that is \[ g(u,u) > 0 \quad\forall u\in T_p\Mf, \]
then $g$ is called a \textit{metric} tensor.
\end{defn}
Regularity is not a necessary property for the metric. It is included in the above definition for reasons of convenience.
We shall be working only with regular (pseudo-)metrics. Furthermore we note that given
a metric according to the above definition the orthogonality of two vectors becomes well-defined.

Given a basis $\CalB_{T_p\Mf}$ one can apply Sylvester's law of inertia to the Gram matrix of $g$. The
number $s$ of negative entries of the Sylvester form is called the index of the metric.

The distinction between a metric and a pseudo-metric may seem as a purely mathematical one, but
its physical implication is the presence or absence of the light-cone. The confusion is increased 
by the following nomenclature
\begin{notation}
A pseudo-metric of index $s=-3$ (or $s=-1$, depending on the convention) on $\RR^4$ is called \textit{Lorentzian metric}. Similarly a metric
of signature $s=0$ is called \textit{Riemannian metric}.
A tuple $(\Mf,g)$ consisting of a manifold $\Mf$ and a metric $g$ is denoted a riemannian/lorentzian manifold if the metric is riemannian/lorentzian.
\end{notation}
The distinction is not always present in physics, \cite{fliesbach_allgemeine_2012} is a prime example where it is not present and where the lorentzian structure
of general relativity is mislabelled as Riemannian.
The distinction of the two shall prove important when analysing the corpus of mathematical publications 
on the topic of isospectral domains (c.f. \ref{sec:Theo_IDsPriorArt}), we shall thus adhere to it.
\begin{defn}[Musical Maps]\label{def:Theo_MusicalMaps}
Let $g$ be a non-degenerate (pseudo)metric. Then a mapping from $T_p\Mf$ to $T^\ast_p\Mf$ can be defined via $g$ as
\begin{align*}
\flat : T_p\Mf &\rightarrow T_p^\ast\Mf\\
v &\mapsto v^\flat := g(v,\cdot)
\end{align*}
and $\sharp := \flat^{-1}$ is the inverse map.
\end{defn}
Thus, given some type of metric on a manifold one may freely transform between tangent and cotangent spaces. Which explains the
general lack of distinction of the two spaces in physics, where the existence of a metric is always assumed and independence of expressions
from the explicit form of the metric is postulated.
\subsubsection{Hodge-Duality}
For practical applications we require the definition of an interior product of differential forms.
\begin{defn}[Interior Product of Forms]\label{def:Theo_IPOfForms}
Let $k\geq l,\quad \iota:\Omega^k\times\Omega^l\rightarrow\Omega^{k-l}$. Then $\iota_{\omega^k}(\omega^l)$ is an interior product of differential forms if it fullfills the following properties
\begin{enumerate}
\item It is bilinear in its arguments.
\item For 1-forms holds: $\iota_\nu(\omega) = \iota_{\nu^\sharp}(\omega) = \omega(\nu^\sharp)$. Where the expression is obviously symmetric in $\nu,\omega$.
\item It is an antiderivation \cite{Tu2010} of order $-1$: $\iota_\nu(\omega^k\wedge\omega^l) = \iota_\nu(\omega^k)\wedge\omega^l + (-1)^{k}\omega^k\wedge\iota_\nu(\omega^l)\;\nu\in\Omega^1$.
\item $\iota_{\nu_1\wedge\nu_2} = \iota_{\nu_2}\circ\iota_{\nu_1},\quad \nu_i\in\Omega^1$.
\end{enumerate}
\end{defn}
The interior product of forms is required for the definition of the second most-important operator,
\begin{defn}[Hodge Operator]\label{def:Theo_HodgeOp}
Again, let $g\in\Tf{0}{2}{\Mf}$ be non-degenerate. If we define $\Tf{0}{0}{\Mf}:=\RR$ then using \ref{def:Theo_VolumeForm} one can define a map
\begin{align*}
\ast: \Omega^l(\Mf) &\rightarrow \text{twisted } \Omega^{n-l}(\Mf)\\
\omega &\mapsto \ast\omega := \iota_{\omega}(\epsilon)\quad.
\end{align*}
The mapping is called Hodge operator and is a bijection with the following important properties
\begin{enumerate}
\item $1\in\Omega^0\Rightarrow \ast 1 = \epsilon$
\item $\ast\circ\ast\bigg\vert_{\Omega^l} = (-1)^{(l\cdot(n-l)+s)} \mathbb{I}_{\Omega^l}$ if $g$ is symmetric with signature $s$.
\item $\ast\epsilon = \iota_\epsilon(\epsilon) = (-1)^s$.
\end{enumerate}
\end{defn}
Where $\epsilon$ is the canonical volume form (top form) defined by
\begin{defn}[Volume Form]\label{def:Theo_VolumeForm}
Let $\Mf$ be an $n$-dimensional manifold equipped with a non-degenerate tensor $g\in\Tf{0}{2}{\Mf}$. We define the canonical volume form
$\epsilon\in\Omega^n(\Mf)$ as:
\begin{equation}
\epsilon := \sqrt{\vert \det g\vert} dx^1\wedge\cdots\wedge dx^n\;,
\end{equation}
where $\det g$ is of course the determinant of the Gram matrix of $g$.
\end{defn}

The expressions for the Hodge-duals of the coordinate 1- and 2-forms on $\RR^4$ for a general metric tensor $g\in\Tf{0}{2}{\RR^4}$ are
provided in anticipation of their practical use.
For 1-forms we have
\begin{equation}
*dx^i = \sqrt{ \left\vert \det g\right\vert } \left[ g^{i0} dx^1\wedge dx^2\wedge dx^3
- g^{i1}dx^0\wedge dx^2\wedge dx^3 + g^{i2} dx^0\wedge dx^1\wedge dx^3
-g^{i3} dx^0\wedge dx^1\wedge dx^2\right]\;,\label{eq:GeneralHodgeDualOfOneForm}
\end{equation}
whereas for a 2-form we obtain
\begin{multline}
*(dx^i\wedge dx^j) = \sqrt{\left\vert \det g \right\vert } \left[ g^{i0}\left(
g^{j1}dx^2\wedge dx^3 - g^{j2}dx^1\wedge dx^3 +g^{j3}dx^1\wedge dx^2\right) \right.\\
 - g^{i1}\left( g^{j0}dx^2\wedge dx^3 - g^{j2} dx^0\wedge dx^3 + g^{j3} dx^0\wedge dx^2\right)
 + g^{i2}\left( g^{j0}dx^1\wedge dx^3 - g^{j1} dx^0\wedge dx^3 + g^{j3} dx^0\wedge dx^1\right)\\
\left. - g^{i3}\left( g^{j0} dx^1\wedge dx^2 - g^{j1} dx^0\wedge dx^2 + g^{j2}dx^0\wedge dx^1\right)
\right]\quad .\label{eq:GeneralHodgeDualOfTwoForm}
\end{multline}
Note that in both cases $g^{ij}$ denote the components of the inverse of the Gram matrix of $g$.
The expressions for $3$-forms are derived trivially from \eqref{eq:GeneralHodgeDualOfOneForm} by application of \ref{def:Theo_HodgeOp}.
The derivation of \eqref{eq:GeneralHodgeDualOfOneForm} is provided in \ref{sec:App_HodgeDual}.

The Hodge operator, by merit of the volume form transforms differential forms into twisted differential forms (or densities) and vice versa.
With the definition of this operator we may now define an additional operator, the co-differential
\begin{defn}[Co-differential]
The co-differential $\delta:\Omega^{m-l}\Mf\rightarrow\Omega^{m-l-1}\Mf$ is defined by $\delta = (-1)^l\ast d\ast^{-1}$. The latter is equivalent to
$\delta = (-1)^{m(l-1)}\ast d \ast$.
\end{defn}
With this we may now make the final definition of the necessary mathematical operators
\begin{defn}[Laplace-Beltrami Operator]\label{def:Theo_LaplaceBeltrami}
The so-called Laplace-Beltrami operator\linebreak $\Delta:\Omega^l\Mf\rightarrow \Omega^l\Mf$ of differential geometry is
\begin{equation}
\Delta := d\delta + \delta d\quad.
\end{equation}
\end{defn}
It is important to note, that for a Riemannian metric, the Laplace-Beltrami operator is the same as the classical Laplace operator, but for a Lorentzian metric
it is equivalent to the d'Alembert operator.

As promised at the beginning of the chapter all prior definitions were used to arrive at the definition of the Laplace-Beltrami operator.
\section[Prior Art]{Isospectral Domains - Prior Art}\label{sec:Theo_IDsPriorArt}

\begin{figure}[bp]
\centering
\subfloat[Orbifolds]{
\includegraphics[scale=0.7]{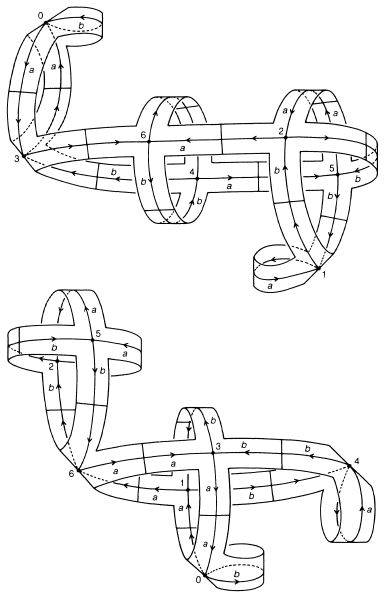}
}\hspace{1em}
\subfloat[Fundamental tile]{
\includegraphics[scale=1]{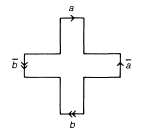}
}
\caption[Orbifolds of \cite{gordon_isospectral_1992}]{The two orbifolds $G_1\setminus\Mf,G_2\setminus\Mf$, along with their fundamental tile, taken from \cite{gordon_isospectral_1992}, from which \ref{subfig:ElaborateIDs} results by cutting along the center plane. Both are "embedded" in $\RR^3$ for visualization purposes only.}
\label{fig:Theo_GWW_QuotientManifolds}
\end{figure}
With the basic definitions given above we may now proceed to the analysis of the origins of the domains presented in \ref{fig:Intro_IsospectralManifolds},
as well as the prior research done by mathematicians, in order to understand whether the question at hand has been answered, in a convoluted way perhaps, by mathematicians.
Another reason for taking a closer look at the origin of the two domains is provided by the fact that 
the domains possess no obvious symmetries or special points which could be used as the base point for the axis of rotation. Hence a closer look at the
creation process of the domains is warranted to see whether any such point may become apparent in due course. Alas, neither hold true.
Furthermore it shall be shown that the proofs presented in 
\cite{gordon_isospectral_1992},\cite{sunada_riemanninan_????},\cite{buser_isospectral_1986},\cite{berard_transplantation_1992} and \cite{gordon_robin_2015}
are not applicable (at least in a straightforward way) to the problem of rotating isospectral domains.

It should be remarked upon, that the mathematical notions introduced so far are not sufficient for a thorough understanding of the literature surveyed in this section.
I shall refrain from making further definitions because the mathematical theory presented here will play no significant role beyond this chapter and a thorough
mathematical analysis does not constitute the main focus of this work. Furthermore only the gist of the publications on the topic of the theory of isospectral domains is presented here,
with publications on the numerical treatment of the domains relegated to \ref{ch:Exp_Introduction}.

The domains of \ref{fig:Intro_IsospectralManifolds} are adapted from \cite{gordon_isospectral_1992}, which thus serves as the point of origin for mathematical research.
I would be remiss in not mentioning at least the gist of the construction of isospectral domains as well as the theorem at the centre of the proofs of their isospectrality.
The proof of isospectrality relies on the theorem of Sunada \cite{sunada_riemanninan_????}, which states that given a finite (abstract) group $G$ and two of its subgroups $G_1,G_2$, which are
almost conjugate\footnote{an example of almost conjugacy is given in \cite[Example 8]{berard_transplantation_1992}.} and which act freely on a manifold $\Mf$ on which $G$ acts by isometries (length-preserving mappings), the two quotient manifolds $G_1 \setminus \Mf,G_2 \setminus \Mf$ are isospectral.

The abstract theorem becomes more palatable when viewed with the pasting method proposed by \citeauthor{buser_isospectral_1986} in \cite{buser_isospectral_1986},
elaborated on in \cite{berard_transplantation_1992} and utilised by \citeauthor{gordon_isospectral_1992}.
Given a fundamental manifold (tile) $\mathcal{T}$ and a transformation group $G$ (i.e., translations) as above an orbifold, or an orbit-manifold, $G_i\setminus\mathcal{T}$ is 
constructed by applying the elements of the transformation group to $\mathcal{T}$ (i.e., translating $\mathcal{T}$) and gluing the images obtained from this action together.
The requirement that the subgroups act freely means that there are no fixed points of the action, that is - no element of $G_i$, except for the identity,
leaves $\mathcal{T}$ in one place, thus ensuring that any point in the resulting orbifold belongs only to finitely many images of $\mathcal{T}$.
Because the action of a group is not limited to a finite set of images a representative manifold is chosen from the set
of all images by choosing only one image if multiple overlap. A result of this prescription is depicted in \ref{fig:Theo_GWW_QuotientManifolds}.
Note that the requirement that $G_i$ be almost-conjugate leads to non-isometric isospectral manifolds,
which in the present context means that there is no isometry $f:(G_1\setminus\Mf,g)\rightarrow(G_2\setminus\Mf,g)$, where $g$ is the same in both cases.

It has to be pointed out that the domains of \ref{fig:Intro_IsospectralManifolds} are quotients of the actual orbifolds under the group of reflections in the plane of the figure.
The domains of \ref{subfig:ElaborateIDs} are obtained by cutting the figures of \ref{fig:Theo_GWW_QuotientManifolds} in the central plane and retaining only half of each.
According to \citeauthor{gordon_isospectral_1992} the resulting domains are Dirichlet and Neumann isospectral, furthermore the authors describe
how to transform these domains into \ref{subfig:SimpleIDs}.
This transformation has a profound influence upon the applicability of Sunada's theorem to the pair \ref{subfig:SimpleIDs}, such that their isospectrality
has to be proven using Buser's method of transplanting eigenfunctions \cite{buser_isospectral_1986}. This technique shows isospectrality by proving that
the eigenfunctions of one orbifold span the same eigenspaces for the other orbifold, it is also intrinsically coupled to the method with which the domains are created.
We will use this method in a very simplified way for a preliminary analysis of the IDs in \ref{sec:Theo_Analysis_IDs}.

Neither the constructions nor the proofs provide a straightforward method to check their applicability to the problem treated in this work. The simplest way to
proceed would be to deny applicability of Sunada's theorem to the present case. This is indeed possible in a simple and rather crude way. Every proof of isospectrality
provided in the aforementioned papers begins with a requirement that $(\Mf,g)$ be a \textit{Riemannian} manifold. Working in a physical setting the basic metric of this thesis is, however, Lorentzian.
This does not preclude the validity of the theorems due to the fact that the restriction of the classical Lorentz pseudo-metric to the $\RR^3$ results in the Euclidean metric,
which is of course a Riemannian metric. Indeed, when the domains are at rest and thus in an inertial coordinate frame they are isospectral even though our basic metric is still
Lorentzian.
It will become apparent in the following chapter, that we may still rely upon the distinction of the metrics
because the motion of the domains will lead to mixing of space and time and thus imbue the Lorentzian structure upon the Laplace-Beltrami operator.

The assumption of a Riemannian manifold explains the reason why the Laplace operator is variously called Laplace-Beltrami (\ref{def:Theo_LaplaceBeltrami}) or Laplace operator in the works cited.
Another peculiarity which ties into the difference between Laplace and Laplace-Beltrami operators is the difference in what is actually sought after by mathematicians and physicists.
An eigenpair $(\mu,u)$ of the Laplacian is defined by the relation
\begin{equation}
\Delta u = \mu u\quad . \label{eq:Theo_MathDefEigenpair}
\end{equation}
Thus whenever the spectrum of the Laplacian is discussed the above equation is taken as a starting point. 
This mathematical definition was obtained by abstracting away the physical meaning of eigenmodes of an oscillating membrane. If we reintroduce physics into the above equation then
an eigenpair of the Laplacian consists of an eigenfrequency and the associated eigenfunction, describing the displacements of an oscillating membrane\footnote{assuming 2D geometries}.
An oscillation however, is a fundamentally time-dependent phenomenon. As such the first step in a physical analysis (in an inertial coordinate frame) constitutes the wave-equation
\begin{equation}
\square\psi = 0\quad,\label{eq:Theo_WaveEquation}
\end{equation}
which yields, along with the assumption $\psi(x,t)=e^{\lambda ct}\phi(x)$
\begin{equation}
\Delta\phi = \lambda^2\phi\quad .\label{eq:Theo_PhysDefEigenpair}
\end{equation}
Here $c$ is the speed of the wave propagation in the medium, it is the speed of light for vacuum.
On performing the identification $\mu\leftrightarrow\lambda^2,u\leftrightarrow\phi$ the difference between mathematics and physics vanishes. 
When operating in non-inertial reference frames it will become obvious that the latter approach is more meaningful. We shall therefore keep to the physical
way of determining the spectrum of the Laplacian by taking a detour along the d'Alembertian route.
A slightly different approach to proving isospectrality of plane domains is taken by \citeauthor{gordon_robin_2015} in \cite{gordon_robin_2015}. There they study Robin-Steklov instead
of Dirichlet-Neumann isospectral plane domains using again Sunada's theorem as well as an integral formulation\footnote{so-called weak form.} of \eqref{eq:Theo_MathDefEigenpair}.
A similar approach was taken by \citeauthor{berard_transplantation_1992} \oc This approach ties in perfectly to the finite element method which will be used as a tool for 
analysing the behaviour of the eigenfrequencies and eigenmodes of rotating domains in the second part of this thesis.
Finally we remark that apart from the two pairs of domains depicted in \ref{fig:Intro_IsospectralManifolds} a plethora of Dirichlet-Neumann isospectral domains is provided in
\cite{buser_planar_2010} and interesting examples of Robin-Steklov isospectral domains are given in \cite{gordon_robin_2015}. 
It should be noted that many of the domains of \cite{buser_planar_2010} are trivially isospectral, being rotations and reflections of one another.

\chapter{Physical Basis} \label{ch:PhysicalBasis}
Although titled "Physical Basis" the concepts presented here will be mainly of geometric nature
w.r.t. the space-time, as mentioned in \cite[Ch. 4]{Gourgoulhon2012}. Here the physical basis for the calculations of
the following chapters is provided. The shortness of the chapter is justified by the assumption of familiarity with
the concepts of electrodynamics and special relativity. Hence only the extensions I wish to address are presented here.
The concepts are not new and have been collected from \cite{hehl_foundations_2003,Schouten2011,Post1962,fliesbach_allgemeine_2012,Gourgoulhon2012,heer_resonant_1964} 
and \cite{mandelstam_elektrodynamik_1926} as well as \cite{minkowski_grundgleichungen_1910}.
\section[Intermezzo in GR]{Intermezzo in General Relativity}\label{sec:Theo_GRIntro}
\subsection{Physical Principles}\label{sec:Theo_Phys_GRPrinciples}
Being primarily concerned with uniformly rotating isospectral domains places the topic formally outside the field of special relativity, which deals with non-accelerated motion only.
As such I will introduce the principles of general relativity upon which the physical justification of this work is resting.
\begin{notation}
Prior to any definitions the notation used should be agreed upon. When working in the 4-dimensional space-time the indices of
coordinates will be denoted by greek letters $\mu,\nu,\dots\in\lbrace 0,1,2,3\rbrace$.
The indices denoting purely spatial coordinates of a frame will be denoted by latin letters $i,j,\dots$ with their range being $\lbrace 1,2,3\rbrace$.
\end{notation}
Next the main physical principle upon which this thesis is build is postulated.
\begin{thm}[Strong Principle of Equivalence]\label{Theo_StrongPrincipleOfEquivalence}
The strong principle of equivalence states that gravitational forces are equivalent to inertial forces. Or equivalently that the laws of special relativity
are valid in a local inertial system.
\end{thm}
A local inertial system is provided by a momentarily co-moving inertial frame (MCIF) which, when one works specifically with gravity, is often referred to as
a locally freely falling frame ($\text{LF}^3$). In the remainder of this work only MCIF's will be required seeing as the question  to be treated is of a type
which can be taken care of by purely geometrical means (in the sense of differential geometry).  The practical implication of the above principle is the
ability of transferring the physical laws (in tensorial form) stated for inertial systems to non-inertial systems, and hence into the domain of general relativity,
by means of appropriate coordinate transformations. The latter, of course, are non-trivial in general. I shall thus, in the following, adhere to the following
simplistic formula \fbox{SR Law + Coordinate Transformation = GR Law}.
It remains to be noted that the formalism used in this thesis makes no reference to Einstein-Hilbert equations. Thus, although formally in the realm of general relativity
a purely geometric view of it is maintained (as stated in \cite{Gourgoulhon2012}).

The last physical assumption made is the well-known continuum hypothesis. Space, time and matter are assumed to be smooth. The only quantized quantity will be the charge, although it too,
will occur only in the smoothed form of a charge density.

\subsection{The Metric}\label{subsec:Theo_Phys_Metric}
Using \ref{def:Theo_MetricTensor} the Lorentzian metric which shall be used henceforth is
\begin{defn}[Minkowski Metric]\label{def:Theo_MinkowskiMetric}
The Minkowski metric tensor $g\in\Tf{0}{2}{\Mf},\;\Mf=\RR^4$ is given by
\begin{equation}
g = \eta_{\mu\nu}dx^\mu\otimes dx^\nu = c^2 dt\otimes dt - \sum_{i=1}^3 dx^i\otimes dx^i\;, \label{eq:DefMinkowskiMetricTensor}
\end{equation}
with $\eta=\diag(c^2,-1,-1,-1)$. With the associated volume form being
\begin{equation}
\epsilon = \sqrt{\vert\det g\vert}\;dt\wedge dx^1\wedge dx^2\wedge dx^3\;.\label{eq:DefVolumeForm4Space}
\end{equation}
\end{defn}
Henceforth, whenever the term \textit{metric} is used sans preposition it shall mean the pseudo-metric $g$ as defined above.
The above definition was made to facilitate the understanding of \cite{hehl_foundations_2003} as well as \cite{Post1962}, who both use
the above convention for the coefficients of the metric. The differences which would arise from choosing $\eta=\diag(1,-1,-1,-1)$ can be traced down trivially using dimensional
analysis as it is described below.

Using \ref{def:Theo_Pullback} it becomes trivial to determine the expressions for the coordinate functions of $g$ in any chart. It is also immediately obvious,
that such coordinate transformations do not affect $\Mf$ itself, such that the metric remains flat.
\subsection{The Matter of Time}\label{subsec:Theo_Phys_Time}
\subsubsection{Proper Time and World Lines}
The distinction between temporal and spatial coordinates will be of utmost importance in the derivation of the Minkowski relations \cite{minkowski_grundgleichungen_1910}
for matter in accelerated motion.
Therefore some time is spent recalling the necessary definitions.
\begin{defn}[Proper Time]\label{def:Theo_ProperTime}
The proper time of a point $p$ in the 3-dimensional space is the time measured by a clock
attached to that point. It is hence the time measured in the coordinate frame in which $p$ is locally and instantaneously at rest, the so called momentarily co-moving inertial frame (MCIF for short).
The proper time coordinate shall be denoted by $\tau$.
\end{defn}
Absent any proof of the contrary one may well assume time to be an intrinsically directed coordinate for any inertial observer. That is, notions of past and future are well defined.
The proper time defined above is a physical quantity in the sense that it is measurable by a clock. Contrast this to a coordinate time, an arbitrary choice of a scale and a label of the temporal
coordinate made by the theoretician. The latter is a purely theoretical construct which has no physical meaning save for the possibility of distinguishing the temporal order of points in space-time,
when the temporal coordinate axis is equipped with an (arbitrary) orientation.

The proper (physical) time of a point $p\in\RR^3$ is the natural parameter of the path $\gamma$ of that point through the space-time $\Mf$, thus:
\begin{align*}
\gamma : \RR &\rightarrow \Mf\\
\tau &\mapsto \gamma(\tau)\quad .
\end{align*}
This path is commonly called a world-line in physics.
On $\RR$ one may define a unit tangent vector field $\Tf{1}{0}{\RR}$ in a straightforward way. This vector field shall be denoted $\del_\tau$. It yields immediately (using the definition of
a covector) a 1-form $d\tau\in\Omega^1(\RR)$. The dual pairing $(\cdot,\cdot):T_p\RR\times T_p^\ast\RR\rightarrow \RR$ is invariant under diffeomorphishms (i.e., coordinate transformations).
Thus we have, using definitions \ref{def:Theo_TangentCotangentSpaces} and \ref{def:Theo_TangentMap}:
\begin{equation}
1 = (\del_\tau\vert_0,d\tau\vert_0) = d\tau(\del_\tau) = d\tau\circ \del_\tau =  d\tau\circ d\gamma^{-1}\circ d\gamma\circ \del_\tau = d(\tau\circ\gamma^{-1})(d\gamma\circ\del_\tau) = \left((\gamma_\ast\del_\tau)\vert_p, (\gamma_\ast d\tau)\vert_p\right)\label{eq:Theo_ProofInvarianceOfDualPairing}
\end{equation}
where in the last equality one uses \ref{def:Theo_Pullback} and \ref{def:Theo_Pushfwd}. This shall prove to be useful when deriving the coordinate
representation of the electric excitation $\CalD$ and magnetic field $\CalH$ for moving media. Note that as the manifold, upon which the theory plays out, is smooth 
the world-lines exist in perpetuity, thus defining a global vector field. The geometry may therefore seem boring, but it shall nevertheless, yield interesting results.

\subsubsection{Time and Space - Foliation}
Next we turn to the special nature of time. As has been noted above, physical - or proper - temporal distances are measured in a manner different from
spatial distances.
The unification of space and time into space-time, the achievement of the theory of special relativity, does not eliminate said difference but circumvents it
by either scaling the temporal coordinate of a point in the space-time manifold by a constant $c$ with a dimension of velocity attached, or by defining the 
distance measure as in \ref{def:Theo_MinkowskiMetric}. The difference is further encoded in the different signs of spatial and temporal components of said metric.
\citeauthor{Post1962}, in his book \cite{Post1962}, refers to this distinction as the "principle of dimensional coordinate individuality", 
assuming implicitly that only a physical time (i.e., proper time) is considered.

Intimately related to the above is the decomposition of the space-time manifold $\Mf$ into 
non-intersecting
3-dimensional hyper-surfaces (space-like surfaces) parametrised by a temporal coordinate $\sigma$. Such a decomposition
is called a foliation of $\Mf$ with the hypersurfaces being called leafs or folia. Borrowing from \cite{Gourgoulhon2012},
we note that in the present case the space-like hypersurfaces would be Cauchy surfaces and $(\Mf,g)$  a globally hyperbolic
manifold, for which the scalar wave equation (the ultimate goal of this part of the thesis) is well posed. 

The $3+1$ decomposition of differential forms and, by extension, of $\Mf$ is taken from the main guidebook of this thesis: \citetitle{hehl_foundations_2003} \cite{hehl_foundations_2003}
by \cite{hehl_foundations_2003}.
The general formulations of chapters B.1.2 and B.1.4 therein will be reformulated in the terms of proper time\footnote{Thus I am effectively skipping
ahead directly to chapter E.1.3 resp. E.4.1 therein.}.
The temporal coordinate $\sigma$ mentioned above will be taken to be the proper time $\tau$ of an arbitrary point in space. The path $\gamma$, as defined above, is parametrised
by $\tau$ and, as has been shown in \eqref{eq:Theo_ProofInvarianceOfDualPairing} transports the unit tangent field $\del_\tau$ onto $\Mf$ by means of a push-forward, 
resulting in a 4-velocity vector field which shall be denoted by $u$.
Furthermore, having a coordinate representation of the world-line is sufficient, in principle, to determine a representation the proper-time 1-form $d\tau$ w.r.t. the coordinate cobasis.

Following \cite[B.1.4]{hehl_foundations_2003} let $\omega\in\Omega^p(\Mf)$. Then $\omega$ can be decomposed into components longitudinal
to $u$ and transverse to it,
\begin{subequations}\label{eq:Theo_3+1DecompositionOfForms}
\begin{align}
 \omega^\perp &:= d\tau\wedge\omega_\perp & \omega_\perp &:= \iota_u(\omega) \label{eq:Theo_3+1DecompLongitudinal}\\
\underline{\omega} &:=\iota_u(d\tau\wedge\omega) & \iota_u( \underline{\omega}) &= 0\quad .\label{eq:Theo_3+1DecompTransverse}
\end{align}
\end{subequations}
Here the remarks on coordinate 1-forms of \ref{subsec:Theo_DG_BareStructure} along with \ref{def:Theo_WedgeProduct} come in handy.
$\iota_u(\omega)\sim\iota_{\del_\tau}(\omega)$ extracts only those parts of $\omega$ which contain contributions from $d\tau$, they
are thus parallel (longitudinal) to $d\tau$.
By nature of the operation the result does not contain $d\tau$ itself anymore, which is remedied by defining $\omega^\perp$.
The complement thereof is the part of $\omega$ "transverse" to the time-like vector field $\del_\tau$.

This is the $3+1$ decomposition of an arbitrary $p$-form on $\Mf$.
Given the proper-time 1-form any coordinate co-frame may be transformed into an adapted coframe by means of the above, resulting in
\begin{equation}
(dt,dx^i)\longrightarrow (d\tau, \ul{dx^i})\;.\label{eq:Theo_AdaptedCoframeDef}
\end{equation}
Alas, to be applicable in \eqref{eq:Theo_3+1DecompositionOfForms} $d\tau$ has to be expressed in the (coordinate) co-basis first,
in which the original $\omega$ has been given.
In principle \eqref{eq:Theo_ProofInvarianceOfDualPairing} provides a method for determining that expression (c.f. \ref{def:Theo_Pushfwd}).
An easier way, provided a metric is given, is to use \ref{def:Theo_MusicalMaps} and the 4-velocity expressed in a coordinate basis.
To do that an intermediate normalization step is necessary due to \[ ( u^\flat , u ) = u^\flat(u) = g(u,u) = c^2 \neq 1\;. \] Thus:
\begin{equation}
d\tau = \frac{1}{c^2} u^\flat\quad .\label{eq:Theo_ProperTimeFormFrom4Velocity}
\end{equation}

\section[Differential Electrodynamics]{Electrodynamics and Differential Forms}\label{sec:Theo_EDDG}
\subsection{Definitions}
From the classical (vectorial) formulation of electrodynamics the vector-fields of the electric field $\VE$, the magnetic induction $\VB$,
electric excitation $\VD$, magnetic field $\VH$ as well as electric current $\Vj$ are known along with the scalar field of electric charge density $\rho$.
The electric field is commonly integrated along a line to obtain the electric potential difference $\Delta\varphi$, similarly the magnetic induction field $\VB$
is integrated over a 2-surface in $\RR^3$ to obtain a flux, thus:
\begin{align*}
\varphi(B)-\varphi(A) &= \int_A^B \VE\cdot d\boldsymbol x = \int_A^B E_i dx^i= \int_\gamma E\\
\Phi &= \int_S \VB\cdot d\boldsymbol A = \int_S \VB\cdot\boldsymbol n dA = \int_S \epsilon_{ijk} B_i dx^jdx^k = \int_S B\quad .
\end{align*}
As has been remarked upon, differential forms are natural objects to be integrated over a manifold. The above therefore suggests the following definition:
\begin{defn}[EM Forms]\label{def:Theo_EMForms}
Let $E\in\Omega^1(\Mf)$ and $B\in\Omega^2(\Mf)$ be such that
\begin{align}
E &:=  E_i dx^i = \VE_i dx^i\label{eq:Def_ElectricField1Form}\\
B &:= \frac{1}{2!}B_{ij} dx^i\wedge dx^j = B_{23}dx^2\wedge dx^3 + B_{31}dx^3\wedge dx^1 + B_{12} dx^1\wedge dx^2\label{eq:Def_MagneticField2Form}\\
&= \VB_1 dx^2\wedge dx^3 + \VB_2 dx^3\wedge dx^1 + \VB_3 dx^1\wedge dx^2\notag,
\end{align}
where $\VE_i,\VB_i$ are the components of the vector fields mentioned above.
Then $E$ is called the \textit{electric field strength 1-form} and $B$ is the \textit{magnetic field strength 2-form}.
\end{defn}
Neither of the above is in the class of twisted differential forms (or densities). They will thus not be invariant under a change of orientation.
In a similar manner the definition of the differential form for the current may be obtained.
\begin{defn}[Current Form]\label{def:Theo_CurrentForm}
Let $\rho\in\Omega^3$ and $j\in\Omega^2$ be given by
\begin{align}
\rho &:= \frac{1}{3!}\rho_{ijk}dx^i\wedge dx^j\wedge dx^k = \rho\;dx^1\wedge dx^2\wedge dx^3\label{eq:Def_ChargeDensity3Form}\\
j &:= \frac{1}{2!}j_{ik}\;dx^i\wedge dx^k = j_{23} dx^2\wedge dx^3 + j_{31} dx^3\wedge dx^1 + j_{12} dx^1\wedge dx^2\label{eq:Def_Current2Form}\\
&= \Vj_1 dx^2\wedge dx^3 + \Vj_2 dx^3\wedge dx^1 + \Vj_3 dx^1\wedge dx^2\notag.
\end{align}
Then $j$ is called the \textit{charge current 2-form} and $\rho$ the \textit{charge 3-form}. Both can be combined into the (twisted)
\textit{current 3-form} $J\in\Omega^3 \text{ twisted }$ by
\begin{equation}
J := -j\wedge dt + \rho\quad .\label{eq:Def_Current3Form}
\end{equation}
\end{defn}
In the remainder of this work the current and charge forms will feature only by their absence, as only situations without 
free charges and currents will be considered. This is due to the fact that the present work
will be mainly concerned with electromagnetic waves in absence of charges and currents. Analogously to the current 3-form one can combine $E,B$ to yield:
\begin{defn}[Faraday 2-Form]\label{def:Theo_FaradayForm}
The so-called Faraday 2-form $F$ is
\begin{equation}
F := E\wedge dt + B\quad .\label{eq:Def_Faraday2Form}
\end{equation}
\end{defn}
Trivially, $E$ is the longitudinal and $B$ the transverse part of $F$. The Faraday 2-form is another way of writing the electromagnetic field tensor $F_{\mu\nu}$
(in the notation common to physics \cite{Schopohl2012}), 
to which it is related by \[ F = \frac{1}{2!}F_{\mu\nu} dx^\mu\wedge dx^\nu\;.\]
Next the remainder of the vector fields of classical electrodynamics,
the fields in matter, are processed into differential forms.
\begin{defn}[Excitation Forms]\label{def:Theo_EMExcitationForms}
Let $\CalH\in\text{twisted }\Omega^1,\CalD\in\text{twisted }\Omega^2$ be given as
\begin{align}
\CalH &:= \CalH_i dx^i = \VH_i dx^i\label{eq:Def_MagneticExcitationTwisted1Form}\\
\CalD &:= \frac{1}{2}\CalD_{ij} dx^i\wedge dx^j = \CalD_{23} dx^2\wedge dx^3 + \CalD_{31}dx^3\wedge dx^1 + \CalD_{12} dx^1\wedge dx^2\label{eq:Def_ElectricExcitationTwisted2Form}\\
&= \VD_1 dx^2\wedge dx^3 + \VD_2 dx^3\wedge dx^1 \VD_3 dx^1\wedge dx^2	\notag\;.
\end{align}
Again with $\VH,\VD$ as above. Then $\CalH$ is the \textit{magnetic excitation 1-form} and $\CalD$ is the \textit{electric excitation 2-form}.
\end{defn}

Both excitation forms $(\CalH, \CalD)$ are twisted, that is they are densities, and are thus the natural objects to be integrated (c.f. \ref{subsec:Theo_DG_BareStructure}).
More so than the pair $(E,B)$. With physical laws being independent of orientation\footnote{At least in classical electrodynamics.} one should work with twisted forms.
As in the case of the Faraday 2-form one may define a twisted excitation 2-form $H$ as follows:
\begin{defn}[Excitation 2-Form]\label{def:Theo_Excitation2Form}
The excitation form $H\in\text{twisted }\Omega^2$ is given by
\begin{equation}
H := dt \wedge \CalH + \CalD\;.\label{eq:Def_Excitation2Form}
\end{equation}
\end{defn}
Again, it is obvious that $\CalH$ is the longitudinal and $\CalD$ the transverse part of $H$. The latter is (mathematically) the object to be integrated.
Its connection to the current form $J$ constitutes the first set of Maxwell's equations  and its connection to the Faraday 2-form is given by the Maxwell-Lorentz 
space-time relation and constitutive equations of the medium (all given below).

It is time to make good on the promise of a simpler dimensional analysis than what \citeauthor{Schouten2011} and \citeauthor{Post1962} present in
their respective works \cite{Schouten2011} and \cite{Post1962}. In short the promise made at the outset of the previous chapter is fulfilled here.
To appreciate the simplicity the definitions of \cite{Post1962} are provided.
\begin{defn}[Dimensions \cite{Post1962}]
Provided $P\indices{^a_{bc}}$ is a tensor in the physical space (physical tensor) it is obtained from a purely arithmetic tensor $P\indices{^\alpha_{\beta\gamma} }$
in the following form:
\begin{equation*}
P\indices{^a_{bc}} = D(P) A^a_\alpha A^\beta_b A^\gamma_c P\indices{^\alpha_{\beta\gamma} }\;,
\end{equation*}
where $A^a_\alpha = \del_\alpha x^a,A^\beta_b = \del_bx^\beta$ are the entries of the Jacobi matrix of the transformation from/to the arithmetic space.
The \textit{dimensional gauge factor} $D(P)$ is meant to introduce dimensions not associated with length or time and is called the \textit{absolute dimension}
of $P\indices{^a_{bc}}$. Note that $\left[P\indices{^\alpha_{\beta\gamma} }\right] = 1$, for the quantity is defined in the (dimensionless) arithmetic space.
The \textit{relative dimension} is obtained from the absolute dimension by multiplying it by $T$ or $L$ for each appropriate contravariant index and $\tfrac{1}{T}$ or
$\tfrac{1}{L}$ for each appropriate covariant index, as well as $\left( \tfrac{1}{TL^3}\right)^k$ for a density of weight $k$.
\end{defn}
The definition is a considerable simplification of the original definition provided by \citeauthor{Schouten2011}, to which it is equivalent.
Nevertheless it is my opinion that it is still rather cumbersome for practical use. This is where the formalism of differential forms provides a nice
simplification.
\begin{defn}[Dimensions]\label{def:Theo_AbsRelDimensions}
The components of the cobasis carry the dimensions of length $L$ or time $T$ and their duals carry the associated inverse dimensions.
The dimension of the differential form, i.e., $\rho$, is called the \textit{absolute dimension} of the form and denoted by $[\rho]$.
The dimension of the \textit{coefficients} of the form is called the \textit{relative dimension} of the form - $[\rho_{abc}]$.
\end{defn}
The operation of integrating a form (or density) over a manifold, if interpreted as a summation, does not contribute to the dimension of the result, thus
one obtains for the charge form $\rho$:
\begin{equation*}
q = \int_{V_3} \rho \Rightarrow  Q = [q] = \left[\int_{V_3}\rho \right] = [\rho]\Rightarrow Q = [\rho] = \left[\frac{1}{3!}\rho_{abc}dx^a\wedge dx^b\wedge dx^c\right]
=\left[\rho_{abc}\right]\cdot\left[dx^a\right]^3 = \frac{Q}{L^3}\cdot L^3\;.
\end{equation*}
Here $q$ is the charge which has the dimension of charge $Q$ and $V_3$ is a volume in $\RR^3$.
Using differential forms the meaning behind the terms \textit{absolute} and \textit{relative} becomes immediately obvious as the dimension of the whole
and the dimension of the components relative to a (co)basis, something which is, in my opinion, lost in the definitions provided by \citeauthor{Post1962}
and \citeauthor{Schouten2011}.
\subsection{Maxwell, Minkowski \& Matter}
Next the formulation of Maxwell's equations in media and vacuum in the language of differential forms is due. The only media which shall be considered
here are so-called "simple" media \cite[Ch. V]{Post1962}. A medium is simple if it is homogeneous, isotropic and responds linearly and instantaneously to
external fields. At every point of the medium, in a MCIF, the constitutive equations known from vectorial electrodynamics hold,
\begin{subequations}\label{eq:VectorialConstEq}
\begin{align}
\VD &= \epsilon_0\epsilon_r \VE\label{subeq:ElectricExcitationConstEq}\\
\VB &= \mu_0\mu_r \VH\;.\label{subeq:MagnFldConstEq}
\end{align}
\end{subequations}
The definitions of the main differential forms of electrodynamics - \ref{def:Theo_FaradayForm} and \ref{def:Theo_Excitation2Form} -
are formally disjoint with the former defining a form and the latter a twisted form (or density).
As has been suggested in \ref{subsec:Theo_DG_MetricStructure} both can be combined using \ref{def:Theo_HodgeOp}. The trivial relation $H=\ast F$ would be, however,
incorrect, as a quick dimensional analysis reveals. Here \ref{def:Theo_AbsRelDimensions} comes in handy,
\begin{align*}
[H] &= \left[dt\wedge\CalH + \CalD\right] = \left[dt\wedge\CalH\right] + \left[\CalD\right] = T\left[\CalH\right] + Q = Q\\
[F] &= \left[E\wedge dt +B\right] = \frac{\Phi}{T}\cdot T + \Phi = \Phi\\
\Rightarrow [H] = Q &= \left[\ast\right]\Phi = \left[\ast\right]\left[F\right]\;.
\end{align*}
One may argue that the Hodge-operator does not contribute to the dimension. If one considers \ref{def:Theo_HodgeOp} and the properties of the operator 
the answer is less clear, for each of the 3 properties listed on p. \pageref{def:Theo_HodgeOp} yields a different dimension
\begin{align*}
[\ast] = [\ast 1] &= [\epsilon] = L^4\\
[\ast]^2 = [\ast\circ\ast] &= [(-1)] = 1\\
[\ast] L^4 = [\ast][\epsilon] = [\ast\epsilon] &= [(-1)^s] = 1\;
\end{align*}
Thus the dimension of the Hodge operator depends on the space it operates on. With $H,F$ being a twisted and an untwisted 2-forms the Hodge operator
will thus indeed be dimensionless\footnote{This can be verified by an analysis in a cobasis using relative dimensions.}. One therefore has
\begin{equation*}
[H] = [dt\wedge\CalH + \CalD] = [\CalD] = Q \neq   \Phi  = [\ast B] = [\ast F]\;.
\end{equation*}
The discrepancy is mended by introduction of the vacuum admittance $\lambda_0:=\sqrt{\tfrac{\epsilon_0}{\mu_0}}\approx \tfrac{1}{377} \left[\tfrac{1}{\Omega}\right]$, thus:
\begin{defn}[Maxwell-Lorentz Relation]\label{def:Theo_MaxwellLorentzRelation}
The connection between the electromagnetic excitation form $H$ and the Faraday 2-form $F$ in vacuum is provided by the \textit{Maxwell-Lorentz
space-time relation}:
\begin{equation}
H = \lambda_0 \ast F\;.\label{eq:Def_MaxwellLorentzVacuumRelation}
\end{equation}
\end{defn}
This connection between a form and a density suggests, that the vacuum itself may be viewed as a form of medium (c.f. \cite[Ch. D.6.1]{hehl_foundations_2003}).
A simple medium, as it has been defined above, is then but a linear map,
\begin{align}
\kappa: \text{twisted }\Omega^2 &\rightarrow \text{twisted }\Omega^2\label{eq:Def_LinearConstitutiveRelations}\\
\ast F &\mapsto \kappa(\ast F)\notag\quad .
\end{align}
Which can be written in matrix-form provided a basis of $\Omega^2$ is chosen. Using the basis given in \cite[A.1.10]{hehl_foundations_2003} one
obtains
\begin{equation*}
\Mf(\kappa) = \begin{pmatrix}
\frac{1}{\mu_r} \mathbb{I}_3 & 0\\
0 & \epsilon_r \mathbb{I}_3
\end{pmatrix}\;.
\end{equation*}
The above is sufficient to determine the response of simple media to electromagnetic fields, that is - to determine $(\VH,\VD)$ from $(\VE,\VB)$. Because
this action will be central to the rest of this thesis it is given here in a concise, almost algorithmic, way:
\begin{enumerate}
\item Begin with the Faraday 2-form in an arbitrary coordinate frame $K$.
\item Determine a coordinate transformation $\Phi:K'\rightarrow K$ from the frame $K'$ of the observer (measuring apparatus) to the frame $K$.
\item Determine the explicit forms of $\Phi^\ast F, \Phi^\ast g$.
\item Determine an adapted coframe $(d\tau_{obs},\underline{\underline{dx^a}})$ for the observer by applying the $3+1$ decomposition \eqref{eq:Theo_3+1DecompositionOfForms}.
\item Determine a transformation $\Psi:\hat{K}\rightarrow K'$ from a coordinate frame $\hat{K}$ attached at a point in the medium to the coordinate frame of the observer.
\item Determine an adapted coframe $(d\tau, \underline{dx^a})$ for the medium.
\item Determine the explicit forms of $\Psi^\ast(\Phi^\ast F),\Psi^\ast(\Phi^\ast g)$.
\item Apply \eqref{eq:Def_MaxwellLorentzVacuumRelation} and \eqref{eq:Def_LinearConstitutiveRelations}.
\item Express $(d\tau_{obs},\underline{\underline{dx^a}})$ in terms of $(d\tau,\underline{dx^a})$.
\item Using the relations obtained in the previous step transform $H$ from the adapted coframe of the medium to the adapted coframe of the observer.
\end{enumerate}
Following the above prescription one obtains the Minkowski relations between $\VH,\VD$ and $\VE,\VB$,  first derived by \citeauthor{minkowski_grundgleichungen_1910} in
\citeyear{minkowski_grundgleichungen_1910} in his paper \cite{minkowski_grundgleichungen_1910} for media moving with constant velocity (in the context of special relativity). 

Note that provided the coordinate transformations between the coordinate systems of the observer and the moving medium are known one can, at every moment,
determine a MCIF of a point in the moving medium. Using this MCIF one can then use classical Lorentz transformations to obtain said relations (c.f. \cite{koks_explorations_2006, minkowski_grundgleichungen_1910}).
The above prescription however, does not use Lorentz transformations and is thus more general.
Its application will be presented in the next chapter.

Finally, with the Maxwell-Lorentz space-time relations and their adaptation to matter the Maxwell equations are given,
\begin{thm}[Maxwell Equations]\label{def:MaxwellEquations}
Let $F,H,J$ be the differential forms of classical electrodynamics as given in definitions \ref{def:Theo_FaradayForm} as well as \ref{def:Theo_CurrentForm} and \ref{def:Theo_Excitation2Form}.
Then Maxwell's equations are
\begin{subequations}\label{eq:Def_MaxwellEqnsDG}
\begin{align}
dF &= 0\label{subeq:Def_MaxwellEqnsHomogeneous}\\
dH &= J\;.\label{subeq:Def_MaxwellEqnsInhomogeneous}
\end{align}
\end{subequations}
\end{thm}
The above can be verified by straightforward calculations. The equations are augmented by the continuity equation $dJ = 0$ and carry two different
gauge transformations, which will not be expounded upon here, but can be found in \cite[B.1.3, B.4.1]{hehl_foundations_2003}.
\clearpage\thispagestyle{plain}

\chapter{Derivation of the Equations}\label{ch:Derivations}
\section{Introduction and Set-Up}\label{sec:Deriv_IntroAndSetup}
The goal of this work is to answer the simple question whether two isospectral domains remain isospectral when subjected to accelerated motion. After the brief
venture into the realm of general relativity in the previous chapter it is apparent that the question is equivalent to whether they remain isospectral when subjected to
external gravitational fields. To investigate the spectrum of the Laplace-Beltrami operator for the domains of \ref{fig:Intro_IsospectralManifolds} 
one has to provide explicit expressions for the coefficient functions of $H$ in a coordinate cobasis.
The physical set-up which will serve as a basis is that of a domain in the $xy$-plane extended infinitely into the $z$ direction and rotated about an axis
parallel to the $z$-axis with uniform angular frequency $\omega$. The observer (or measuring device) will be co-rotating with the medium and positioned at the axis of rotation. The coordinate basis is
centred at the axis.
As has been hinted in the previous chapter, it is assumed that the domains are manufactured from a simple medium \eqref{eq:Def_LinearConstitutiveRelations}.

Due to the geometry being infinitely extended in the $z$-direction with the boundary of the domain being made of a perfect conductor the $z$-component
of the electric field will vanish. I.e. only TM modes of the field are considered. Ultimately this will yield eqn. (92) of \cite{anderson_electromagnetic_1969},
a differential equation for the z-component of the electric field. The publication mentioned above shall, along with the literature and the equations from previous chapters, serve as
the guide for the computations performed in this chapter. If one were, in the spirit of Kac' original question, to consider the domains as two-dimensional drums,
then $E_z$ may be considered as the displacement of the membrane, an interpretation we shall return to in \ref{sec:Exp_Discussion}.
The use of the electric field $E_z$ suggests the use of a stationary point-charge as the "observer" of the oscillations.

\section{3+1 Decomposition}\label{sec:Deriv_SpacetimeFoliations}
For rotational motion around a fixed spatial axis cylindrical coordinates in space present a natural choice of coordinates. The physical set-up
presents two distinct coordinate frames, the laboratory $K$, where the motion takes place and which is assumed to be inert, and the rotating medium, to which
the coordinate frame $K'$ is attached s.t. in both frames the axis of rotation coincides with the $z$-axis of the spatial coordinates.
The coordinates in $K$ will be $(t,r,\varphi,z)$ and in $K'\; (t',r',\varphi',z')$. The Minkowski metric $g$ of \ref{def:Theo_MinkowskiMetric}
is provided therein in Cartesian coordinates, an expression for $g$ in cylindric spatial coordinates is stated here without derivation (which is trivial):
\begin{equation}
g = c^2 dt\otimes dt - dr\otimes dr - r^2 d\varphi\otimes d\varphi-dz\otimes dz\;.\label{eq:MinkowskiMetricISCylCoord}
\end{equation}
It is perhaps surprising that since the time of Einstein no transformation to a rotating coordinate frame has been found, which fulfils all requirements of 
special relativity \cite{Rizzi2004}. Due to its importance for further derivations the coordinate transformations to and from rotating frames have been
investigated in the framework of this thesis. Unfortunately the only two transformations which may be remotely relevant are the
Trocheris-Takeno (TT) transformation and its modified version (MTT) presented in \cite{herrera_relativistic_1999}. 
Both, TT and MTT transformations, yield lengthy coefficient
functions for the metric tensor which make their use in calculations cumbersome. To conform with \cite{anderson_electromagnetic_1969} and others the Galilei-Newton
coordinate transformation is therefore chosen for further derivations:
\begin{align}
\Phi: K' &\rightarrow K\label{eq:Deriv_GalileiNewtonCoordinateTrafo}\\
\begin{pmatrix}t'\\r'\\\varphi'\\z'\end{pmatrix} &\mapsto \begin{pmatrix}t\\r\\\varphi\\ z\end{pmatrix} = \begin{pmatrix}t'\\r'\\\varphi' +\omega t'\\ z'\end{pmatrix} \notag.
\end{align}

Using the above transformation one obtains the following expression for the coefficients of the metric w.r.t. the coordinate cobasis $(dt',dr',d\varphi',dz')$:
\begin{multline}
g = c^2\left( 1-\left( \frac{r'\omega}{c}\right)^2 \right)dt'\otimes dt' -dr'\otimes dr' -(r')^2d\varphi'\otimes d\varphi' - dz'\otimes dz'\\
 - (r')^2\omega\left[ dt'\otimes d\varphi' + d\varphi'\otimes dt'\right]\ = \left[\frac{c}{\gamma(r')}\right]^2 dt'\otimes dt' - dr'\otimes dr' - (r')^2d\varphi'\otimes d\varphi'
 -dz'\otimes dz'\\
  - (r')^2\omega\left[ dt'\otimes d\varphi' + d\varphi'\otimes dt'\right]\label{eq:Deriv_MinkowskiMetricInRotatingCoordCoframe}
\end{multline}
Note that this cobasis does not reflect the space-time foliation associated with the motion. Also,
\begin{equation}
\gamma(r') = \frac{1}{\sqrt{1-\left(\frac{r'\omega}{c}\right)^2}}\;.\label{eq:Deriv_GammaFactor}
\end{equation}

In the coordinate frame $K'$ every point of the matter domain is at rest. The infinitesimal displacement $(\Delta s)^2$
is therefore
\begin{align}
(\Delta s)^2 = g(u,u) &= c^2\left( 1-\left( \rocf\right)^2\right) (\Delta t')^2\label{eq:Deriv_InfinitesimalDisplacementRotSys}\\
\intertext{with $u=(1,0,0,0)^t$ and the proper-time increment is defned as}
(\Delta\tau)^2 = \frac{1}{c^2}\Delta s^2 &= \left[\frac{1}{\gamma(r')}\right]^2(\Delta t')^2\;.\label{eq:Deriv_InfinitesimalProperTimeIncrement}
\end{align}
The coordinate transformation from the proper-time domain to the world-line in $\Mf$ thus becomes
\begin{align}
\gamma: \RR &\rightarrow \Mf\\
\tau &\mapsto \gamma(\tau) = \begin{pmatrix}
x^0(\tau)\\
x^1(\tau)\\
x^2(\tau)\\
x^3(\tau)
\end{pmatrix}
= \begin{pmatrix}
\gamma(r')\tau\\
r'\\
\varphi'\\
z'
\end{pmatrix}\notag\;.
\end{align}
Which then yields the four-velocity by push-forward
\begin{equation}
\gamma_\ast u = \gamma_\ast\del_\tau = \gamma(r')\del_t'\;.\label{eq:Deriv_VelocityPushforwardObserverFrame}
\end{equation}
Using \ref{def:Theo_MusicalMaps} and \eqref{eq:Theo_ProperTimeFormFrom4Velocity}  results in the following expression for the proper-time 1-form
in the coordinate coframe of the rotating medium
\begin{equation}
d\tau = \frac{1}{\gamma}dt' - \gamma\left(\frac{r'^2\omega}{c^2}\right) d\varphi'\;.\label{eq:Deriv_ProperTimeFormInCoordCoframe}
\end{equation}
Now the coordinate cobasis of $K'$ may be $3+1$ decomposed, with the help of \eqref{eq:Theo_3+1DecompositionOfForms}, w.r.t. the 4-velocity $u$ to obtain an adapted cobasis of $\hat{K}$, the natural
coframe of the moving medium. This yields
\begin{align*}
d\tau &= \frac{1}{\gamma}dt' - \gamma\left(\frac{r'^2\omega}{c^2}\right) d\varphi' &\Rightarrow dt' &= \gamma d\tau + \gamma^2\frac{r'^2\omega}{c^2}d\varphi'\\
dr'^\perp &= d\tau\wedge\iota_u(dr') = 0 &\Rightarrow \ul{dr} &= dr'\\
d\varphi'^\perp &= d\tau\wedge\iota_u d\varphi' = 0 &\Rightarrow \ul{d\varphi} &= d\varphi'\\
dz'^\perp &= d\tau\wedge\iota_u(dz') = 0 &\Rightarrow \ul{dz} &= dz'\;.
\end{align*}
And for the adapted cobasis $(d\tau,\ul{dx^a})$ one obtains
\begin{equation}
\hat{\CalB}^\ast =(d\tau,\ul{dr},\ul{d\varphi},\ul{dz})= (d\tau,dr',d\varphi',dz')\;.\label{eq:Deriv_AdaptedCoframeAndRotCoordCoframe}
\end{equation}
The relations between the coordinate cobasis of $K'$ and the adapted cobasis are
\begin{subequations}\label{eq:Deriv_CoordCoframeToAdaptedCoframeTransform}
\begin{align}
d\tau &= \frac{1}{\gamma}dt' - \gamma\left(\frac{r'^2\omega}{c^2}\right) d\varphi' & dt' &= \gamma d\tau + \gamma^2\left(\frac{r'\omega}{c^2}\right)\ul{d\varphi}\\
\ul{dt} &= dr' & dr' &= \ul{dr}\\
\ul{d\varphi} &= d\varphi' & d\varphi' &= \ul{d\varphi}\\
\ul{dz} &= dz' & dz' &= \ul{dz}\;.
\end{align}
\end{subequations}
Substituting the above expressions for the coordinate coframe of $K'$ into \eqref{eq:Deriv_MinkowskiMetricInRotatingCoordCoframe} yields
\begin{subequations}\label{eq:Deriv_MetricInAdaptedCoframe}
\begin{align}
g &= c^2 d\tau\otimes d\tau -\ul{dr}\otimes \ul{dr} - (\gamma r)^2\ul{d\varphi}\otimes\ul{d\varphi} - \ul{dz}\otimes\ul{dz}\label{subeq:Deriv_MetricInAdaptedRotatingCoframe}\\
G = \Mf(g) &= \diag (c^2 , -1, -(r\gamma)^2,-1)
\label{seq:Deriv_GramMatrixInAdaptedRotatingCoframe}\\
\det G &= -(cr\gamma)^2\label{subeq:Deriv_MetricDeterminantInAdaptedRotatingCoframe}\\
G^{-1} &= \diag \left(\frac{1}{c^2}, -1,-\left(\frac{1}{r\gamma}\right)^2,-1\right)\;.\label{subeq:Deriv_GramMatrixInverseInAdaptedRotatingCoframe}
\end{align}
\end{subequations}
\section{Derivation}\label{sec:Deriv_Derivation}
\subsection{Minkowski Relations}
Next one has to determine the coefficients of the excitation form $H$ as functions of the coefficients of the EM forms $(E,B)$ in the rotating frame $K'$ of the observer. It is assumed that the observer
is positioned at the axis of rotation, thus at the origin of the coordinate frame of the rotating medium. Following the prescription provided at the end of the previous chapter one begins
with $F=E\wedge dt'+ B$, transforms it by use of \eqref{eq:Deriv_CoordCoframeToAdaptedCoframeTransform} to the adapted coframe of the medium $\hat{K}$, applies 
\eqref{eq:Def_MaxwellLorentzVacuumRelation} using \eqref{eq:Deriv_MetricInAdaptedCoframe} in conjunction with \eqref{eq:Def_LinearConstitutiveRelations} to obtain $H$, which is then transformed back to the coordinate coframe
of the observer $K'$. Thus
\begin{align*}
F &=E\wedge dt' + B \\
&= E_1'dr\wedge dt'+ E_2'd\varphi'\wedge dt' + E_3' dz'\wedge dt' + B_{23}'d\varphi'\wedge dz' + B_{31}'dz'\wedge dr' + B_{12}'dr'\wedge d\varphi'\\
&= \left[ E_1'\ul{dr} + E_2'\ul{d\varphi}+E_3'\ul{dz}\right]\wedge\left(\gamma d\tau + \gamma^2\frac{r^2\omega}{c^2}\ul{d\varphi}\right)\\
&+ B_{23}' \ul{d\varphi}\wedge\ul{dz} + B_{31}'\ul{dz}\wedge\ul{dr} + B_{12}\ul{dr}\wedge\ul{d\varphi}\\
&= \gamma\left[E_1'\ul{dr} +E_2'\ul{d\varphi}+E_3'\ul{dz}\right]\wedge d\tau + \left[B_{23}'-\gamma^2\frac{r^2\omega}{c^2}E_3'\right]\ul{d\varphi}\wedge\ul{dz}\\
&+ B_{31}'\ul{dz}\wedge\ul{dr} + \left[B_{12}'+\gamma^2\frac{r^2\omega}{c^2}E_1'\right]\ul{dr}\wedge\ul{d\varphi}\\
&= \left[\ul{E}_1 \ul{dr} +\ul{E}_2\ul{d\varphi} + \ul{E}_3\ul{dz}\right]\wedge d\tau + \ul{B}_{23}\ul{d\varphi}\wedge\ul{dz}
+ \ul{B}_{31}\ul{dz}\wedge\ul{dr} + \ul{B}_{12}\ul{dr}\wedge\ul{d\varphi}\;.
\end{align*}
The coefficients of the Faraday 2-form are then
\begin{subequations}\label{eq:Deriv_EMCoeffInAdaptedCoframe}
\begin{align}
\ul{E}_1 &= \gamma E_1' & \ul{E}_2 &= \gamma E_2' & \ul{E}_3 &= \gamma E_3'\\
\ul{B}_{23} &= B_{23}'-\gamma^2\frac{r^2\omega}{c^2}E_3' & \ul{B}_{31} &= B_{31}' & \ul{B}_{12} &= B_{12}' + \gamma^2\frac{r^2\omega}{c^2}E_1'\;.
\end{align}
\end{subequations}
Next one applies the Hodge operator \ref{def:Theo_HodgeOp} using \eqref{eq:GeneralHodgeDualOfTwoForm} and \eqref{eq:Deriv_MetricInAdaptedCoframe}, resulting in
\begin{align*}
\ast F &= -\ul{E}_1\ast\left(d\tau\wedge\ul{dr}\right) -\ul{E}_2\ast\left(d\tau\wedge \ul{d\varphi}\right) -\ul{E}_3\ast\left( d\tau\wedge\ul{dz}\right)\\
&+ \ul{B}_{23}\ast\left(\ul{d\varphi}\wedge\ul{dz}\right) + \ul{B}_{31}\ast\left(\ul{dz}\wedge\ul{dr}\right) +\ul{B}_{12}\ast\left(\ul{dr}\wedge\ul{d\varphi}\right)\\
&= \frac{r\gamma}{c}\ul{E}_1\ul{d\varphi}\wedge\ul{dz} +\frac{1}{cr\gamma}\ul{E}_2\ul{dz}\wedge\ul{dr} + \frac{r\gamma}{c}\ul{E}_3\ul{dr}\wedge\ul{d\varphi}\\
&+ \frac{c}{r\gamma}\ul{B}_{23} d\tau\wedge\ul{dr} + cr\gamma\ul{B}_{31}d\tau\wedge \ul{d\varphi} + \frac{c}{r\gamma}\ul{B}_{12}d\tau\wedge\ul{dz}\;.
\end{align*}
A further application of the linear constitutive relations for a simple medium yields the coefficient functions of the excitation form $H$ in the adapted coframe
\begin{subequations}\label{eq:Deriv_ExcitationCoeffInAdaptedCoframe}
\begin{align}
\ul{\CalH}_1 &= \frac{\lambda_0 c}{\mu r\gamma}\ul{B}_{23}, & \ul{\CalH}_2 &= \frac{\lambda_0 cr\gamma}{\mu}\ul{B}_{31}, & \ul{\CalH}_3 &= \frac{\lambda_0 c}{\mu r\gamma}\ul{B}_{12}\\
\ul{\CalD}_{23} &= \frac{\lambda_0 \epsilon r\gamma}{c}\ul{E}_1, & \ul{\CalD}_{31} &= \frac{\lambda_0 \epsilon}{cr\gamma}\ul{E}_2, & \ul{\CalD}_{12} &= \frac{\lambda_0\epsilon r\gamma}{c}\ul{E}_3\;.
\end{align}
\end{subequations}
The excitation form $H$ is then expressed again in the coordinate coframe $K'$ of the observer by utilizing \eqref{eq:Deriv_CoordCoframeToAdaptedCoframeTransform}:
\begin{align*}
H &= \ul{\CalH}_1d\tau\wedge\ul{dr} +\ul{\CalH}_2 d\tau\wedge\ul{d\varphi} + \ul{\CalH}_3d\tau\wedge\ul{dz} + \ul{\CalD}_{23}\ul{d\varphi}\wedge\ul{dz} + \ul{\CalD}_{31}\ul{dz}\wedge\ul{dr} + \ul{\CalD}_{12}\ul{dr}\wedge\ul{d\varphi}\\
&= dt'\wedge\left[ \frac{1}{\gamma} \ul{\CalH}_1 dr' +\frac{1}{\gamma} \ul{\CalH}_2 d\varphi' + \frac{1}{\gamma}\ul{\CalH}_3 dz' \right] + \left[ \ul{\CalD}_{23} - \gamma\left(\frac{r^2\omega}{c^2}\right) \ul{\CalH}_3 \right]d\varphi'\wedge dz' \\
&+ \ul{\CalD}_{31}dz'\wedge dr'
+\left[\ul{\CalD}_{12}+\gamma\left(\frac{r^2\omega}{c^2}\right)\ul{\CalH}_1\right]dr'\wedge d\varphi'\;.
\end{align*}
From which one can extract the coefficients of the excitation form in the coordinate coframe of the observer
\begin{subequations}\label{eq:Deriv_ExcitationCoeffInCoordCoframe}
\begin{align}
\CalH_1' &= \frac{1}{\gamma}\ul{\CalH}_1 & \CalH_2' &= \frac{1}{\gamma}\ul{\CalH}_2 & \CalH_3' &= \frac{1}{\gamma}\ul{\CalH}_3\\
\CalD_{23}' &= \ul{\CalD}_{23} - \gamma\left(\frac{r^2\omega}{c^2}\right)\ul{\CalH}_3 & \CalD_{31}' &= \ul{\CalD}_{31} &
\CalD_{12}' &= \ul{\CalD}_{12} + \gamma\left(\frac{r^2\omega}{c^2}\right)\ul{\CalH}_1\;.
\end{align}
\end{subequations}
Upon using \eqref{eq:Deriv_ExcitationCoeffInAdaptedCoframe} and \eqref{eq:Deriv_EMCoeffInAdaptedCoframe} one obtains
\begin{subequations}\label{eq:Deriv_ExcitationCoeffMinkowskiRelations}
\begin{align}
\CalH_1' &= \frac{\lambda_0 c}{\mu r' \gamma^2}\left(B_{23}' - \gamma^2 \frac{r'^2\omega}{c^2}E_3'\right)
& \CalH_2' &= \frac{\lambda_0 c r'}{\mu} B_{31}'\\
 \CalH_3' &= \frac{\lambda_0 c}{\mu r'\gamma^2}\left(B_{12}'+\gamma^2\frac{r'^2\omega}{c^2}E_1'\right)
& \CalD_{23}' &= \gamma^2 \left[ \frac{\lambda_0\epsilon r'}{c}-\frac{\lambda_0 c}{\mu}\frac{r'^3\omega^2}{c^4}\right]E_1'-\frac{\lambda_0c}{\mu}\frac{r'\omega}{c}B_{12}\\
 \CalD_{31}' &= \frac{\lambda_0\epsilon}{cr'}E_2'
& \CalD_{12}' &= \gamma^2\left[\frac{\lambda_0\epsilon r'}{c}-\frac{\lambda_0 c}{\mu}\frac{r'^3\omega^2}{c^4}\right]E_3' + \frac{\lambda_0c}{\mu r'}\frac{r'^2\omega}{c^2}B_{23}'\;.
\end{align}
\end{subequations}
\subsection{The Equation}
With the relations for the coefficient functions of the excitation form derived the remainder of the way to the desired partial differential equation for $E_z$
is but an exercise in application of the definitions of \ref{ch:MathematicalMinimum} and \ref{ch:PhysicalBasis}.

Expanding \eqref{subeq:Def_MaxwellEqnsHomogeneous} into $dF = dE\wedge dt' + E\wedge d^2 t' + dB$ and using \ref{def:Theo_ExteriorDerivative} yields
\begin{subequations}\label{eq:Deriv_HMaxwellEqnsRotCoordCoframe_ClassForm}
\begin{align}
\del_{t'}B_{23}' + \del_{\varphi'}E_3' - \del_{z'}E_2' &= 0\label{subeq:Deriv_HMClassForm_1}\\
\del_{t'}B_{31}' + \del_{z'}E_1' - \del_{r'}E_3' &= 0\label{subeq:Deriv_HMClassForm_2}\\
\del_{t'}B_{12}' + \del_{r'}E_2' - \del_{\varphi'} E_1' &= 0\label{subeq:Deriv_HMClassForm_3}\\
\del_{r'}B_{23}' + \del_{\varphi'} B_{31}' + \del_{z'} B_{12}' &= 0\;.\label{subeq:Deriv_HMClassForm_4}
\end{align}
\end{subequations}

From \ref{def:MaxwellEquations}, using \ref{def:Theo_MaxwellLorentzRelation} and assuming $J=0$, it follows that
\begin{equation}
0 = \lambda_0\Delta F = \lambda_0\left[\delta d+d\delta\right]F = \lambda_0 d \delta F\;.\label{eq:Deriv_LaplaceBeltramiInVacuum}
\end{equation}
Where $\Delta$ is the Laplace-Beltrami operator of \ref{def:Theo_LaplaceBeltrami}. Substituting back $H = \lambda_0\ast F$ and 
applying the exterior derivative yields
\begin{align*}
0 = dH &= \left[\del_{t'}\CalD_{12}' - \del_{r'}\CalH_2' + \del_{\varphi'}\CalH_1'\right] dt'\wedge dt'\wedge d\varphi'
+ \left[ \del_{t'}\CalD_{31}' - \del_{z'}\CalH_1' + \del_{r'}\CalH_3 \right] dt'\wedge dz'\wedge dt'\\
&+ \left[ \del_{t'}\CalD_{23}' - \del_{\varphi'}\CalH_3' + \del_{z'}\CalH_2'\right] dt'\wedge d\varphi'\wedge dz'
+ \left[ \del_{r'}\CalD_{23}' + \del_{\varphi'}\CalD_{31}' + \del_{z'}\CalD_{12}'\right] dr'\wedge d\varphi'\wedge dz' .\\
0 &= \ast dH = \left[\del_{t'}\CalD_{12}' - \del_{r'}\CalH_2' + \del_{\varphi'}\CalH_1'\right] \ast \left( dt'\wedge dt'\wedge d\varphi' \right)\\
&+ \left[ \del_{t'}\CalD_{31}' - \del_{z'}\CalH_1' + \del_{r'}\CalH_3 \right] \ast \left( dt'\wedge dz'\wedge dt' \right)
+ \left[ \del_{t'}\CalD_{23}' - \del_{\varphi'}\CalH_3' + \del_{z'}\CalH_2'\right] \ast \left( dt'\wedge d\varphi'\wedge dz' \right)\\
&+ \left[ \del_{r'}\CalD_{23}' + \del_{\varphi'}\CalD_{31}' + \del_{z'}\CalD_{12}'\right] \ast \left( dr'\wedge d\varphi'\wedge dz' \right)\\
&= \left[\del_{t'}\CalD_{12}' - \del_{r'}\CalH_2' + \del_{\varphi'}\CalH_1'\right] dz'
+ \left[ \del_{t'}\CalD_{31}' - \del_{z'}\CalH_1' + \del_{r'}\CalH_3 \right] d\varphi' \\
&+ \left[ \del_{t'}\CalD_{23}' - \del_{\varphi'}\CalH_3' + \del_{z'}\CalH_2'\right] dr'
+ \left[ \del_{r'}\CalD_{23}' + \del_{\varphi'}\CalD_{31}' + \del_{z'}\CalD_{12}'\right] dt'\;.
\end{align*}

A final application of the exterior derivative will finally yield the following second-order partial differential equations for the coefficient functions of
H in the coordinate coframe of the co-rotating observer
\begin{subequations}\label{eq:Deriv_LaplaceBeltrami_Coefficients}
\begin{align}
0 &= \del_{t'}^2 \CalD_{23}' - \del_{r'}^2\CalD_{23}' + \del_{z',t'}^2\CalH_2' - \del_{\varphi',t'}^2\CalH_3'-\del_{r',\varphi'}^2\CalD_{31}' - \del_{z',r'}^2\CalD_{12}'
\label{subeq:Deriv_LaplaceBeltramiDfld_1}\\
0 &= \del_{t'}^2 \CalD_{31}' - \del_{\varphi'}^2\CalD_{31}' + \del_{r',t'}^2\CalH_3' - \del_{z',t'}^2\CalH_1'-\del_{r',\varphi'}^2\CalD_{23}' - \del_{z',\varphi'}^2\CalD_{12}'
\label{subeq:Deriv_LaplaceBeltramiDfld_2}\\
0 &= \del_{t'}^2 \CalD_{12}' - \del_{z'}^2\CalD_{12}' + \del_{\varphi',t'}^2\CalH_1' - \del_{r',t'}^2\CalH_2'-\del_{z',r'}^2\CalD_{23}' - \del_{z',\varphi'}^2\CalD_{31}'
\label{subeq:Deriv_LaplaceBeltramiDfld_3}\\
\intertext{and}
0 &= \del_{r'}^2 \CalH_3' + \del_{\varphi'}^2 \CalH_3' - \del_{r',z'}^2 \CalH_1' + \del_{r',t'}^2 \CalD_{31}' - \del_{\varphi',z'}^2\CalH_2' - \del_{\varphi',t'}^2\CalD_{23}'
\label{subeq:Deriv_LaplaceBeltramiHfld_1}\\
0 &= \del_{\varphi'}^2 \CalH_1' + \del_{z'}^2 \CalH_1' - \del_{\varphi',r'}^2 \CalH_2' + \del_{\varphi',t'}^2 \CalD_{12}' - \del_{r',z'}^2\CalH_3' - \del_{z',t'}^2\CalD_{31}'
\label{subeq:Deriv_LaplaceBeltramiHfld_2}\\
0 &= \del_{z'}^2 \CalH_2' + \del_{r'}^2 \CalH_2' - \del_{\varphi',z'}^2 \CalH_3' + \del_{z',t'}^2 \CalD_{23}' - \del_{\varphi',r'}^2\CalH_1' - \del_{r',t'}^2\CalD_{12}'
\label{subeq:Deriv_LaplaceBeltramiHfld_3}
\end{align}
\end{subequations}

Seeing as the goal of this section is to derive an equation for the $z$ component of the electric field - thus $\CalD_{12}'$ - only \eqref{subeq:Deriv_LaplaceBeltramiDfld_3} is
necessary, the other equations are provided for the sake of completeness. The procedure is explained and executed in a different way in \cite{anderson_electromagnetic_1969}, which
served as a guide for the derivation performed here.
Assuming, as \citeauthor{anderson_electromagnetic_1969} did, that none of the coefficient functions $E_i',B_{ij}'$ depend on $z$ and 
using \eqref{eq:Deriv_HMaxwellEqnsRotCoordCoframe_ClassForm} as well as \eqref{eq:Deriv_ExcitationCoeffMinkowskiRelations} results in (dropping the primes 
to improve readability):
\begin{align*}
0 &= \del_{t}^2 \CalD_{12} - \del_{z}^2\CalD_{12} + \del_{\varphi,t}^2\CalH_1 - \del_{r,t}^2\CalH_2-\del_{z,r}^2\CalD_{23} - \del_{z,\varphi}^2\CalD_{31}\\
&= \del_{t}^2 \CalD_{12} + \del_{\varphi,t}^2\CalH_1 - \del_{r,t}^2\CalH_2\\
&= \del_t^2 \left[ \gamma^2\left[\frac{\lambda_0\epsilon r}{c}-\frac{\lambda_0 c}{\mu}\frac{r^3\omega^2}{c^4}\right]E_3 + \frac{\lambda_0c}{\mu r}\frac{r^2\omega}{c^2}B_{23}\right]
+ \del_{\varphi,t}^2\left[ \frac{\lambda_0 c}{\mu r \gamma^2}\left(B_{23} - \gamma^2 \frac{r^2\omega}{c^2}E_3\right) \right]
-\del_{r,t}^2\left[\frac{\lambda_0 c r}{\mu} B_{31}\right]\\
&= \frac{\lambda_0 cr}{\mu}\left[\left(\frac{\gamma}{c}\right)^2\left(n^2-\left(\rocf\right)^2\right)\del_{t}^2 E_3
+ \frac{\omega}{c^2}\left(-\del_\varphi \del_tE_3\right) + \frac{1}{r^2}\left[1-\left(\rocf\right)^2\right] \left(-\del_\varphi^2 E_3\right) \right.\\
&- \left.\frac{\omega}{c^2}\del_\varphi \del_t E_3 -\frac{1}{r}\del_r E_3 -\del_r^2 E_3\right]\;.\
\end{align*}
After multiplying the last equation with $-\tfrac{\mu}{\lambda_0 c r}$ and some rearrangement one obtains
\begin{equation}
\boxed{0 = \Delta_{r,\varphi} E_3 - \left( \frac{\omega}{c} \right)^2 \del^2_\varphi E_3 + 2 \frac{\omega}{c^2} \del_\varphi \del_t E_3
- \left[ n^2 - \left( \frac{r\omega}{c}\right)^2 \right] \left[ \frac{\gamma}{c}\right]^2 \del_t^2 E_3 }\quad .\label{eq:10}
\end{equation}

This is the central equation of this work. It encodes the influence of motion on the observations of a co-rotating observer through a 
Coriolis-Zeeman term $\propto 2\tfrac{\omega}{c^2}\del_{\varphi,t}^2$ as well as the terms of higher order in the angular velocity of rotation.

In the case of a stationary system the equation obviously reduces to the classical wave equation (save for the index of refraction), since
$\tfrac{\omega}{c}\rightarrow 0,\;\gamma\rightarrow1$:
\begin{equation*}
0 = \Delta_{r,\varphi} E_3 -  n^2 \frac{1}{c^2}\del_t^2 E_3 \quad .
\end{equation*}

Following the above derivation it becomes obvious that the wave equation for the $z$ component of the electric field acquired an additional term just
by changing the metric by a coordinate transformation. The Coriolis (or Zeeman) like term couples the spatial and temporal parts of the equation.
Using the same operator as Kac and others but considering the dynamics of rotating domains adds terms to the original wave equation.
One can already hypothesize that the additions will not go unnoticed by the eigenfrequencies of the domains in question.

\section{Verification}\label{sec:Deriv_Verification}

Consider the ansatz proposed by \citeauthor{anderson_electromagnetic_1969} in  \cite{anderson_electromagnetic_1969}:
\begin{equation*}
E_3(r,\varphi; t) = E_3(r) e^{i(\kappa\varphi - \Omega t)}\quad\kappa,\Omega \in \RR\quad .
\end{equation*}
This yields
\begin{equation}
\begin{split}
\del_t E_3 &= -i\Omega E_3\\
\del_t^2 E_3 &= (-i\Omega)^2 E_3  = -\Omega^2 E_3\\
\del_\varphi E_3 &= i\kappa E_3\\
\del^2_\varphi E_3 &= -\kappa^2 E_3\quad .
\end{split}\label{eq:12}
\end{equation}
Plugging these into \eqref{eq:10} we obtain
\begin{equation*}
0 = \del^2_r E_3 +\frac{1}{r}\del_r E_3 - \left(\frac{\kappa}{r}\right)^2 E_3 + \left( \frac{\kappa\omega}{c}\right)^2 E_3
- \frac{2i^2\Omega\omega\kappa}{c^2} E_3 + \left[ n^2 - \left( \frac{r\omega}{c}\right)^2 \right]\left[ \frac{\Omega\gamma}{c}\right]^2 E_3\;,
\end{equation*}
and assuming $c= 1$ results in:
\begin{equation}
0 = \left( \del^2_r + \frac{1}{r}\del_r -\left(\frac{\kappa}{r}\right)^2 + (\kappa\omega)^2 + 2\omega\Omega\kappa 
+\left[n^2 -(r\omega)^2\right] [\Omega\gamma]^2\right) E_3\; .\label{eq:13}
\end{equation}

Let us compare this result to \cite[eq. (92)]{anderson_electromagnetic_1969} with constants $\alpha,\beta,\lambda$ taken from Tbl. III (Case 3) therein and $v=r\omega$:

\begin{align*}
0 &= \left[  \frac{1}{n^2}\del_r r\del_r +\left( \Omega^2\alpha r-2\lambda\kappa\Omega - \beta\frac{\kappa^2}{r}\right)\right] E_3\\
&= \left[ \frac{1}{n^2}\del_r r\del_r +\left( r\Omega^2\left[ 1 + \left( 1-\frac{1}{n^2}\right) \gamma^2(r\omega)^2\right]
-2\kappa\Omega\left[ -\frac{r\omega}{n^2}\right] -\frac{\kappa^2}{r}\left[ \frac{1}{n^2}\left( 1-(r\omega)^2\right)\right]\right)\right]E_3\\
\overset{n^2}{\Rightarrow}\quad 0 &= \left[ \del_r + r\del^2_r +\left( r\Omega^2\left[ n^2+\left(n^2-1\right)\gamma^2(r\omega)^2\right]
+2\kappa r\Omega\omega - \frac{\kappa^2}{r}\left(1 - (r\omega)^2\right)\right)\right] E_3\\
\overset{\tfrac{1}{r}}{\Rightarrow}\quad 0 &= \left[ \del^2_r + \frac{1}{r}\del_r +\left( \Omega^2\left[ n^2+(n^2-1)\gamma^2(r\omega)^2\right]
+ 2\kappa\omega\Omega - \left[\frac{\kappa}{r}\right]^2 \left(1-(r\omega)^2\right)\right)\right] E_3\\
 &= \left[ \del^2_r + \frac{1}{r}\del_r + \left(n\Omega\right) +(n^2-1)(\gamma\Omega)^2(r\omega)^2
+ 2\kappa\omega\Omega - \left[\frac{\kappa}{r}\right]^2 + (\kappa\omega)^2\right] E_3\\
 &= \left[ \del^2_r + \frac{1}{r}\del_r + 2\kappa\omega\Omega - \left[\frac{\kappa}{r}\right]^2 + (\kappa\omega)^2
+n^2\Omega^2\left(1+\gamma^2(r\omega)^2\right) -(\gamma\Omega r\omega)^2\right] E_3\\
 &= \left[ \del^2_r + \frac{1}{r}\del_r + 2\kappa\omega\Omega - \left[\frac{\kappa}{r}\right]^2 + (\kappa\omega)^2
+n^2\Omega^2\gamma^2 -\gamma^2\Omega^2(r\omega)^2 \right] E_3\\
 &= \left[ \del^2_r + \frac{1}{r}\del_r + 2\kappa\omega\Omega - \left[\frac{\kappa}{r}\right]^2 + (\kappa\omega)^2
+(n^2-(r\omega)^2)(\gamma\Omega)^2 \right] E_3\quad .
\end{align*}
Where the last equation is obviously equivalent to \eqref{eq:13}.

Consider now the ansatz made by \cite{sunada_sagnac_2006}:
\begin{equation}
E_3(r,\varphi;t) = E_3(r,\varphi)e^{-ikct}\;.\label{eq:SHTemporalDependenceAnsatz}
\end{equation}
It obviously yields for \eqref{eq:10}
\begin{equation}
0 = \Delta E_3 - \left(\frac{\omega}{c}\right)^2\del^2_\varphi E_3 -\frac{2ik\omega}{c}\del_\varphi E_3
+ \left[ n^2 -\left(\frac{r\omega}{c}\right)^2\right] \left[ k\gamma \right]^2 E_3\quad .
\end{equation}
Neglecting terms $\mathcal{O}\left( \left(\tfrac{r\omega}{c}\right)^2 \right)$ yields $\gamma\approx 1$ and:
\begin{equation*}
0 = \Delta E_3 -\frac{2ik\omega}{c}\del_\varphi E_3 + n^2k^2 E_3\quad.
\end{equation*}
Transforming the above to Cartesian coordinates  yields:
\begin{equation}
0 = \Delta_{x,y} E_3 -\frac{2ik\omega}{c}\left(-y\del_x + x\del_y\right)E_3 + n^2k^2 E_3\quad .\label{eq:SHDerived}
\end{equation}
This equation, with the substitution $-ik\rightarrow\lambda$ shall be referred to as the linearised version of \eqref{eq:10}.
Note that I choose, again in an abuse of notation, to use the same symbol $E_3$ when working in polar and in Cartesian coordinates.
This is done for the sake of better readability as the coordinate system of interest will always be obvious. Compare to \cite[eq. (1)]{sunada_design_2007}  or \cite[eq. (2)]{sunada_sagnac_2006}:
\begin{equation*}
0 = \Delta\psi + n^2k^2 \psi - 2ik\vec{h}\cdot\vec{\nabla}\psi 
\end{equation*}
with $\psi = E_3,\ \vec{h}:= \tfrac{1}{c} \vec{r}\times\vec{\omega},\ \vec{\omega}:=\omega \vec{e}_z,\quad \omega\in\RR_+$. Thusly:
\begin{equation}
0 = \Delta\psi + n^2k^2\psi -\frac{2ik\omega}{c}\begin{pmatrix}
y\\
-x\\
0
\end{pmatrix}\cdot\begin{pmatrix}
\del_x\\
\del_y\\
\del_z
\end{pmatrix}\psi
= \Delta\psi +n^2k^2\psi -\frac{2ik\omega}{c}\left( y\del_x - x\del_y\right)\psi\quad . \label{eq:SHReference}
\end{equation}
One can see that \eqref{eq:SHDerived} $\neq$ \eqref{eq:SHReference} unless the additional sign of the term
linear in $\omega$ is absorbed into $\omega$, thereby reversing the direction of rotation. This discrepancy between
the above derivation and \cite{sunada_sagnac_2006} is due to the choice of the sign in the ansatz $e^{-ickt}$
for the temporal dependence of $E_z$, which is equivalent to the choice of the sense of rotation of the domain (clock-wise 
vs. counter clock-wise).

The comparison to \cite{sunada_design_2007} may seem arbitrary, provided that the correctness of \eqref{eq:10} has been verified by comparing 
its re-formulation \eqref{eq:13} to the results obtained by \citeauthor{anderson_electromagnetic_1969}. It is nevertheless necessary, for the
microdisc discussed in \cite{sunada_design_2007} is used in \ref{ch:Exp_TheExperiment} as a verification of the correctness of
the numerical method for the solution of \eqref{eq:10} chosen here.

\chapter{Analysis}\label{ch:TheoAnalysis}
\section{The Unit Square}\label{sec:Theo_UnitSquare}
Consider the wave equation for the unit square $\Cr = [0,1]^2$ in euclidean space (inertial frame). The solutions to
\begin{equation}
\square\Psi = \left(\Delta\Psi - \frac{1}{c^2}\del_t^2\right)\psi = 0\label{eq:HelmholtzEquation}
\end{equation}
are known to be products of trigonometric functions. Assuming Dirichlet boundary conditions a complete set
of solutions consists of 
\begin{equation}
\Psi_{k,m}(x,y) = \sin(k\pi x)\sin(m\pi y)\quad k,m\in\NN_+\quad .
\end{equation}
These are classically derived by assuming the time dependence of the fields to be of the form $e^{i\alpha ct},\;\alpha\in\RR$.
Assuming instead $e^{\lambda ct},\;\lambda\in\CC$ one obtains
\begin{equation}
\lambda_{k,m} = \pm i \alpha_{k,m}\quad .
\end{equation}
This is as good a place as any to notice that by using the above assumption we fix the dimension of $\lambda$ to be $\left[ \lambda\right] = \tfrac{1}{L} \overset{\scriptsize \text{SI}}{=} \tfrac{1}{m}.$
Using \eqref{eq:SHDerived} with the above assumption yields (in Cartesian coordinates)
\begin{equation*}
0 = \Delta\Psi + 2\frac{\omega}{c}\lambda\left(-y\del_x+x\del_y\right)\Psi - \lambda^2n^2\Psi\quad.
\end{equation*}
For $n=1$ and stationary domain one obtains \eqref{eq:HelmholtzEquation}. One therefore may expand the eigenfunctions of \eqref{eq:10} 
on $\Cr$ in a Fourier series:
\begin{equation}
u(x,y) = \sum_{k,m=1}^\infty c_{k,m}\Psi_{k,m}(x,y)\quad c_{k,m}\in\RR.
\end{equation}
One may now reformulate the problem in the following way ($n=1$):
\begin{align*}
0 &= \Delta u + \frac{2\omega}{c}\lambda(-y\del_x+x\del_y)u - \lambda^2 u\\
 &= \sum_{k,m=1}^\infty c_{k,m} \bigl[ -\pi^2\left(k^2+m^2\right)\Psi_{k,m} + \frac{2\omega}{c}\lambda
\left(-y\del_x+x\del_y\right)\Psi_{k,m} - \lambda^2\Psi_{k,m} \bigr]\quad .
\end{align*}
Next one multiplies the above with $\Psi_{n,l}$ from the left and integrates over $\Cr$. In doing so one obtains:
\begin{align*}
0 &= \sum_{k,m=1}^\infty c_{k,m}\left[ -\pi^2\left(k^2+m^2\right) \brak{\Psi_{n,l} }{ \Psi_{k,m} } + \frac{2\omega}{c}\lambda
\brak{ \Psi_{n,l} }{ (-y\del_x+x\del_y)\Psi_{k,m} } - \lambda^2 \brak{ \Psi_{n,l} }{ \Psi_{k,m} } \right]\\
&\overset{*}{\approx} \sum_{k,m=1}^N c_{k,m}\left[ -\pi^2\left(k^2+m^2\right) \brak{\Psi_{n,l} }{ \Psi_{k,m} } + \frac{2\omega}{c}\lambda
\brak{ \Psi_{n,l} }{ (-y\del_x+x\del_y)\Psi_{k,m} } - \lambda^2 \brak{ \Psi_{n,l} }{ \Psi_{k,m} } \right]\\
&= \left[ \Amat + \frac{2\omega}{c}\lambda \Cmat - \lambda^2 \frac{1}{4} \mathbb{I}\right]\cx\quad,\refstepcounter{equation}\tag{\theequation}\label{eq:ModalDecompositionQEVP}\\
\Amat_{(n,l),(k,m)} &= -\frac{\pi^2}{4} ( n^2 + l^2)\delta^n_k \delta^l_m\\
\Cmat_{(n,l),(k,m)} &= \begin{cases}
0 & k=m=n=l\wedge k=m\neq n=l\\
\frac{2(-1+(-1)^{l+m})(-1+(-1)^{m+n} )lm^2 n}{(l^2-m^2)\pi^2}\left[\frac{1}{m^2-n^2}+\frac{1}{(m^2-n^2)^2}\right] & k\neq m\neq n=l \\
\frac{2(-1+(-1)^{l+m})(-1+(-1)^{m+n} )klmn}{(l^2-m^2)\pi^2}\left[\frac{1}{k^2-n^2}+\frac{1}{(k^2-n^2)^2}\right] & k\neq m\neq n\neq l
\end{cases}
\end{align*}
where in $*$ one introduces a truncation mode number $N$ to make the problem tractable on a computer.
The above is a decomposition of the eigenfunctions of $\mathcal{R}$ for \eqref{eq:SHDerived} into Fourier modes - thus it is a modal decomposition.
Equation \eqref{eq:ModalDecompositionQEVP} is an eigenvalue problem where the eigenvalue $\lambda$ occurs to linear and quadratic order. The eigenvalue problem (EVP for short)
is thus commonly referred to as a quadratic eigenvalue problem, or QEVP. This type of eigenvalue problems is discussed further in \ref{sec:ExpIntro_QEVP}.
Note that the matrix $\Cmat$ is,by construction, antisymmetric. This will have major implications in \ref{ch:Exp_TheExperiment}, where
\eqref{eq:ModalDecompositionQEVP} will be solved numerically.

This shall be used in \ref{sec:Exp_Validation} to validate a part of the numerical method chosen to investigate the isospectral domains.
The expression $\brak{\cdot }{ \cdot}$ in the above derivations denotes the standard $L^2$ inner product on $\Cr$.

An important point to note in the above derivation is the expression for $\Cmat_{(n,l),(k,m)}$ where $k=m=n=l$ or $k=m\neq n=l$. These coefficients represent the 
first order contribution of the Coriolis term to the eigenvalues/eigenfrequencies of the system. Indeed, if one were to perform calculations necessary
for first-order perturbation theory, with $\tfrac{\omega}{c}$ as the parameter, one would realise that the integrals vanish due to the integration of an odd
function over a symmetric domain. Thus one may expect a contribution of the rotation to the eigenfrequencies of the square at the earliest to second order in $\tfrac{\omega}{c}$.
This is in stark contrast to the results obtainable analytically for a (unit) disc, as shown in the next section.

\section{The Unit Disc}\label{sec:Theo_UnitDisc}
\subsection*{Static Analysis}
An analytically tractable case is given by a disc in $\RR^2$. For simplicity the unit disc $\Cd_1=\bigl\{\xx\in\RR^2\; :\; \Vert \xx\Vert_2 \leq 1\bigr\}$
is chosen as the domain with $n=1$. Starting with \eqref{eq:10} and assuming
\begin{align}
E_3(r,\varphi;t) &= \Psi(r,\varphi)e^{c\lambda t}\qquad\lambda\in\CC \notag\\
0 &= \Delta_{r,\varphi}\Psi - \left[\ocf\right]^2\del_\varphi^2\Psi + 2\ocf\lambda\del_\varphi\Psi - \lambda^2 \Psi\quad .\label{eq:UnitDisc_PDE_QEVP}
\end{align}
Clearly, to remain in the realm of physically meaningful equations one has to require $\tfrac{R_0\omega}{c}\leq 1$, where $R_0=1$ is the upper bound on the extent of
the domain from the origin, through which goes, by derivation, the axis of rotation. Considering the above equation in one can see, that a sufficient
condition for $\lambda$ to remain unaffected by rotation is for the associated eigenmode of the disc to be invariant under continuous rotations.

It is trivial to verify that the lowest eigenmode of the disc fulfils this requirement.
Using the above equation and assuming $\omega=0$ one may follow the procedure for deriving the common Bessel's equation. Due to $\lambda\in\CC$ and the resulting difference in signs
the result will be a modified Bessel's equation, whose solutions are combinations of modified Bessel functions $I_m(x),K_m(x)$. By asymptotic analysis
$K_m$ can be neglected and using $I_\nu(x) = e^{\pm i\nu\tfrac{\pi}{2}}J_\nu\left(xe^{\pm i\tfrac{\pi}{2}}\right)$ one returns to the common Bessel's
equation.
The static eigenvalues are then
\begin{equation}
\lambda_{m,s} = \pm j_{m,s}\;,
\end{equation}
where $j_{m,s}$ is the s-th root of the Bessel function of the first kind $J_m$ and is real \cite{abramowitz_handbook_????}.

The analysis of the unit disc is carried out in full in \ref{sec:App_UnitDisc}

\subsection*{Dynamic Analysis}
If one drops the assumption $\omega = 0$ and follows the same steps as for the stationary case, one will arrive at the following precursor of 
a modified Bessel equation
\begin{equation*}
0 = x^2 \tilde{R}'' +x\tilde{R}' - \left[x^2\left(1\mp\frac{2i\omega}{c\lambda} m -\left(\frac{\omega}{c\lambda}\right)^2m^2\right) +m^2\right] \tilde{R}\;.
\end{equation*}
Using \eqref{eq:App_UnitDiscSecondSubstitution}, specifically defining
\begin{equation*}
u(x) := x\sqrt{ 1\mp 2\frac{i\omega}{c\lambda} m -\left(\frac{\omega}{c\lambda}\right)^2}\; ,
\end{equation*}
will then result in a modified Bessel equation for $\hat{R}(u)$. From the latter it then follows that following may be set
\begin{equation}
\lambda_{m,s}(\omega) := i \left[ j_{m,s} +\frac{m\omega}{c}\right]\;.\label{eq:TheoAn_UnitDiscEVEvolution}
\end{equation}

This form confirms the conclusions at which \citeauthor{sunada_sagnac_2006} arrived in \cite{sunada_sagnac_2006}, that the eigenmodes which represent co- and counter-propagating 
waves on a disc diverge linearly with increasing angular velocity of rotation. Their statement is improved here in so far that for vacuum the change is purely linear in $\omega$
and no higher order terms are present.

If preference is given to the wave number $k$ instead of $\lambda$\footnote{Or the angular frequency $\Omega = 2\pi ck$.}, then using the relation $-ik=\lambda$ one obtains
\begin{equation}
k_{m,s} = \frac{m\omega}{c} + \bar{k}_{m,s}\;,\label{eq:TheoAn_UnitDiscWaveNumberChange}
\end{equation}
where $\bar{k}_{m,s}$ again denotes the static wave number.

\section{Isospectral Domains}\label{sec:Theo_Analysis_IDs}
\begin{figure}[tb]
\centering
\begin{tikzpicture}
	\begin{axis}[plot box ratio=1 1]
		\addplot[mark=none,patch, patch type=triangle,mesh,blue] file {Plots/Geometries/ID1_1_pgfmesh.dat};
	\end{axis}
\end{tikzpicture}
\hspace{0.15cm}
\begin{tikzpicture}
	\begin{axis}[plot box ratio=1 1]
		\addplot[mark=none,patch, patch type=triangle,mesh,blue] file {Plots/Geometries/ID1_2_pgfmesh.dat};
	\end{axis}
\end{tikzpicture}
\caption[Triangulations of the domains of fig. \ref{subfig:SimpleIDs}.]{The triangulations of the domains of Fig. \ref{subfig:SimpleIDs}. The make-up of the domains is apparent. These triangulations serve
as the starting meshes for the numerical solution of \eqref{eq:10} using finite element methods \ref{ch:Exp_AlgebraicFormulation}. }.
\label{fig:ID1_Triangulations}
\end{figure}
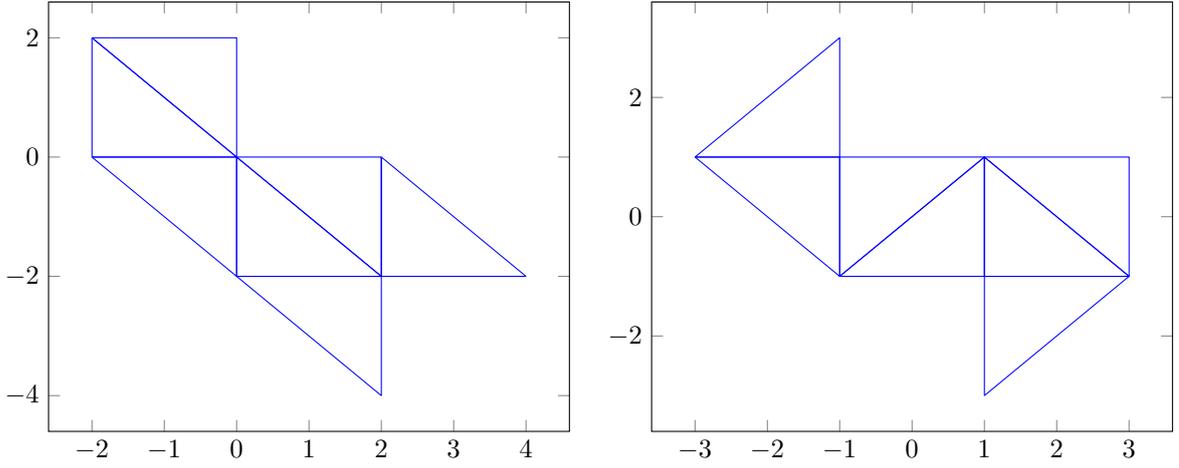
\subsection*{Static Analysis}
It has already been mentioned, that the eigenvalues and eigenmodes of the domains of \ref{fig:Intro_IsospectralManifolds} are not analytically known, save for a few.
The proofs mentioned in \ref{sec:Theo_IDsPriorArt} do not require explicit knowledge of either. Nevertheless, a select set of eigenvalues and eigenfunctions may be determined analytically.
This shall be demonstrated below for domains of fig. \ref{subfig:SimpleIDs} (resp. \ref{fig:ID1_Triangulations}) assuming Dirichlet boundary conditions. 
The parameters of the domains are given in \ref{tbl:SimParamID1} on p. \pageref{tbl:SimParamID1}. 
The dimensions were chosen to facilitate the comparison of computed eigenvalues to those provided by \citeauthor{driscoll_eigenmodes_1997} in \cite{driscoll_eigenmodes_1997}.

Upon closer inspection it becomes obvious that the domains are composed of isosceles triangles (of common side length $2$). This is clearly visible in \ref{fig:ID1_Triangulations}, where
coarse triangulations of the domains are provided. Some eigenmodes of each domain may therefore be composed of eigenmodes of isosceles triangles, provided the latter are contained completely
in said triangles. This is equivalent to requiring that Dirichlet boundary conditions are fulfilled for each triangle and then covering the domains with the triangular tiles.
The eigenfunctions of a (fundamental) tile are then transplanted onto each tile and thus onto the domain. This is a practical application of the transplantation method of  \citeauthor{buser_isospectral_1986}
mentioned in \ref{sec:Theo_IDsPriorArt}. It should be noted that \citeauthor{sridhar_experiments_1994} provide a different and accessible explanation of the transplantation method in \cite{sridhar_experiments_1994}. 
For a treatment of isospectral domains it is therefore necessary to obtain eigenvalues and eigenfunctions of right triangles. This is easily done
by noticing that a square is composed of two such triangles. Eigenfunctions of the triangle will therefore be linear combinations of the eigenfunctions of a square.
A deeper analysis of eigenfunctions of triangles is provided in \cite{mccartin_laplacian_2011}.
The eigenfunctions provided in \ref{sec:Theo_UnitSquare} yield for the present case
\begin{align}
\Psi_{k,m}(x,y) &= \sin\left(\frac{k\pi}{L}x\right)\sin\left(\frac{m\pi}{L}y\right) = \sin\left(\frac{k\pi}{2}x\right)\sin\left(\frac{m\pi}{2}y\right)\label{eq:Analysis_Eigenfunc2Square}\\
\alpha_{k,m}^2 &= -(k^2+m^2)\frac{\pi^2}{L^2} = -(k^2+m^2)\frac{\pi^2}{4}\notag\\
\lambda_{k,m}^2 &= i(k^2+m^2)\frac{\pi^2}{L^2} = i(k^2+m^2)\frac{\pi^2}{4}\;.\label{eq:Analysis_Eigenval2Square}
\end{align}
Here $L$ is the length of a side of the square.
As only triangles with Dirichlet boundary conditions are considered an a linear combination of eigenfunctions of the square may be an eigenfunction of the triangle
if and only if it has a node-line along each edge of the triangle. This eliminates highly symmetric eigenfunctions ($k=m$) of the square and only eigenfunctions with $k\neq m$ 
may be considered as building blocks for eigenfunctions of the triangles
\begin{equation}
\phi^{\pm}_{k,m}(x,y) = \frac{1}{\sqrt{2}}\left(\Psi_{k,m} \pm \Psi_{m,k}\right)\;.\label{eq:Analysis_EigenfunctionsOfTriangles}
\end{equation}
The functions are locally eigenfunctions of a triangle but, provided the triangulation is appropriate - the squares provide some freedom - one or the other will be a global eigenfunction of a domain.
The first (real) eigenvalue of the domains obtained in such fashion is therefore
\begin{equation}
\alpha_9^2 =\alpha_{1,2}^2 = -\frac{\pi^2}{4}\left(1^2+2^2\right) = -\frac{5\pi^2}{4}\;,\label{eq:Analysis_IDFirstAnaEigenval}
\end{equation}
where the index $9$ is taken from \cite{driscoll_eigenmodes_1997} and indicates that this is not the lowest eigenvalue of the domains (c.f. \ref{sec:Exp_Validation}).
Note that the next eigenvalue of the static isospectral domains known analytically is $\alpha_{21} = -\tfrac{10\pi^2}{4}$ which suggests an increase in the number of
eigenvalues located between two consecutive analytically known eigenvalues.
In deference to the later discussion I choose an abuse of notation and denote the real eigenvalues $\alpha_{i,j}$ by the common place-holder $\lambda_{i,j}$. It will be
apparent whether a complex or a real value is being considered.
\subsection*{Dynamic Analysis}
With an analytic expression of an eigenfunction one may immediately attempt to apply first order perturbation theory to determine the changes in the associated eigenvalue.
Alas, as has been explained at the end of \ref{sec:Theo_UnitSquare} the eigenfunctions of an isospectral domain, as given in \eqref{eq:Analysis_EigenfunctionsOfTriangles},
will suffer the same fate as those of a square. As such the Coriolis term in \eqref{eq:10} will not, to first order in $\tfrac{\omega}{c}$, contribute to the change
of the eigenvalues of a triangle.

First order stationary perturbation theory therefore provides a lower bound on the degree of the dependence of the
eigenvalues on rotation. The result of this brief analysis is that, if the eigenvalues of the isospectral domains change, they do so at least $\sim \left( \tfrac{\omega}{c}\right)^2$.


Let therefore $\lambda_k^{(i)},i = 1,2$ be the $k-$th eigenvalue of the first and second domain of an isospectral pair. Due to the above analysis of $\lambda_9$ 
using first order perturbation theory one may assume that the $\omega$-dependence of the eigenvalues is of the following form
\begin{equation}
\lambda_k(\omega) = \lambda_{k,0} + \alpha_k \omega^2 + \mathcal{O}(\omega^3)\label{eq:EVasFunctionOfAngFreq}
\end{equation}
where $\alpha_k\in\RR$ is a proportionality constant and $\lambda_{k,0}$ is the stationary eigenvalue, i.e., the eigenvalue of 
the common Laplacian. With this one obtains the following expressions for the anisospectrality of
both domains depending on $\omega$:
\begin{align}
\lambda_k^{(1)}(\omega) - \lambda_k^{(2)}(\omega) &= \lambda_{k,0}^{(1)} +\alpha_k \omega^2 - \lambda_{k,0}^{(2)} -\beta_k \omega^2 + \mathcal{O}(\omega^3)\notag\\
&= (\alpha_k-\beta_k)\omega^2 + \mathcal{O}(\omega^3)\label{eq:EV_PredictedChange}\\
\left( \lambda_k^{(1)}\right)^2(\omega) - \left(\lambda_k^{(2)}\right)^2(\omega) &= 
\left[ \left( \lambda_{k,0}^{(1)}\right)^2 +2\alpha_k \lambda_{k,0}^{(1)}\omega^2 + \alpha_k^2 \omega^4 \right]
-\left[ \left( \lambda_{k,0}^{(2)}\right)^2 +2\beta_k \lambda_{k,0}^{(2)}\omega^2 + \beta_k^2 \omega^4 \right]\notag\\
&= 2\lambda_{k,0}\left(\alpha_k-\beta_k\right) \omega^2 + \mathcal{O}(\omega^3)\quad .\label{eq:EV_Square_PredictedChange}
\end{align}
Here  $\mathcal{O}(\omega^3)$ was omitted in the third equation for reasons of brevity but was reintroduced in the last equation
to emphasize, that it is the next-leading-order term when the squares of the eigenvalues are compared.
Higher order perturbation theory can not be applied because analytical expressions for all eigenfunctions of the static domains are not known.
This does not preclude the possibility of applying some form of higher-order perturbation theory to numerically determined eigenfunctions of the domains. Such an approach 
may be used to verify the conclusions made above and those made in the second part of this work.
Note that further analysis has been deferred to \ref{sec:Exp_Results}.

\part{Numerical Experiment}

\chapter{Introduction}\label{ch:Exp_Introduction}
This part of the thesis is dedicated to the practical efforts of determining an answer to the question whether
domains which are isospectral at rest remain that way when accelerated, specifically when they are subject to uniform rotation.

As has been shown in the first part of this work (\ref{sec:Theo_IDsPriorArt}) analytical statements about the spectrum of such domains are not trivial, even in the case of flat surfaces embedded in the three dimensional physical space.
As the derivation of \eqref{eq:10} has shown, subjecting a domain to uniform rotation requires, in the context of geometric electrodynamics, its treatment as a (part of)
Lorentzian manifold. To be able to make a statement aout the properties of the spectra of rotating isospectral drums I will resort to
determining said spectra numerically.

In \ref{sec:ExpIntro_MethodOverview} different ways of tackling a partial differential equation numerically are presented, which could be used in the
present case. Their advantages and disadvantages are discussed and the choice of the finite element method is deliberated.
The latter is then elaborated upon in the \ref{ch:Exp_AlgebraicFormulation}, where the formulae necessary for an implementation of the FEM are derived
and the theoretical and practical aspects of the method are discussed in the present context.
As has been shown in \ref{sec:Theo_UnitSquare} a discrete formulation of \eqref{eq:10} will yield a non-standard eigenvalue problem.
The peculiarities of quadratic eigenvalue problems (QEVP), specifically the problem resulting from \eqref{eq:10}, are discussed in \ref{sec:ExpIntro_QEVP}.
Therein the method for solving the QEVP at hand is presented, too.

Due to the fact that the results of the present work rely heavily on the numerical solution of \eqref{eq:10},
the software written for the purpose warrants an extensive validation. The results thereof are presented
in \ref{sec:Exp_Validation} with further demonstrations of correctness not presented here due to being of secondary importance to the problem at hand.

The results of this thesis and the answer to the ever present question are provided in \ref{ch:Exp_TheExperiment}, specifically in \ref{sec:Exp_Results} and
\ref{sec:Exp_Discussion}.
\section{Discretization Methods}\label{sec:ExpIntro_MethodOverview}
In the present case, unfortunately, the domains of interest have, by construction, no apparent symmetry and
an attempt at treatment of \eqref{eq:10} using a separation of variables ansatz fails\footnote{Henceforth all considerations take place in $\RR^2$.}.
To tackle a PDE, that is an equation for a continuous variable, on a computer, which can only represent a finite subset of $\RR$, the PDE has to be discretized. 
The simplest approach to a discretization of \eqref{eq:10} is to revert to finite differences wherever
derivatives are required. Generally this is done by covering the computational domain with a tensor-product mesh
and approximating the derivatives at each point by finite differences. These are the premises of finite-differences
methods, whose main advantage is their simplicity. The disadvantages outweigh the advantages of FD methods, for apart from posessing a low order 
of consistency in general and a low convergence rate (c.f. \cite[Ch. 11]{Quarteroni2002}) the methods are saddled with the problem of a TP mesh.
The simplicity of implementation of methods relying on a TP grid (or mesh) is paid for by a severe limit  on the geometries which may be considered.
Generally only rectangular domains yield themselves to TP methods, although by application of clever transformations this class can be extended.
The limits on geometry may be circumvented by embedding an irregular domain into a TP mesh, which requires aligning the boundaries of the domain to the mesh and
in general is similar to the problem of squaring the circle.
A sufficient resolution of the finer details of the domain or a function on it introduces, in general, additional mesh nodes which may not be necessary.
These superfluous nodes bloat the system matrices (representing differential operators) and lead to an increase in runtime and memory consumption, whilst contributing
little meaningful information to the results.
In the case of stationary isospectral domains the use of finite differences methods
has been attempted, as \citeauthor{driscoll_eigenmodes_1997} reports in \cite{driscoll_eigenmodes_1997}, with very modest results.

The geometric problems of the TP mesh are also shared by methods with global ansatz functions \cite[Ch. 4.2.2]{Deuflhard2013} and methods which
rely on interpolation on a TP grid as described in extenso in \cite{Kopriva2008}. 
So called collocation methods are commonly but not universally implemented on a TP mesh. This class of methods approximates the solution to a PDE by interpolation using
orthogonal polynomials as a basis. The advantage of these methods is that it is generally possible to determine the discrete representations of differential operators
analytically, by exploiting suitable identities of the chosen basis functions, and that the order of convergence of such methods tends to be very high.
Their downside is that the resulting matrices tend to be quite dense.
A remedy to this could be the use of matrix-free methods. This would not circumvent the problem of having a quadratic eigenvalue problem (c.f. \eqref{eq:ModalDecompositionQEVP}) in the present case.
No method is known to me to solve the latter, for eigenvalues smallest in magnitude, without providing an explicit matrix. 
Collocation methods are often in the category of spectral methods\footnote{Methods with exponential convergence \cite{Kopriva2008}.} but are not its sole constituents.

The collocation approach was  taken by \citeauthor{driscoll_eigenmodes_1997} as well as \citeauthor{betcke_reviving_2005}, with very good results for static domains (isospectral domains at rest).
As has been remarked by \citeauthor{driscoll_eigenmodes_1997} in \cite{driscoll_eigenmodes_1997} there exists an analytical solution of the Poisson equation at an infinitely extended
wedge.
One might therefore try to make an ansatz for the solution of \eqref{eq:10} as a truncated Fourier-Bessel series. Alas, the boundaries of the domains are not infinitely
extended and multiple corners have to be considered.
Nevertheless \citeauthor{betcke_reviving_2005} exploit this idea in their paper \cite{betcke_reviving_2005} to compute the eigenvalues of multiple domains with reentrant corners
by using collocation with a Fourier-Bessel basis around each obtuse corner of the domain.
They enforce homogeneous Dirichlet boundary conditions at a set of boundary collocation nodes whilst also requiring the solution to be bounded away from $0$
in the interior of the domain. The search for the eigenvalues of the domain then amounts to a 1D rootfinding problem for the
smallest singular value of the resulting overdetermined linear system. The results reported by \citeauthor{betcke_reviving_2005}
are an improvement on those reported by \citeauthor{driscoll_eigenmodes_1997},
which are used here as reference values for the eigenvalues of the Laplacian for static isospectral domains (IDs).
The method yields eigenvalues with an impressive precision and for one re-entrant corner, as well as simple boundaries its implementation is simple. It becomes
a formidable task for multiple corners and for multiple domains (i.e., the pairs provided in \cite{buser_planar_2010}) even more so.
Further complexity is added when the $\del_\varphi$ term of \eqref{eq:10} has to be considered. For practical reasons the preference is therefore given to a
different method.

Finally there are finite element methods (FEM) for the solution of PDEs. The main difference from the method of global ansatz functions is
the requirement, that the solution be expressed in a basis , where each ansatz function has compact support in $\CalD$ - the domain of the
PDE (here: isospectral domains). 
The discrete versions of the differential operators are then very sparse matrices. A further bonus of requiring the ansatz functions to have
compact support is that one can require the latter to consist of certain geometric shapes, e.g. a triangle, which allow
for simple tiling of the computational domain. There is, in general, no restriction on the size of the FE, but shape restrictions exist 
\cite[Ch.4.4.3]{Deuflhard2013}.
It has been mentioned in \ref{sec:Theo_Analysis_IDs} that isospectral domains considered here are composed of triangular tiles, this suggests
triangular finite elements for the solution of \eqref{eq:10}.
Using such FE one fully covers the domain and thus avoids the problems (loss of information on the boundary) one would have were one to use a TP mesh.
FE shapes are, of course, not limited to triangles though triangles are most common finite elements. Rectangles and hexagons follow suit.
With rectangles as a possible choice of the fundamental tile FEM may be married to spectral methods on TP meshes by introducing a TP mesh on each
element, thus producing so-called \textit{spectral element methods} (c.f. \citeauthor{Kopriva2008} \oc).
Furthermore, by means of their construction, FEM are well suited for curved surfaces. This suggests them as a good choice for the present work.
Although here I am concerned only with euclidean geometry the derivations provided should carry over to non-euclidean geometry as well.

Considering the prior emphasis of this thesis on differential forms in conjunction with \cite{desburn_discrete_2008} and \cite{Kopriva2008}
the choice of the finite element method for the numerical solution of \eqref{eq:10} seems natural.
The major drawback of using FE is that their implementation is far more difficult than of any method presented so far. Fortunately a
basic implementation of the FEM was available to me due to prior work, which facilitated the use of FEM in the present work.

\section[The QEVP]{The Quadratic Eigenvalue Problem}\label{sec:ExpIntro_QEVP}
The purpose of all methods presented in the preceding section is to obtain an approximation of the true (continuous) solution to a partial differential equation
by transforming the problem into an algebraic problem and solving the latter, preferably guaranteeing convergence towards the exact solution in the
limit of decreasing mesh width (increasing resolution).
As such, all methods presented so far will ultimately yield a matrix eigenvalue problem as a discrete version of \eqref{eq:UnitDisc_PDE_QEVP} (c.f. \ref{sec:Theo_UnitSquare})
\begin{equation}
\left( \Amat + \lambda \Cmat + \lambda^2 \Mmat \right) \vpsi = 0\quad .\label{eq:QEVP}
\end{equation}
Where $\Cmat$ is, by merit of being the discrete version on $\del_\varphi$, antisymmetric, $\Amat$ corresponds to the $2^{\text{nd}}$ spatial derivative (Laplacian),
$\Mmat$ to the constant term of \eqref{eq:UnitDisc_PDE_QEVP} and $\vpsi$ is the discrete approximation of the eigenfunction of the equation on a domain $\CalD$.

The PDE \eqref{eq:10} can be regarded as an ordinary differential equation in time, which allows one to draw a parallel to
the classical (damped) harmonic oscillator. This is the origin of the nomenclature for the matrices of \eqref{eq:QEVP}.
The matrix $\Amat$ is called the "stiffness matrix" due to a discrete representation of the second spatial derivative and thus being associated with
the stiffness of a spring. $\Mmat$ is the "mass matrix" and $\Cmat$ is the "damping matrix", due to its occurrence as a factor in the term linear in $\del_t$ which is commonly associated
with damping \cite{tisseur_quadratic_2001}. As stated above the damping in this case is non-dissipative. In such cases $\Cmat$ is referred to as "gyroscopic"
and the problem is then categorized as a gyroscopic quadratic eigenvalue problem.

The eigenvalues of such problems always occur in quadruplets $\lbrace \lambda,-\lambda,\bar{\lambda}, -\bar{\lambda} \rbrace$
possibly degenerating into complex-conjugate pairs \cite{lancaster_strongly_1999}. An overview of quadratic eigenvalue problems is provided in
\cite{tisseur_quadratic_2001}.
The stability of gyroscopic systems is discussed at length in \cite{lancaster_strongly_1999}. \citeauthor{lancaster_strongly_1999}
discusses the geometric behaviour of the eigenvalues of a gyroscopic system, where he states that the eigenvalues move on circles in $\CC$ and 
split only at $\lbrace \RR,i\RR\rbrace$ (c.f. also \cite{Marsden1987}).

There exist multiple ways of solving the quadratic eigenvalue problem (henceforth QEVP) \eqref{eq:QEVP}.
A simple way would be to consider the problem of determining an eigenpair $(\lambda,\vpsi)$ as a rootfinding problem and using, e.g.
Newton's method to solve it. For this to be successful \eqref{eq:QEVP} has to be augmented by a normalization condition
$||\vpsi||_2=1$.

Alternatively, as done here, one may consider the substitution $\vec{v}:=\lambda\vpsi$ thus transforming \eqref{eq:QEVP}
into a modified system. This so called linearisation is shown in \cite{tisseur_quadratic_2001,mehrmann_structure-preserving_2001}. \citeauthor{mehrmann_structure-preserving_2001}
also mention, that a trivial linearisation of a gyroscopic QEVP does not preserve the structure of the eigenvalue distribution. A structure-preserving
linearisation is given to be \cite[eq. 2.8]{mehrmann_structure-preserving_2001}:
\begin{equation}
\left[
\begin{pmatrix}
0 & -\Amat\\
\Mmat & 0
\end{pmatrix}
- \lambda \begin{pmatrix}
\Mmat & \Cmat\\
0 & \Mmat
\end{pmatrix}\right] \begin{pmatrix}
\psi\\ v
\end{pmatrix}0\quad.\label{eq:H_SH_Pencil}
\end{equation}
The matrix $\Kmat = \begin{pmatrix}
0 & -\Amat\\ \Mmat & 0\end{pmatrix}$ is a Hamiltonian matrix, meaning $(KJ)^t=KJ$ where $J=\begin{pmatrix}
0 & \mathbb{I}\\
-\mathbb{I} & 0
\end{pmatrix}$
whereas the matrix $\Bmat = \begin{pmatrix}
\Mmat & \Cmat\\
0 & \Mmat
\end{pmatrix}$ is skew-Hamiltonian ($(\Bmat\Jmat)^t=-\Bmat\Jmat$).
The matrix $\Bmat$ is non-singular and permits the following decomposition
\begin{equation}
\Bmat = \Bmat_1 \Bmat_2 = \begin{pmatrix}
\Mmat & \frac{1}{2}\Cmat\\
0 & \mathbb{I}
\end{pmatrix}
\begin{pmatrix}
\mathbb{I} & \frac{1}{2}\Cmat\\
0 & \Mmat
\end{pmatrix}\quad.
\end{equation}
This in turn permits one to rewrite \eqref{eq:H_SH_Pencil} as a standard EVP:
\begin{equation}
0= \Bmat_1^{-1}\Kmat \Bmat_2^{-1} - \lambda \mathbb{I} =: \Wmat -\lambda\mathbb{I} \label{eq:HamiltonianSTDEVP}
\end{equation}
where $\Wmat\in\RR^{2N\times 2N}$ (provided $\Amat,\Cmat,\Mmat\in\RR^{N\times N}$) is again Hamiltonian. 
The form \eqref{eq:H_SH_Pencil} is also called a Hamiltonian/skeq-Hamiltonian matrix pencil (H/SH pencil for short).

If one is to solve \eqref{eq:HamiltonianSTDEVP} numerically a method which preserves the structure of the eigenvalue distribution
should be preferred as it will yield the \textit{exact} eigenvalues of a slightly perturbed H/SH pencil (or alternatively a slightly perturbed
eigenvalues of the exact H/SH pencil).
Seeing as we shall be interested only in a few eigenvalues, small in magnitude, of a very large and very sparse matrix
the choices as to the numerical method for the eigenvalue search are limited. One may either use the inverse (shifted) iteration to sequentially obtain
the eigenvalues nearest to a set of chosen shifts $\sigma_i\in\CC$, or one may use an implicitly restarted Arnoldi method (IRA)
with a shift-and-invert strategy. The latter method is chosen due to the availability of tried and tested software with very good performance.

The choice for the numerical eigenvalue solver library is ARPACK (\textbf{Ar}noldi \textbf{Pack}age) with an interface to C++ (ARPACK++ \cite{_reuter_arpackpp}).
The library requires one to provide the number of eigenvalues to be determined as well as the target shift $\sigma$. Furthermore it allows one to provide a user-defined
shift operation $y\leftarrow S(\sigma)x$.
As is noted in \cite{mehrmann_structure-preserving_2001} the trivial shift $S(\sigma)=\left( \Wmat -\sigma \mathbb{I}\right)^{-1}$ is not Hamiltonian.
As a remedy \citeauthor{mehrmann_structure-preserving_2001} propose to use
\begin{equation}
R_2(\sigma,\Wmat) = \left( \Wmat -\sigma \mathbb{I}\right)^{-1}\left( \Wmat +\sigma \mathbb{I}\right)^{-1}\label{eq:HamiltonianShiftR2}
\end{equation}
as a shift operator $S$, which is skew-Hamiltonian, in case $\sigma\in\lbrace\RR,i\RR\rbrace$. Note that \citeauthor{tisseur_quadratic_2001} \cite{tisseur_quadratic_2001}
do not mention the fact that the shift may be purely imaginary for \eqref{eq:HamiltonianShiftR2} to be usable, which may lead one to use the much more complex shift $R_1(\sigma,\Wmat)$, 
described in \cite{mehrmann_structure-preserving_2001} and reproduced in \cite{tisseur_quadratic_2001}.

Due to the presence of the factor $\tfrac{\omega}{c}\leq 1$ and its powers in either \eqref{eq:10} or \eqref{eq:SHDerived}
the eigenvalues $\lambda_k$ of the system are not expected (\textit{a  priori}) to deviate far from the eigenvalues $\lambda_{k,0}$ of the classical wave equation,
to which both PDEs reduce if one sets $\omega=0$, if we are to work with physically meaningful equations.
It is therefore reasonable to choose a shift $\sigma_i \in \lbrace\lambda_{k,0}\rbrace_{k=1}^m$ which
in turn means that $\sigma_i\in i\RR$ and one may safely use \eqref{eq:HamiltonianShiftR2}.

According to \cite{mehrmann_structure-preserving_2001} a skew-Hamiltonian shift operator will, in theory, generate isotropic Krylov subspaces. In practice, due to finite
precision arithmetic being used, it is necessary to additionally $\Jmat$-orthogonalize\footnote{Orthogonalize w.r.t. the inner product
$\langle u,v\rangle_J := u^t\Jmat v$.} the Arnoldi-vectors to uphold the isotropy of the subspaces.
Such a method is then referred to as skew-Hamiltonian implicitly restarted Arnoldi method (SHIRA).
The result of a shift-and-invert SHIRA are the eigenvalues of largest modulus of the shift operator.

To obtain the eigenvalues of the original Hamiltonian matrix $\Wmat$ and thus of the H/SH pencil
\citeauthor{mehrmann_structure-preserving_2001} suggest to use the computed Arnoldi basis to determine the Ritz values \cite[Ch. 7]{Dahmen2008}
$\mu_i$ of $\Wmat^2$ and to take $\pm\sqrt{\mu_i}$ to obtain the eigenvalues of $\Wmat$. Note that the latter is inadvisable due to the loss of precision
when taking the square root. As has been shown in \ref{sec:Theo_Analysis_IDs} the eigenvalues are not strictly necessary, their squares may suffice
for an investigation of the change of the isospectrality.

I am therefore content with determining $\lambda_i^2$ for each domain and computing their difference. To avoid the computation of the Ritz values of $\Wmat^2$
a vanishing shift ($\sigma = 0$) is chosen. Thus the eigenvalues of smallest magnitude are determined
This permits one to make use of ARPACK without additional
post-processing of the eigenvalues. The shift operator \eqref{eq:HamiltonianShiftR2} then reduces to
\begin{equation}
R_2(0,\Wmat) = \Wmat^{-2} = \left( \Nmat_1^{-1}\Kmat\Nmat_2^{-1}\right)^{-2} = \Nmat_2 \Kmat^{-1} \Nmat \Kmat^{-1} \Nmat_1\quad .\label{eq:HamiltonianNullShift}
\end{equation}
Furthermore I abstain from implementing an additional $\Jmat$-orthogonalization due to lack of time and immediate necessity.

Summa-summarum this means that essentially a classical shift-and-invert Arnoldi method, as it is implemented in ARPACK, is used with a shift more
fitting to a H/SH matrix pencil \eqref{eq:H_SH_Pencil}.

It should be noted that the SHIRA method mentioned above is not the only way to compute the eigenvalues of a H/SH matrix pencil.
A different method is presented in \cite{benner_krylov-schur-type_2008} and requires a reformulation of the QEVP as a symplectic eigenvalue problem. 
The method is mentioned here only for completeness' sake as there was no immediate need to implement it because the method outlined above
worked reasonably well (as shall be shown hereafter).

\section*{On Software}
More methods for discretization of the PDE and the solution of the QEVP can be easily imagined. 
Alas, it should be equally obvious that not every method suggested by theory is also of practical relevance. 
Between a theoretical method and numerical results is the highly non-trivial step of implementation of the methods.
Here questions of viability, stability and performance contend for attention.

It stands therefore to reason, that one could and should rely on exterior software for the solution
of the problem. For there this step has already been performed by experts.
Examples of such software are \textsc{Matlab}\footnote{www.mathworks.com}, \textsc{Mathematica}\footnote{www.wolfram.com}, \textsc{Comsol}\footnote{www.comsol.com}
and \textsc{FEM++}\footnote{www.freefem.org}, the latter being a finite elements library.
All have their strength's and weaknesses. Whilst proprietary software packages make the handling of
the PDE fairly easy their major shortcoming is that they are generally geared towards a common class of problems (such as a Helmholtz equation).
The closest such class which fits \eqref{eq:10} is that of a Helmholtz equation with a non-constant transport term added and is not a
"standard" case covered by the software packages. It thus requires an adaptation of said software, which is in general highly non-trivial (if not impossible).
Additionally almost all codes use some form of ARPACK to determine eigenvalues of sparse matrices, thus being a middleman between the problem and the methods used in ARPACK.

By using own software it was possible for me to circumvent said middlemen and interface directly with the library, which helped to improve performance and precision
of the results.
\chapter{Algebraic Formulation}\label{ch:Exp_AlgebraicFormulation}
\section{Finite Element Discretisation}\label{sec:AlgForm_FEMDiscretization}
Consider again eq. \eqref{eq:10} and make an ansatz for the temporal dependence of the electric field as described in \ref{sec:Theo_IDsPriorArt}:
\begin{equation}
E_3(r,\varphi;t) = \Psi(r,\varphi)e^{c\lambda t}\quad \lambda\in\CC.\label{eq:E3_TemporalAnsatz}
\end{equation}
Again note that $\left[\lambda\right] = \tfrac{1}{L}$.
Using this in \eqref{eq:10} one obtains
\begin{equation*}
0 = \left( \Delta\Psi - \left[\frac{\omega}{c}\right]^2\del_\varphi^2\Psi + 2\frac{\omega}{c}\lambda \del_\varphi\Psi
-\left[n^2 - \left(\frac{r\omega}{c}\right)^2\right]\gamma^2\lambda^2 \Psi\right) e^{c\lambda t}\quad.
\end{equation*}
Because the above equation has to be valid for all time following holds true
\begin{equation}
0 = \Delta\Psi - \left[\frac{\omega}{c}\right]^2\del_\varphi^2\Psi + 2\frac{\omega}{c}\lambda \del_\varphi\Psi
-\left[n^2 - \left(\frac{r\omega}{c}\right)^2\right]\gamma^2\lambda^2 \Psi\label{eq:EV_PDE_FullEqn}
\end{equation}
In a way similar to the one taken in \ref{sec:Theo_UnitSquare} the time-dependent PDE \eqref{eq:10} has thus been transformed into
an eigenvalue problem. To reiterate: setting $\omega=0$ will yield a Helmholtz equation.
W.l.o.g $\Psi$ may be assumed to be real-valued due to $E_3$ being a measurable physical field
and therefore necessarily real.
For further considerations \eqref{eq:EV_PDE_FullEqn} has to be rewritten in Cartesian coordinates.
To this end we apply the transformation law for basis vectors of a vector space $(\tilde{b}_1,\dots,\tilde{b}_n ) = (b_1,\dots,b_n)\Tmat_{\CalB,\tilde{\CalB} }$
with $\Tmat_{\CalB,\tilde{\CalB}} = J_{ \text{cyl} \rightarrow \text{cart} }$, obtaining $\del_\varphi = -y\del_x + x\del_y$ which is then used to get
\begin{multline}
0 = \Delta\Psi - \left[\frac{\omega}{c}\right]^2 (-y\del_x+x\del_y)^2\Psi + 2\frac{\omega}{c}\lambda(-y\del_x+x\del_y)\Psi
-\left[n^2-\left(\frac{r\omega}{c}\right)^2\right]\gamma^2\lambda^2\Psi\\
= \Delta\Psi - \left[\frac{\omega}{c}\right]^2 \left(\vbeta\cdot\nabla\right)^2\Psi + 2\frac{\omega}{c}\lambda\vbeta\cdot\nabla\Psi
-\left[n^2-\left(\frac{r\omega}{c}\right)^2\right]\gamma^2\lambda^2\Psi\label{eq:AlgForm_EV_PDE_vectorialform}
\end{multline}
where $\vbeta=\vbeta(\xx):=(-y,x)^t,\Psi=\Psi(\xx),r=\sqrt{x^2+y^2}$.

Next one relaxes the differentiability assumptions imposed onto $\Psi$ by \eqref{eq:EV_PDE_FullEqn} and looks instead for 
so-called "weak solutions", that is solutions of the integral version of the differential equation.
What follows is a formal derivation of the finite element discretization of \eqref{eq:EV_PDE_FullEqn}.
Let therefore $U$
be a Banach space and $V$ be a Hilbert space. We seek $\Psi\in U$ s.t. following holds
\begin{equation}
0 = \int_\CalD \phi\left(\Delta\Psi - \left[\frac{\omega}{c}\right]^2 \left(\vbeta\cdot\nabla\right)^2\Psi + 2\frac{\omega}{c}\lambda\vbeta\cdot\nabla\Psi
-\left[n^2-\left(\frac{r\omega}{c}\right)^2\right]\gamma^2\lambda^2\Psi\right) d(x,y)\qquad\forall\phi\in V\quad.\label{eq:EV_PDE_FullEqn_WFStatement}
\end{equation}
Here $\CalD$ is the domain on which \eqref{eq:EV_PDE_FullEqn} is to be solved.
Note that dot above represents the standard inner product on $\RR^2$.
Before rewriting this in a way suitable for further processing, note
\begin{subequations}
\begin{gather}
\nabla\cdot(f\vec{K}) = \nabla\cdot\vec{K} f + \nabla f\cdot\vec{K}\label{eq:DivFuncVFProd}\\
\nabla\cdot\vbeta(\xx) = \del_x(-y)+\del_y(x) = 0\;.\label{eq:DivBeta}
\end{gather}
\end{subequations}
The last equation is rather obvious if one recalls that the vector-field $\vbeta$ is the cartesian representation of a conservative vector-field $\del_\varphi$.
Next one proceeds rewriting \eqref{eq:EV_PDE_FullEqn_WFStatement}:
\begin{align*}
0 &= \int_\CalD \phi\Delta\Psi d(x,y)-\left[\frac{\omega}{c}\right]^2 \int_\CalD\phi (\vbeta\cdot\nabla)^2\Psi d(x,y)\\
&+ 2\left[\frac{\omega}{c}\right]\lambda \int_\CalD \phi (\vbeta\cdot\nabla)\Psi d(x,y) 
-\lambda^2\int_\CalD \left[n^2-\left(\frac{r\omega}{c}\right)^2\right]\gamma^2\phi\Psi d(x,y)\\
&= \int_\CalD \nabla\cdot(\phi\nabla\Psi) - \nabla\phi\cdot\nabla\Psi d(x,y)
-\left[\frac{\omega}{c}\right]^2 \int_\CalD \phi\left( \nabla\cdot \left( \vbeta (\vbeta\cdot\nabla\Psi\right)-\vbeta\cdot\nabla\Psi\nabla\cdot\vbeta\right) d(x,y)\\
&+ 2\left[\frac{\omega}{c}\right] \lambda \int_\CalD \phi\left(\nabla\cdot(\vbeta\Psi)-\Psi\nabla\cdot\vbeta\right) d(x,y)
-\lambda^2\int_\CalD \left[n^2-\left(\frac{r\omega}{c}\right)^2\right]\gamma^2\phi\Psi d(x,y)\\
&= \int_{\del\CalD} \phi\nabla\Psi\cdot\vec{dA} -\int_\CalD \nabla\phi\cdot\nabla\Psi d(x,y)
-\left[\frac{\omega}{c}\right]^2 \int_\CalD \phi\nabla\cdot\left(\vbeta(\vbeta\cdot\nabla\Psi)\right) d(x,y)\\
&+ 2\left[\frac{\omega}{c}\right]\lambda \int_\CalD\phi\nabla\cdot(\vbeta\Psi) d(x,y)
-\lambda^2\int_\CalD \left[n^2-\left(\frac{r\omega}{c}\right)^2\right]\gamma^2\phi\Psi d(x,y)\\
&= \int_{\del\CalD} \phi\nabla\Psi\cdot\vec{dA} -\int_\CalD \nabla\phi\cdot\nabla\psi d(x,y)
-\left[\frac{\omega}{c}\right]^2 \int_\CalD \nabla\cdot\left[\phi\vbeta(\vbeta\cdot\nabla\Psi)\right] - \nabla\phi\cdot\left[\vbeta(\vbeta\cdot\nabla\Psi)\right] d(x,y)\\
&+ 2\left[\frac{\omega}{c}\right]\lambda \int_\CalD \nabla\cdot\left[\phi\vbeta\Psi\right]-\nabla\phi\cdot\vbeta\Psi d(x,y)
-\lambda^2\int_\CalD \left[n^2-\left(\frac{r\omega}{c}\right)^2\right]\gamma^2\phi\Psi d(x,y)\\
&= \int_{\del\CalD} \phi\nabla\Psi\cdot\vec{dA} -\int_\CalD \nabla\phi\cdot\nabla\psi d(x,y)
-\left[\frac{\omega}{c}\right]^2 \int_{\del\CalD} \phi(\vbeta\cdot\nabla\Psi)\vbeta\cdot\vec{dA} 
+ \left[\frac{\omega}{c}\right]^2\int_\CalD (\nabla\phi\cdot\vbeta)(\vbeta\cdot\nabla\Psi) d(x,y)\\
&+ 2\left[\frac{\omega}{c}\right]\lambda \int_{\del\CalD}\phi\Psi\vbeta\cdot\vec{dA}
-2\left[\frac{\omega}{c}\right]\lambda\int_\CalD (\nabla\phi\cdot\vbeta)\Psi d(x,y)
-\lambda^2\int_\CalD \left[n^2-\left(\frac{r\omega}{c}\right)^2\right]\gamma^2\phi\Psi d(x,y)\\
&= \int_{\del\CalD} \phi\left( \nabla\Psi - \left[\frac{\omega}{c}\right]^2(\vbeta\cdot\nabla\Psi)\vbeta + 2\left[\frac{\omega}{c}\right]\lambda\Psi\vbeta\right)
\cdot\vec{dA}\\
&- \int_\CalD \nabla\phi\cdot\nabla\Psi -\left[\frac{\omega}{c}\right]^2(\vbeta\cdot\nabla\phi)(\vbeta\cdot\nabla\Psi)
+2\left[\frac{\omega}{c}\right]\lambda \Psi (\vbeta\cdot\nabla\phi) +\lambda^2\left[n^2-\left(\frac{r\omega}{c}\right)^2\right]\gamma^2\phi\Psi d(x,y)\quad .
 \refstepcounter{equation}\tag{ \theequation }\label{eq:EV_PDE_FullEqn_WeakFormIntegrated}
\end{align*}
For theoretical and practical purposes the function spaces $U,V$ have to be specified. From the equation above it becomes apparent that both, the solution $\Psi$
and the testing functions $\phi$ should have square-integrable derivatives of order $1$ at least. It is thus permissible to set:
\begin{equation}
U=V=H^1(\CalD):=\lbrace f\in L^2(\CalD)\vert \brak{f}{f}_{L^2}+\brak{\nabla f}{\nabla f}_{L^2} <\infty\rbrace\label{eq:H1AnsatzSpace}
\end{equation}
Functions in $H^1(\CalD)$ fulfil the integrability requirements
posed by \eqref{eq:EV_PDE_FullEqn_WeakFormIntegrated}.
This specification suffices for considerations regarding existence and uniqueness of solutions of inhomogeneous equations\footnote{Which I consider
only as tests for the software used to solve the problem at hand.}.
To approximate $\Psi$ numerically we assume Dirichlet boundary conditions, thus limiting $U,V$ further to
\begin{equation}
H^1_0(\CalD):=\lbrace f\in H^1(\CalD) ~:~ f\vert_{\del\CalD} = 0\rbrace\quad. \label{eq:H10ANsatzSpace}
\end{equation}
Furthermore define the following.
\begin{defn}[Triangulation]\label{def:Triangulation}
The set $\Ctvh :=\bigcup_{i\geq 1} T_i^{(h)}$ of triangles $T_i^{(h)}$ with a size-parameter \cite[Ch. 4.3.1]{Deuflhard2013} $h$ which fulfill certain regularity properties
and cover the domain $\CalD$ is called a \textit{triangulation of} $\CalD$.
\end{defn}

\begin{defn}[Linear FE]\label{def:LinearFESpace}
Let
\begin{equation}
\tilde{\Cuvh}^{(1)} :=\lbrace u_h\in\mathcal{C}(\bar{\CalD})~:~ u_h\vert_{T_i}\in \Cpv_1(T_i)\ \forall T_i\in\Ctvh\rbrace\quad.
\end{equation}
The space $\tilde{\Cuvh}^{(1)}$ is the space of piecewise linear functions on a triangulation of $\CalD$ with a mesh-width $h$.
The subspace of piecewise linear functions on $\CalD$ vanishing on $\del\CalD$ is denoted
\begin{equation*}
\Cuvh^{(1)} := \lbrace u_h \in\tilde{\Cuvh}^{(1)}~:~  u_h\vert_{\del\CalD} = 0\rbrace
\end{equation*}
\end{defn}
Henceforth all considerations assume Dirichlet boundary conditions, unless stated otherwise. 
Dirichlet boundary conditions were chosen in part due to their simplicity and in part to facilitate the comparison of the results obtained for 
a micro-disc (taken from \cite{sunada_design_2007}) to the results obtained by \citeauthor{sunada_design_2007} using a Fourier-Bessel ansatz.
The limitation to $\Cpv_1$ is only a formal one, the practical implementation utilizes polynomials of orders $2$ or $3$.

\begin{defn}[Nodal Basis]\label{def:NodalBasis}
Let $\lbrace a_i\rbrace_{i=1}^N$ be the nodes of the triangles defining $\Ctvh$. Define 
\begin{equation*}
\phi_i\in\tilde{\Cuvh}^{(1)}:\quad \phi_i(a_j):=\delta_i^j\quad\forall i,j\in\lbrace 1,\dots,N\rbrace
\end{equation*}
Then $\CalB = \lbrace \phi_i\rbrace_{i=1}^N$ is designated a nodal basis of $\tilde{\Cuvh}^{(1)}$ with the obvious property
that $\tilde{\Cuvh}^{(1)}=\text{span}\lbrace \phi_i\rbrace$.
\end{defn}

From \ref{def:LinearFESpace} it follows that $\Cuvh^{(1)}\subset H^1_0$. The finite element discretization of \eqref{eq:EV_PDE_FullEqn_WeakFormIntegrated}
is then as follows:
\paragraph*{FE Problem} Determine $\Psi_h\in\Cuvh^{(1)}$ such that $\forall\phi_h\in\Cuvh^{(1)}$ holds
\begin{equation}
0 = \int_\CalD \underbrace{\nabla\phi_h\cdot\nabla\Psi_h}_{\text{I}} -\left[\frac{\omega}{c}\right]^2\underbrace{(\vbeta\cdot\nabla\phi_h)(\vbeta\cdot\nabla\Psi_h)}_{\text{II}}
+2\left[\frac{\omega}{c}\right]\lambda \underbrace{\Psi_h (\vbeta\cdot\nabla\phi_h)}_{\text{III}}
+\lambda^2 \underbrace{\left[n^2-\left(\frac{r\omega}{c}\right)^2\right]\gamma^2\phi_h\Psi_h}_{\text{IV}} d(x,y)
\label{eq:EV_FullPDE_FE_Problem}
\end{equation}
where the boundary integral vanishes due to the definition of $\Cuvh^{(1)}$.

By definition:
\begin{subequations}\label{eq:FE_FunctionToVectorMapping}
\begin{align}
\Psi_h &= \sum_{i=1}^N \psi^i\phi_i\equiv \vpsi_\CalB=\left(\psi^1,\dots,\psi^N\right)_\CalB\\
\phi_h &= \sum_{j=1}^N \phi^j\phi_j\equiv \vphi_\CalB = \left(\phi^1,\dots,\phi^N\right)_\CalB\;,
\end{align}
\end{subequations}
where $\psi^i,\phi^i\in\RR,\phi_i\in\CalB$. With this correspondence in mind consider the constituents I-IV of \eqref{eq:EV_FullPDE_FE_Problem}:
\begin{subequations}\label{eq:FE_MatrixEntries}
\begin{align}
\text{I}:\quad \int_\CalD \nabla\Psi_h\cdot\nabla\phi_h d(x,y) &= \sum_{i,j=1}^N \psi^i\phi^j\int_\CalD \nabla\phi_i\cdot\nabla\phi_j d(x,y)
= \sum_{i,j=1}^N \psi^i\phi^j \Amat_{ij} = \vpsi^t \Amat\vphi\notag\\
\Amat_{ij} &:= \int_\CalD \nabla\phi_i\cdot\nabla\phi_j d(x,y)\\
\text{II}:\quad \int_\CalD(\vbeta\cdot\nabla\phi_h)(\vbeta\cdot\nabla\Psi_h)d(x,y) &= \sum_{i,j=1}^N \psi^i\phi^j
\int_\CalD (\vbeta\cdot\nabla\phi_i)(\vbeta\cdot\nabla\phi_j)d(x,y) =\sum_{i,j=1}^N \psi^i\phi^j \tilde{\Amat}_{ij}
= \vpsi^t\tilde{\Amat}\vphi\notag\\
\tilde{\Amat}_{ij} &:= \int_\CalD (\vbeta\cdot\nabla\phi_i)(\vbeta\cdot\nabla\phi_j)d(x,y)\\
\text{III}:\quad \int_\CalD \Psi_h\vbeta\cdot\nabla\phi_h d(x,y) &= \sum_{i,j=1}^N \psi^i\phi^j \int_\CalD\phi_i\vbeta\cdot\nabla\phi_j d(x,y)
= \sum_{i,j=1}^N \psi^i\phi^j \Cmat_{ij} = \vpsi^t\Cmat\vphi\notag\\
\Cmat_{ij} &:= \int_\CalD\phi_i\vbeta\cdot\nabla\phi_j d(x,y)\\
\text{IV}:\quad \int_\CalD \left[n^2-\left(\frac{r\omega}{c}\right)^2\right] \gamma^2\Psi_h\phi_h d(x,y) &=
\sum_{i,j=1}^N \psi^i\phi^j\int_\CalD \left[n^2-\left(\frac{r\omega}{c}\right)^2\right] \gamma^2\phi_i\phi_j d(x,y)\notag\\
&= \sum_{i,j=1}^N \psi^i\phi^j \Mmat_{ij} = \vpsi^t\Mmat\vphi\notag\\
\Mmat_{ij} &:= \int_\CalD \left[n^2-\left(\frac{r\omega}{c}\right)^2\right] \gamma^2\phi_i\phi_j d(x,y)
\end{align}
\end{subequations}

Using the above results in the following algebraic equivalent of \eqref{eq:EV_FullPDE_FE_Problem}:
\begin{equation}
0 = \vpsi^t\left(\Amat -\left[\frac{\omega}{c}\right]^2 \tilde{\Amat} + 2\left[\frac{\omega}{c}\right]\lambda \Cmat
+\lambda^2 \Mmat\right)\vphi\qquad \forall\vphi\in\RR^N\label{eq:EV_FullPDE_FE_AlgebraicForm}
\end{equation}

Again one may ignore terms of order $\left(\tfrac{\omega}{c}\right)^2$, provided $\tfrac{R_0\omega}{c}\ll 1$, for some characteristic length $R_0$ of the domain $\CalD$\footnote{Here $R_0$ will
be the radius of the largest circumcircle of a pair of domains.}. In which case
$n^2-\left(\tfrac{r\omega}{c}\right)^2\approx n^2,\;\gamma^2\approx 1$ and \eqref{eq:EV_FullPDE_FE_AlgebraicForm} reduces to
\begin{align}
0 &= \vpsi^t\left(\Amat + 2\left[\frac{\omega}{c}\right]\lambda \Cmat +\lambda^2 n^2\tilde{\Mmat}\right)\vphi\qquad \forall\vphi\in\RR^N\label{eq:EV_LinPDE_FE_AlgebraicForm}\\
\tilde{\Mmat}_{ij} &:= \int_\CalD \phi_i\phi_j d(x,y)\quad.\notag
\end{align}
Furthermore, by construction (c.f. the derivation on p. \pageref{eq:EV_PDE_FullEqn_WeakFormIntegrated}), the matrix $\Cmat$ is antisymmetric,
assuming Dirichlet boundary conditions. Indeed
\begin{align*}
\Cmat_{ij} &= \int_\CalD \phi_i\vbeta\cdot\nabla\phi_j d(x,y) = \int_\CalD \nabla\cdot(\phi_i\vbeta\phi_j) - \nabla\phi_i\cdot\vbeta\phi_j d(x,y)\\
&= \int_{\del\CalD} \phi_i\phi_j\vbeta\cdot\vec{dA} - \int_\CalD \phi_j\vbeta\cdot\nabla\phi_i d(x,y)\\
&= - \int_\CalD \phi_j\vbeta\cdot\nabla\phi_i d(x,y) = -\Cmat_{ji}\quad.
\end{align*}
The assumption of Dirichlet boundary conditions is vital here, for if either Neumann or Robin conditions are used additional boundary terms show up,
complicating the  treatment of the equation. This is a part of the reason why Dirichlet conditions are assumed. It does not pose a strong limitation on the
investigation carried out in this thesis for, as the proofs referred to in \ref{sec:Theo_IDsPriorArt} state, isospectral domains are Dirichlet and Neumann
isospectral when at rest.

Furthermore, due to its origins as a discretization of the $\varphi$-derivative in polar coordinates the matrix is conservative, i.e., it represents
the discrete version of an area-preserving vector-field $\del_\varphi$, which makes the name "damping matrix" a misnomer, somewhat undermining the
analogy to a damped oscillator. Here the other common name for $\Cmat$ - the "transport matrix" - is a more fitting designation.


\section[Practical Aspects of FEM]{Practical Aspects of Finite Elements}\label{sec:AlgForm_PracticalAspects}
The implementation of the finite element and SHIRA methods has to be discussed, for they are the fundamental
tools which permit me to obtain the results presented in the following chapter.

As has been alluded to in \ref{sec:ExpIntro_MethodOverview}, the implementation of finite element methods is non-trivial. For this work I could build upon
a code developed previously for a lecture on numerical methods for stationary partial differential equations. This code implements a
so-called continuous Galerkin method\footnote{As opposed to discontinuous Galerkin methods} (continuous FEM) and is provided, along with the data obtained from numerical experiments,
on the accompanying medium (DVD at the end of the thesis).
Due to relying on Euler's formula for planar graphs the software used to obtain the results presented hereafter is suitable 
only for simply connected domains without holes, if a mesh resolution study is to be attempted with the in-built (red-green) refinement
routines. The latter refine a mesh uniformly by dividing the edges of the triangles into two and, with the help of the introduced nodes, 
partitioning each triangle into 4 congruent triangles \cite[Ch. 6.2.2]{Deuflhard2013}.
Domains with holes may be used if the user provides the mesh. Further documentation of the software may be found on the accompanying medium (c.f. also \ref{ch:App_Software}).

The degree of the interpolating polynomials has been limited to $3$ due to the necessity for quadrature rules of high order to avoid introducing errors when
computing the integrals in \eqref{eq:FE_MatrixEntries}. Should a higher order interpolation become necessary one will have to either resort to balancing discretization and
quadrature errors, or to using quadrature rules of a sufficiently high order, such that the integrals in \eqref{eq:FE_MatrixEntries} can be computed
exactly. The viability of the latter approach may be limited by the growth in the number of necessary quadrature nodes \cite[Ch. 4.3]{Deuflhard2013}.
An increase in the degree of the polynomials should be easily possible due to the necessary data structures being present for $\Cpv_3$ FE. This
was not the case of $\Cpv_1$ and $\Cpv_2$ finite elements because the orientation of the mesh did not manifest itself with the interpolation nodes
necessary for these elements.

It has been mentioned in \ref{ch:Exp_Introduction} that the full SHIRA method was not implemented here. Instead the ARPACK routines for a
standard eigenvalue problem were used to search for eigenvalues of smallest magnitude. The required shift operator \eqref{eq:HamiltonianNullShift} 
for a null-shift was taken from \cite{mehrmann_structure-preserving_2001}. \citeauthor{mehrmann_structure-preserving_2001} provide
a reformulation of $R_2$ which reduces the factorizations necessary for the shift to one factorization of a matrix with the dimensions of the QEVP,
cutting the number of necessary operations in half.
The choice of a vanishing shift simplified the reformulated shift operator even further. As a result only a symmetric positive definite matrix $\Amat$ had to be factorized once,
instead of factorizing a general matrix for each application of the shift operator. As is known \cite{Engeln-Mullges2004, Quarteroni2002} a s.p.d. matrixd may be factorized using a Cholesky
decomposition instead of an LU decomposition. The former is half as expensive as the latter (in terms of computation cost).
The linear algebra functionality required was provided by the Eigen C++ linear algebra library v. 2.9.3 \cite{_eigen_library}.
A straightforward implementation of the shifts proposed in \cite{tisseur_quadratic_2001, mehrmann_structure-preserving_2001} is possible but
discouraged due to their computational intensity.

As noted in \cite{mehrmann_structure-preserving_2001} the classical IRA introduces impurities, whose magnitudes are on the order of $10^{-15}$, into the Arnoldi basis and thus the
eigenvalues.
This has been observed with the implementation described above and has been neglected during the numerical experiments. 
Indeed it only matters for the second pair of isospectral domains (c.f. fig. \ref{subfig:ElaborateIDs}), where at the beginning one can clearly observe random 
fluctuations in the deviation of the spectra - henceforth denoted as \textit{anisospectrality} -  when it is $ < 10^{-12} $ (c.f. \ref{fig:Exp_ID2AnisoAndEVFullVacuum}).

\chapter{The Experiment}\label{ch:Exp_TheExperiment}
\section{Validation}\label{sec:Exp_Validation}

\subsection{The Unit Disc}
\begin{figure}[!b]
\centering
\includegraphics[width=0.75\linewidth,keepaspectratio]{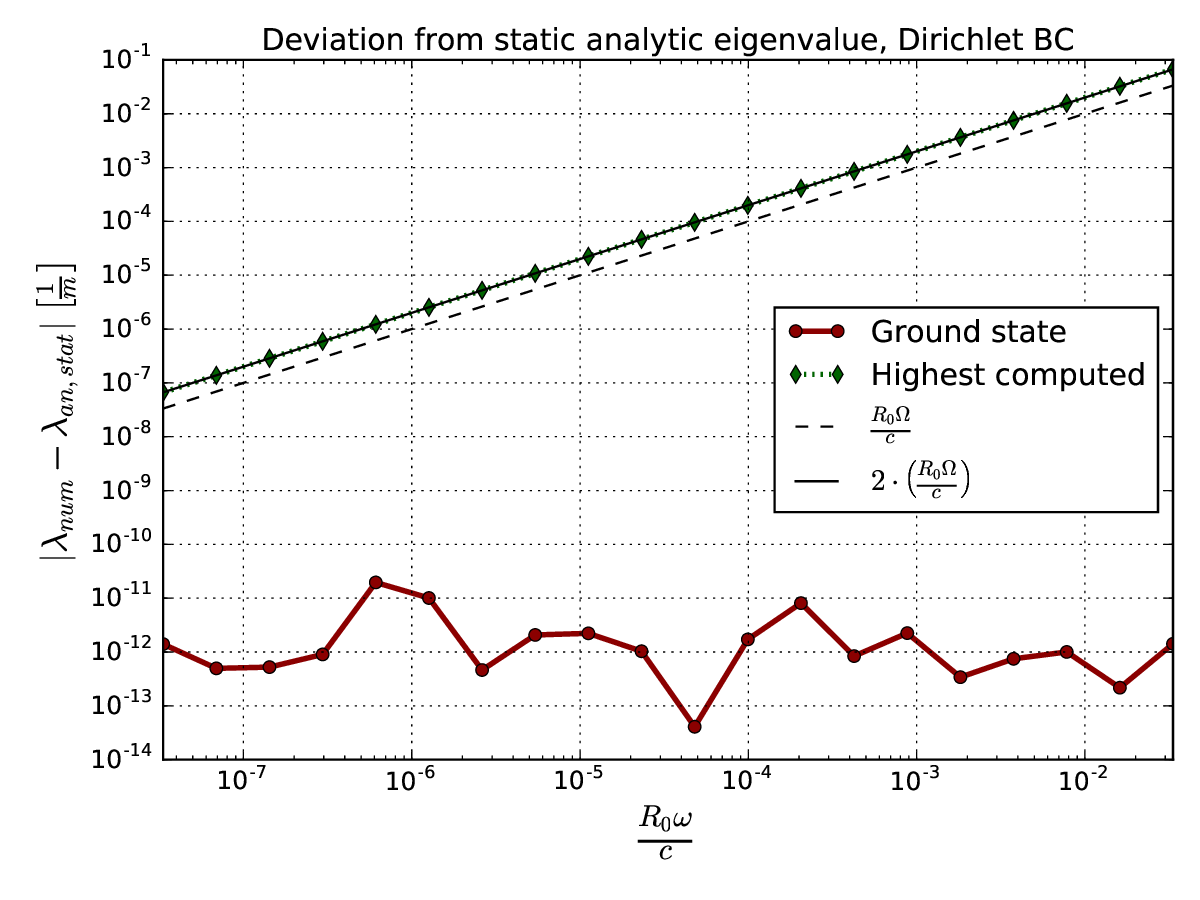}
\caption[Eigenvalue evolution of the unit disc.]{Deviation of the eigenvalues of a rotating disc $\CalD_1$ from their static counterparts computed using COMSOL 5.2 using the
linearized PDE \eqref{eq:SHDerived} with $n=1$.}
\label{fig:Exp_Validation_UnitDisc}
\end{figure}
It has been shown in \ref{sec:Theo_UnitDisc}, that an analytical statement on the dependence of the eigenvalues of a disc 
is possible. The geometry thus provides an
excellent testing ground for a preliminary analysis  of a method for solving the equation.
To this end a solution was determined using the proprietary software \textsc{Comsol} 5.2 which implements the finite element method and
for all means and purposes may be assumed to yield correct results\footnote{See however the remarks at the end of the current section.}. 
Hereafter, whenever a solution was obtained with said software the generation of the mesh was done using the advancing front algorithm.
It was empirically determined in preliminary studies that results obtained using meshes generated with this algorithm are slightly superior
to those obtained with meshes created by Delaunay triangulation.
The advancing front algorithm however, possesses the caveat that it generates meshes which are imbued with a certain structure.

The result for the lowest eigenmode and the $2^\text{nd}$ excited mode are shown in \ref{fig:Exp_Validation_UnitDisc}, where the deviation
of the eigenvalues from the eigenvalues of the static domain is drawn. It is immediately obvious that the obtained results fit
perfectly to the theoretical predictions made in \ref{sec:Theo_UnitDisc}, with the maximal deviation from the analytical prediction
for the 2nd eigenvalue being $|\Delta \lambda_2| \approx 0.0334\left[\tfrac{1}{m}\right]$. This shows that the numerical method chosen here is, 
at least in principle, suitable for the solution of the full equation \eqref{eq:10} and its linearized descendent \eqref{eq:SHDerived} and that the 
analytical prediction of a linear dependence on $\tfrac{\omega}{c}$ for the unit disc is correct.
The oscillating behaviour of the ground state deviation, being less than the $\epsilon = 10^{-11}$ error tolerance requested, may be safely neglected.

\subsection{The Unit Square}
\begin{figure}[!tp]
\centering
\subfloat[Modal ansatz]{
\includegraphics[width=0.45\linewidth]{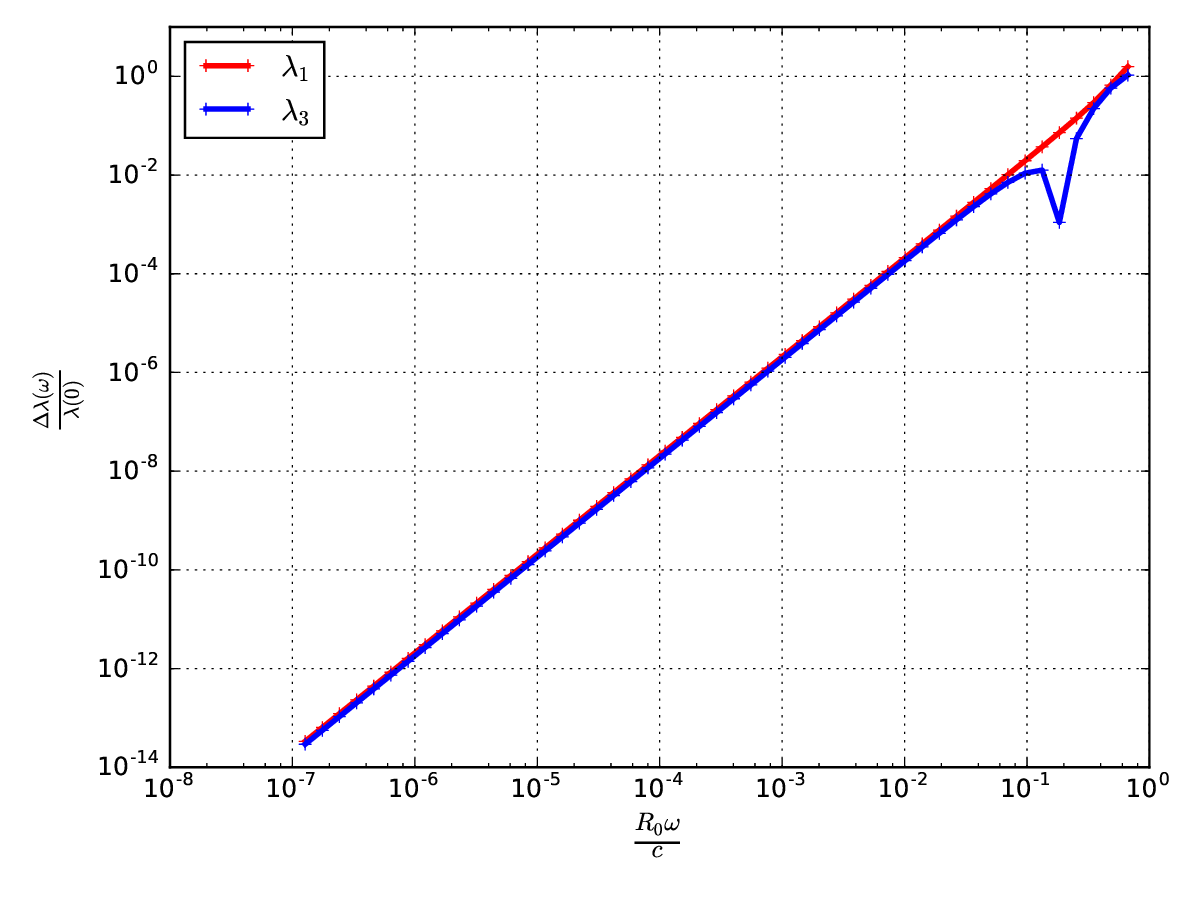}
\label{subfig:Exp_ModalVsNodal_UnitSquare_Modal}
}\;
\subfloat[Nodal ansatz (FEM)]{
\includegraphics[width=0.45\linewidth]{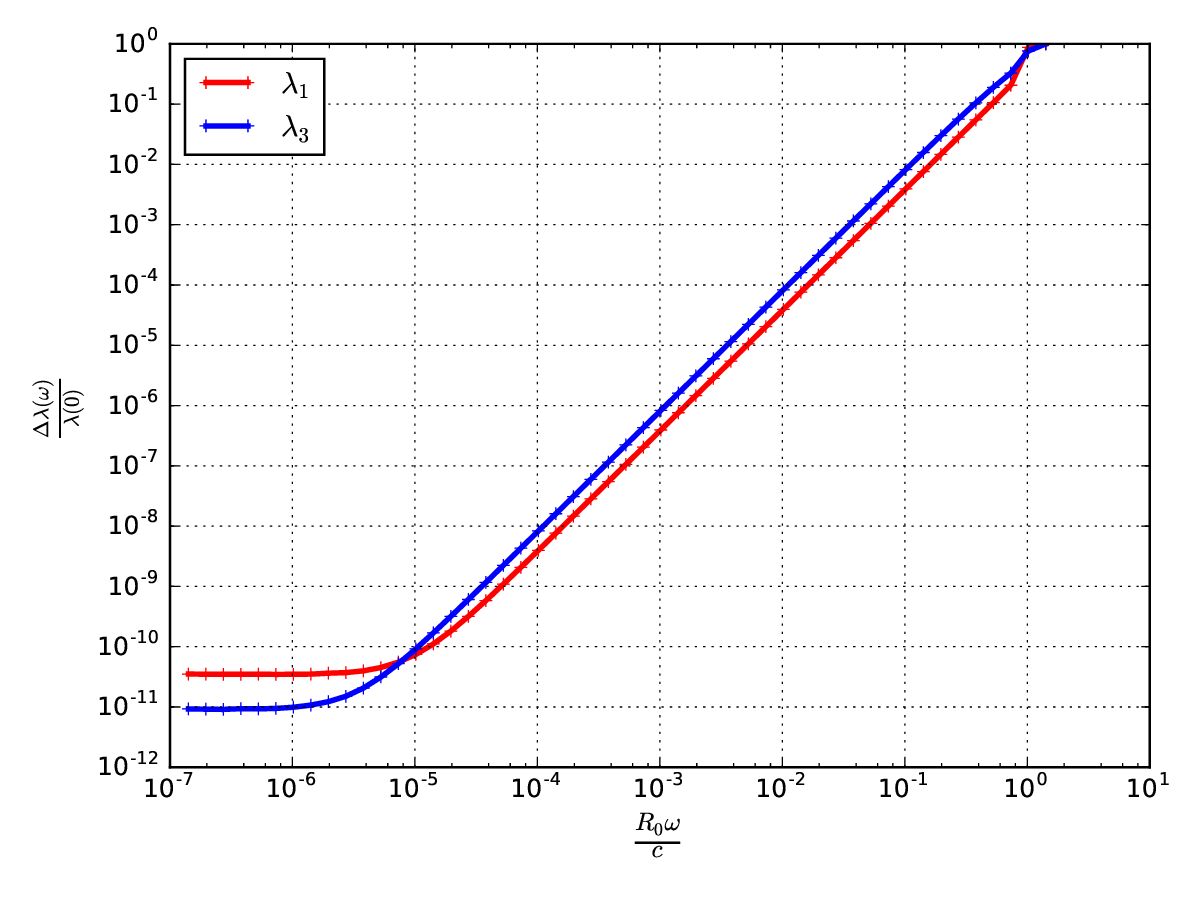}
\label{subfig:Exp_ModalVsNodal_UnitSquare_Nodal}
}
\caption[Modal vs. nodal approximation on the square.]{Comparison of the change of the eigenvalues of the unit square obtained with the modal decomposition of
\eqref{eq:ModalDecompositionQEVP} and FE discretization of the linearized PDE \eqref{eq:EV_LinPDE_FE_AlgebraicForm}.
Both computed for vacuum ($n=1$). The dip for $\lambda_3$ in fig \ref{subfig:Exp_ModalVsNodal_UnitSquare_Modal} is due to a numerical error and may be neglected on
the same basis as the dips encounted in \ref{fig:PolyDegConvergenceStudy}. The flat region for FEM is due to the limited precision of the method.}
\label{fig:Exp_ModalVsNodal_UnitSquare}
\end{figure}
Next the second-simplest case of \ref{ch:TheoAnalysis}, the unit square, has been analysed. Here the results obtained
with self-written FEM code are compared to the results obtained using the modal decomposition of \ref{sec:Theo_UnitSquare} and solving
\eqref{eq:ModalDecompositionQEVP} for $N=20$ modes using the linearization \eqref{eq:H_SH_Pencil}. Modal decomposition was
implemented in \textsc{Mathemaica} v. 10.4.1  whereas the accompanying FEM code is written in C++. The linearized
modal QEVP was solved using the eigenvalue routines provided by \textsc{Mathematica}.

The results are presented in \ref{fig:Exp_ModalVsNodal_UnitSquare}, where the change of the eigenvalues from their static counterparts
is shown as a function of the non-dimensional parameter $\tfrac{R_0\omega}{c}$. Both sets of results were obtained using the linearized version of \eqref{eq:10}.
The domains were rotated around the origin of the coordinate system, wherefrom it follows that the radius of the circumcircle is $R_0=\sqrt{2}$, or the length of the
diagonal.
In fig. \ref{subfig:Exp_ModalVsNodal_UnitSquare_Modal} and fig. \ref{subfig:Exp_ModalVsNodal_UnitSquare_Nodal}
the deviations for $\lambda_1,\lambda_3$ are shown as examples.
The results shown in fig. \ref{subfig:Exp_ModalVsNodal_UnitSquare_Modal} are in line with the predictions of 
a quadratic dependence on $\frac{\omega}{c}$ (resp. $\tfrac{r\omega}{c}$) made in \ref{sec:Theo_UnitSquare}.
By comparing both one can see that the results obtained with self-made code reproduce the correct quadratic growth
rate of the eigenvalues with $\tfrac{\omega}{c}$, which indicates that my implementation of FEM is indeed correct.

If one is to use the square as a test geometry, one should be aware that its high order of symmetry may be detrimental to
results obtained by direct numerical solution. 
The choice of a rectangle instead of a square would be more beneficial as it has fewer symmetries, while the eigenvalues and eigenfunctions
of the Laplacian of the stationary domain are known.

\begin{table}[b]
\centering
\caption{The first ten eigenvalues of the isospectral domains depicted in \ref{subfig:SimpleIDs} as computed by \citeauthor{driscoll_eigenmodes_1997} in \cite[Tbl. 3.1]{driscoll_eigenmodes_1997}.
These eigenvalues have been chosen as reference values for the validation of the finite element method in \ref{subsec:Exp_StaticIDs}. }
$\begin{array}{|c|c|c|c|c|}\hline\hline
2.53794399980 & 3.65550971352 & 5.17555935622 & 6.53755744376 & 7.24807786256\\
9.20929499840 & 10.5969856913 & 11.5413953956 & 12.3370055014 & 13.0536540557\\\hline\hline
\end{array}$
\label{tbl:Driscoll_ID_Eigenvalues}
\end{table}
\subsection{Static Isospectral Domains}\label{subsec:Exp_StaticIDs}
\begin{figure}[!b]
\centering
\includegraphics[width=\linewidth,height=0.4\textheight]{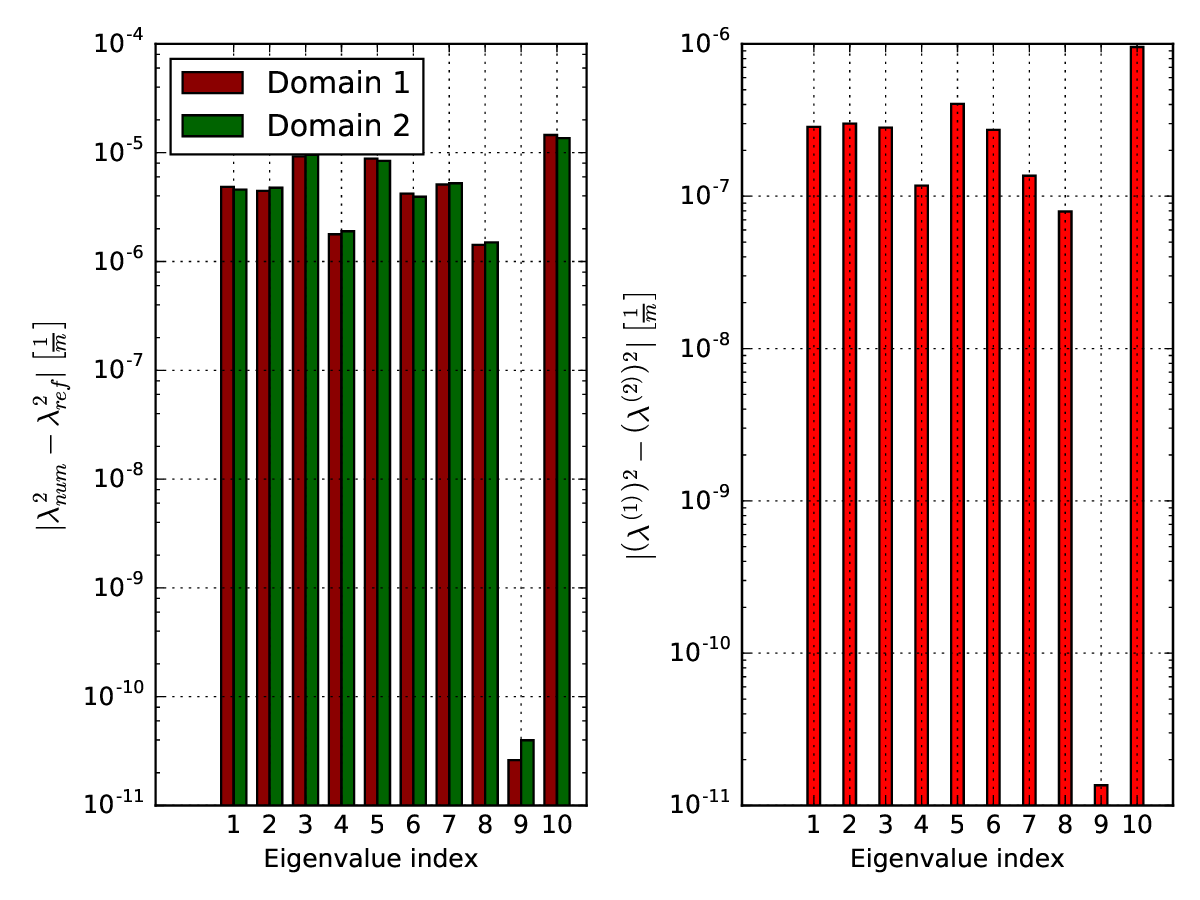}
\caption[Static IDs. Eigenvalues obtained with \textsc{Comsol} and their differences.]{ \texttt{(left panel)} Comparison of the eigenvalues obtained with COMSOL 5.2 for the 
domains depicted in  \ref{subfig:SimpleIDs} to those provided by \citeauthor{driscoll_eigenmodes_1997} \oc \texttt{(right panel)} Difference of the eigenvalues
of the two domains of the first isospectral pair.}
\label{fig:Exp_ID1Stat_ComsolVsDriscoll}
\end{figure}
The third and last test case of \ref{ch:TheoAnalysis} is provided by the simple isospetral pair whose  triangulation is shown in 
\ref{fig:ID1_Triangulations} (or \ref{subfig:Exp_ID1Meshes}). 
As has been noted above, the eigenvalues computed by \citeauthor{driscoll_eigenmodes_1997} \cite{driscoll_eigenmodes_1997} serve as reference values here,
such that the quality of eigenvalues obtained using FEM is measured by their deviation from the reference \textit{and} from each other.
These deviations are shown in \ref{fig:Exp_ID1Stat_ComsolVsDriscoll} for eigenvalues computed with \textsc{Comsol} 5.2. The reference eigenvalues
are reproduced in \ref{tbl:Driscoll_ID_Eigenvalues}.
The better approximation of the eigenvalue $\lambda_9$  is due to it corresponding to \eqref{eq:Analysis_Eigenval2Square}, thus being
associated with a very regular function. The residual discrepancy is, of course, due to the relative
error settings of $\epsilon=10^{-10}$. This eigenvalue serves as a check on the quality of the eigenvalues obtained for the static domains.
It also provides an excellent reference value for determining whether any change of eigenvalues occurs at all when the
domains are subjected to uniform rotation. The reason being that due to the very small deviation of $\lambda_9$ for both domains even a minuscule difference
will be easy to pick up.

\begin{figure}[p]
\centering
\subfloat[$N_{ref} = 3$ - Thrice refined mesh.]{
\includegraphics[width=\linewidth,height=0.4\textheight]{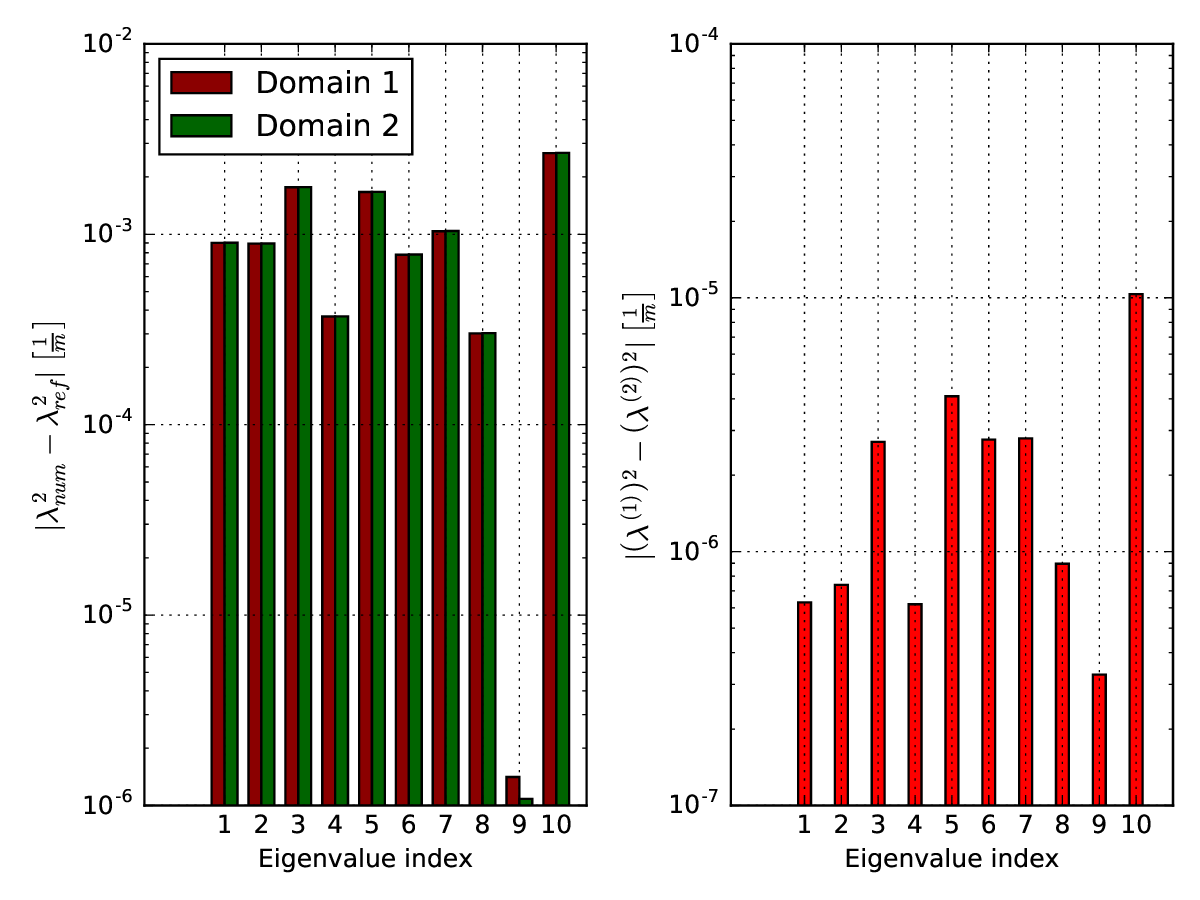}
\label{subfig:Exp_ID1Stat_OwnVsDriscoll_Nref3}
}\;
\subfloat[$N_{ref} = 7$ - Mesh refined 7 times.]{
\includegraphics[width=\linewidth,height=0.4\textheight]{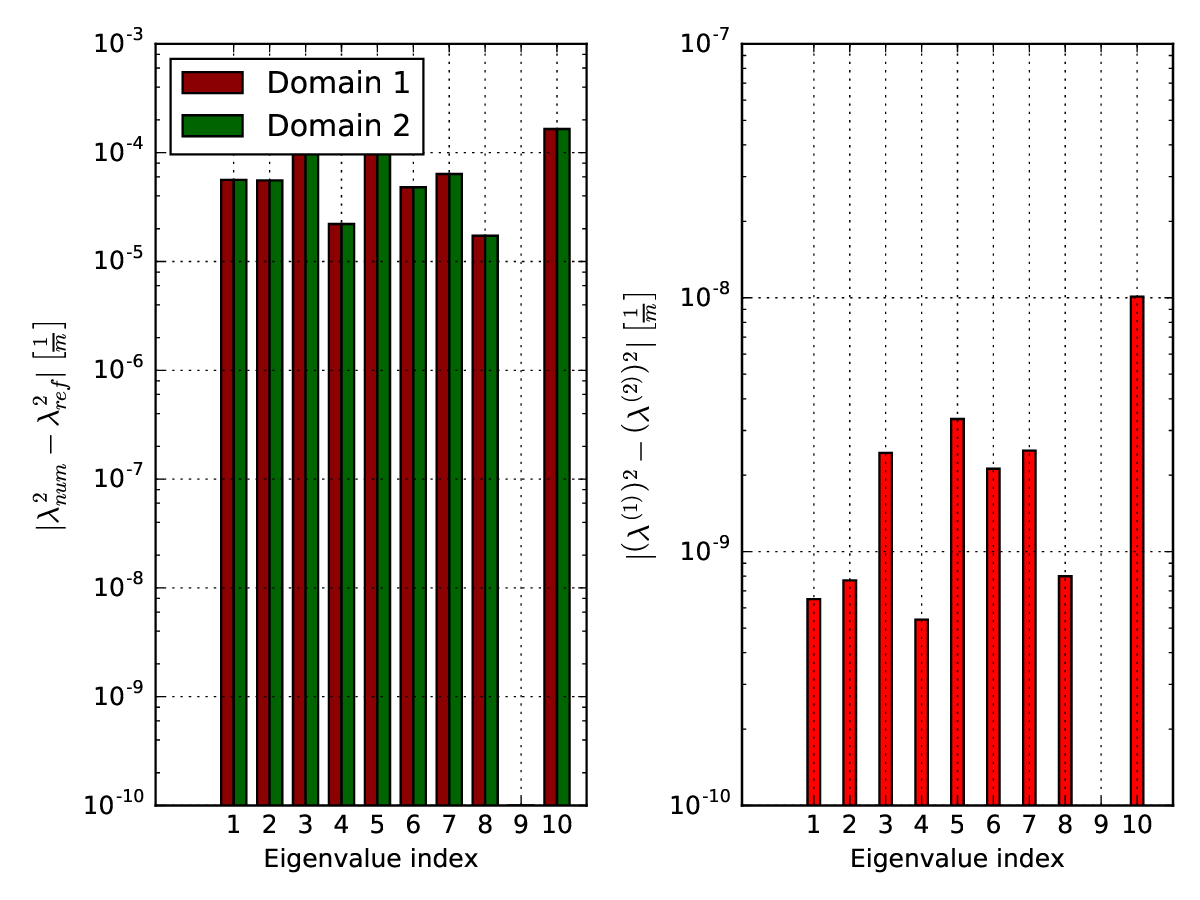}
\label{subfig:Exp_ID1Stat_OwnVsDriscoll_Nref7}
}
\caption[Static IDs. Eigenvalues obtained with self-made software.]{\texttt{(left panel)} Comparison of the eigenvalues obtained with self-made FEM, with the mesh of fig. \ref{subfig:Exp_ID1Meshes} refined $N_{ref}$ times, to those provided by \citeauthor{driscoll_eigenmodes_1997} \oc \texttt{(right panel)} Difference of the eigenvalues of the two domains of the first isospectral pair.}
\label{fig:Exp_ID1Stat_OwnVsDriscoll}
\end{figure}

\paragraph*{A few remarks.}
The results obtained using finite elements may be improved by utilizing FE meshes adapted to the geometry \cite{Deuflhard2013, Gockenbach2006}.
This was not attempted here due to time-constraints.
One may very well wonder why I chose to write a dedicated piece of software for this thesis, instead of using i.e., \textsc{Comsol}.

The answer to this question is twofold. For one, the results obtained with \textsc{Comsol}, though they compare favourably to those
obtained by \citeauthor{driscoll_eigenmodes_1997}, show a higher initial deviation between the eigenvalues of the
two domains of an isospectral pair \ref{fig:Exp_ID1Stat_ComsolVsDriscoll}. This is detrimental to the principal goal of this work.
The self-made FEM software yields a worse approximation of the eigenvalues obtained by \citeauthor{driscoll_eigenmodes_1997}, but the deviation of the
eigenvalues of the two domains of a pair is, provided the mesh refinement level is sufficiently high, smaller, as can be seen in \ref{fig:Exp_ID1Stat_OwnVsDriscoll}.
A second reason for developing a dedicated software is the reproducibility of the results obtained here. With the code and data provided
on the accompanying medium, I hope to guarantee (simple) reproducibility.

As has been noted in the first part of this work neither domain possesses a (geometrical) special point which could be used as the origin for the axis of
rotation. From a physical point of view all posses at least one special point, the centre of mass. It has been determined assuming a uniform mass distribution
throughout the domain. The axis of rotation has therefore been chosen such that
it intersects the domains in their centres of mass and is orthogonal to the domain. To facilitate comparison by de-dimensionalization of $\tfrac{\omega}{c}$
the radii of circumcircles with a centre in the centre of mass of each domain were determined by
\begin{equation*}
R_0 := \sup_{x\in\CalD}\Vert x - x_{com}\Vert_2\quad .
\end{equation*}
Here $x_{com}$ is the coordinate of the centre of mass and $\CalD\subset\RR^2$ the domain.
The parameters are collected in \ref{tbl:SimParamID1}. The difference in the radii of the circumcircles is expected to emerge in the numerical results.
\clearpage

\section{Results}\label{sec:Exp_Results}
\subsection{Microdisc of \cite{sunada_design_2007}}\label{subsec:Exp_SHMicrodisc}
\begin{figure}[!t]
\centering
\subfloat[Results obtained using FEM.]{
\includegraphics[width=0.8\linewidth,keepaspectratio]{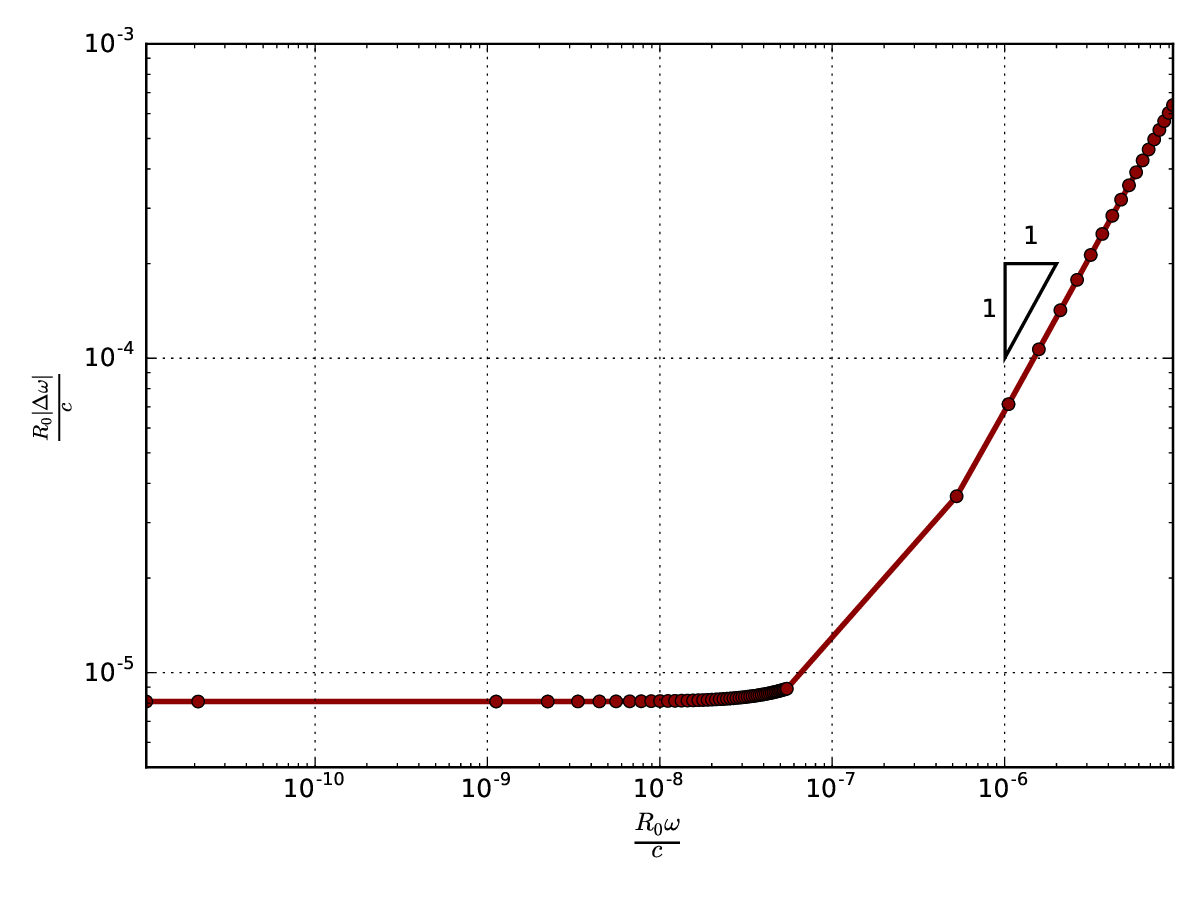}
\label{subfig:Exp_SH_EigenvalueDeviation_Own}
}\;
\subfloat[Results obtained by Sunada and Harayama.]{
\includegraphics[scale=0.7]{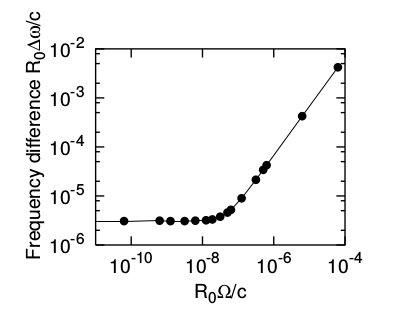}
\label{subfig:Exp_SH_EigenvalueDeviation_Ref}
}
\caption[Results of \citeauthor{sunada_design_2007}]{Frequency difference (non-dimensional) of the two almost-degenerate 
eigenmodes of the football-stadium cavity studied in \cite{sunada_design_2007} as a function of the non-dimension frequency of rotation of the cavity.
The results obtained by Sunada and Harayama are provided for comparison.}
\label{fig:SH_EigenvalueDeviation}
\end{figure}
Having verified the correctness of the method and the code by solving the three reference cases of \ref{ch:TheoAnalysis} numerically we can now proceed
to check whether it is possible to reproduce the results obtained by \citeauthor{sunada_design_2007} for the football-stadium cavity of \cite{sunada_design_2007}.
To this end the linearized equation \eqref{eq:SHDerived} was solved on the two-dimensional domain centred at the origin of the coordinate system and bounded by a curve parametrized by
the polar angle $\theta\in[0,2\pi)$ as $R(\theta) = R_0(1+\epsilon\cos(2\theta) )$ with $R_0=6.2866[\mu m],\epsilon=0.12,n=1$. 
The solution was obtained using \textsc{Comsol} with the shift for the eigenvalue search being $\sigma = \frac{49.3380585}{R_0}\approx 7.84813\cdot 10^{6}\left[\tfrac{1}{m}\right]$
(taken from the \cite{sunada_design_2007}) and the change of the deviation of the two almost-degenerate eigenmodes of the micro-disc is shown in \ref{fig:SH_EigenvalueDeviation}.

Upon comparison to \cite[Fig. 3]{sunada_design_2007}, reproduced in \ref{subfig:Exp_SH_EigenvalueDeviation_Own}, one immediately sees that the results presented by \citeauthor{sunada_design_2007}
have been reproduced qualitatively and quantitatively. For a slightly deformed micro-disc almost-degenerate eigenmodes stay that way
for a large range of rotation frequencies (up to a (non-dimensional) threshold frequency $\tfrac{R_0\omega_{th}}{c}\approx 6\cdot 10^{-8}$), after which
the frequencies of the eigenmodes diverge linearly, in a manner similar to the (unit) disc discussed above. 
The results were obtained by running two simulations with $20$ angular frequencies, distributed logarithmically, each in the range $[0,2386027]\left[\frac{rad}{s}\right]$
and $[2386027,477205484]\left[\frac{rad}{s}\right]$. This resulted in the apparent sharp break at $\tfrac{R_0\omega}{c}\approx 5\cdot 10^{-8}$.
Nevertheless, by extrapolating both halves of the curve one can surmise that there is indeed a change in the behaviour of the
deviation of the two frequencies at the threshold frequency given above.

\citeauthor{sunada_design_2007} treat the eigenmodes for $\omega < \omega_{th}$ using perturbation theory for non-degenerate states,
stating that in this regime a shift of the frequencies of the standing waves does not occur. The statement is, of course only partially true.
The graphical representation chosen in \ref{fig:SH_EigenvalueDeviation} masks the minute changes in the frequency.
For $\omega > \omega_{th}$ they apply perturbation theory for degenerate eigenstates, concluding a linear change of the eigenfrequencies of the domain.

A possible alternative (qualitative) interpretation of these results is as follows:
The stationary micro-disc has 4-fold discrete symmetry (2-fold w.r.t. each $x$- and $y$-axis) and is thus "similar" to a square in the sense that the eigenmodes
chosen by \citeauthor{sunada_design_2007} and shown in \cite[Fig. 2]{sunada_design_2007} impinge on the boundary at locations where the latter is essentially flat and
thus similar to the square.
The absence of an observable quadratic growth of the difference in frequencies for $\omega < \omega_{th}$ may be attributed to the fact that the eigenmodes used
are in fact non-degenerate, such that the slight but constant difference of eigenfrequencies dominates the contributions made by their change.
With increasing rotation frequency $\omega$ the shape of the cavity becomes more similar to that of a disc, for which a linear change of the eigenfrequencies
is expected due to \eqref{eq:TheoAn_UnitDiscWaveNumberChange} (resp. \eqref{eq:TheoAn_UnitDiscEVEvolution}).

The authors back-up  their analytic claims by numerical calculations performed using an expansion of the eigenmodes of the cavity in a Fourier-Bessel series.
The details are provided in \cite{sunada_sagnac_2006}. Here their results have been reproduced with a method which is agnostic to the shape of the domain.
The above considerations further strengthen the confidence in the suitability of the finite element method for the solution of \eqref{eq:10} on non-trivial domains.
Furthermore they show that the equation itself is well-behaved, at least for physically meaningful parameters.
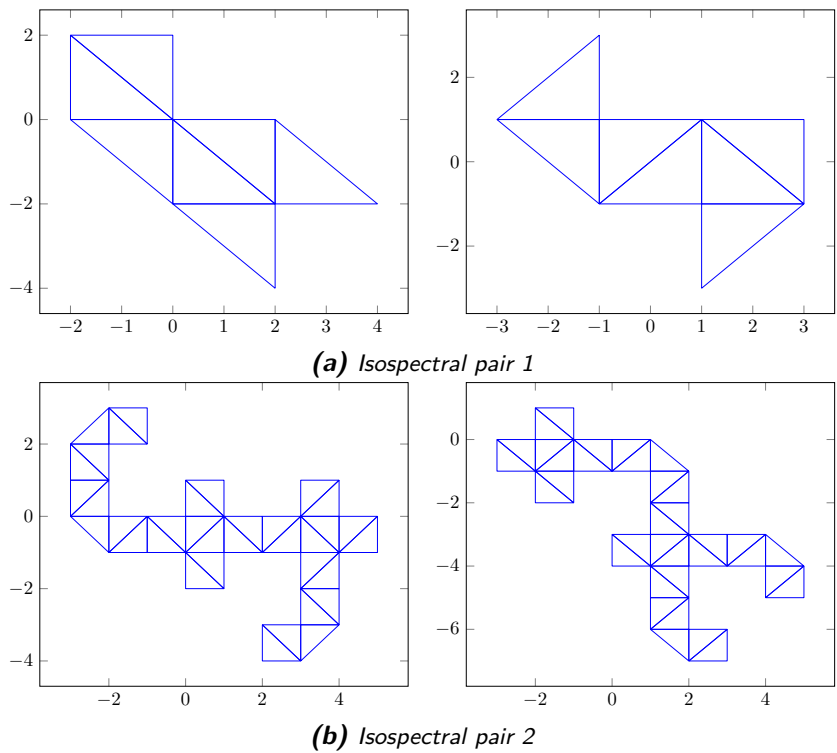
\begin{figure}[!b]
\centering
\subfloat[Isospectral pair 1]{
\resizebox{0.7\linewidth}{!}{
\begin{tikzpicture}
	\begin{axis}[plot box ratio=1 1]
		\addplot[mark=none,patch, patch type=triangle,mesh,blue] file {Plots/Geometries/ID1_1_pgfmesh.dat};
	\end{axis}
\end{tikzpicture}
\hspace{0.15cm}
\begin{tikzpicture}
	\begin{axis}[plot box ratio=1 1]
		\addplot[mark=none,patch, patch type=triangle,mesh,blue] file {Plots/Geometries/ID1_2_pgfmesh.dat};
	\end{axis}
\end{tikzpicture}
}
\label{subfig:Exp_ID1Meshes}
}\;
\subfloat[Isospectral pair 2]{
\resizebox{0.7\linewidth}{!}{
\begin{tikzpicture}
	\begin{axis}[plot box ratio=1 1]
		\addplot[mark=none,patch, patch type=triangle,mesh,blue] file {Plots/Geometries/ID2_1_pgfmesh.dat};
	\end{axis}
\end{tikzpicture}
\hspace{0.15cm}
\begin{tikzpicture}
	\begin{axis}[plot box ratio=1 1]
		\addplot[mark=none,patch, patch type=triangle,mesh,blue] file {Plots/Geometries/ID2_2_pgfmesh.dat};
	\end{axis}
\end{tikzpicture}
}
\label{subfig:Exp_ID2Meshes}
}
\caption[FE meshes of the domains]{The coarse triangulations of the isospectral domains shown in \ref{fig:Intro_IsospectralManifolds} and taken from \cite{gordon_isospectral_1992} as used in this work.
The meshes have been generated by hand. The left mesh of fig. \ref{subfig:Exp_ID1Meshes} is deliberately not "symmetric" w.r.t. the re-entrant corners.}
\label{fig:Exp_FEMMeshes}
\end{figure}

\subsection{Convergence Studies}\label{subsec:Exp_ConvergenceStudies}
\begin{table}[tb]
\centering
\caption{Simulation parameters for the first pair of isospectral domains. The parameters are identical for
the simulations of the linearized PDE \eqref{eq:EV_LinPDE_FE_AlgebraicForm} and the full PDE \eqref{eq:EV_FullPDE_FE_AlgebraicForm}. }
\label{tbl:SimParamID1}
\begin{tabular}{|l|c||cc|cc|}\firsthline
\textsc{Property} & Unit  & ID 1.1 & ID 1.2 & ID 2.1 & ID 2.2\\\hline\hline
Centre of mass & $[m]$ & $(\tfrac{10}{21}, \tfrac{2}{3})^t$ & $ (\tfrac{7}{21}, \tfrac{1}{21})^t $ & $(\frac{313}{210}, -\frac{46}{05})^t$  & $(\frac{317}{210}, -\frac{284}{105})^t$ \\
$R_0$ & $[m]$ & $3.6651$ & $3.4667$ & $ 4.55 $ & $4.8365$\\\hline
index of refraction & $[1]$ & \multicolumn{4}{c|}{$1,\ 1.5,\ 2.42$}\\
$\omega$-range & $[\tfrac{rad}{s}]$ & \multicolumn{4}{c|}{$3\cdot 10^1 - 3\cdot 10^8$}\\
$\omega$-samples & $[1]$ & \multicolumn{4}{c|}{$100$}\\\hline
mesh width & $[m]$ & \multicolumn{2}{c|}{ \scriptsize $1-\tfrac{1}{2^6}$} & \multicolumn{2}{c|}{\scriptsize $\tfrac{1}{2}-\tfrac{1}{2^5}$}\\\hline
\end{tabular}
\end{table}

\begin{figure}[!b]
\centering
\subfloat[Vacuum]{
\includegraphics[width=0.7\linewidth]{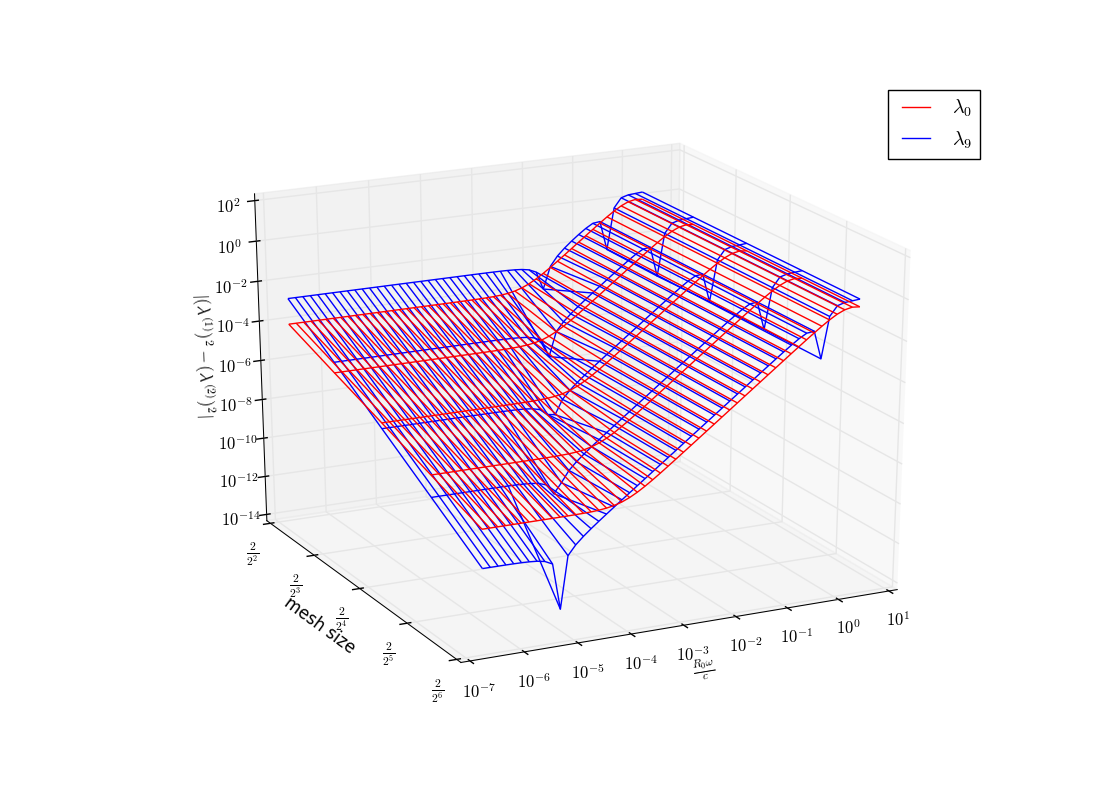}
}\;
\subfloat[Diamond]{
\includegraphics[width=0.5\linewidth]{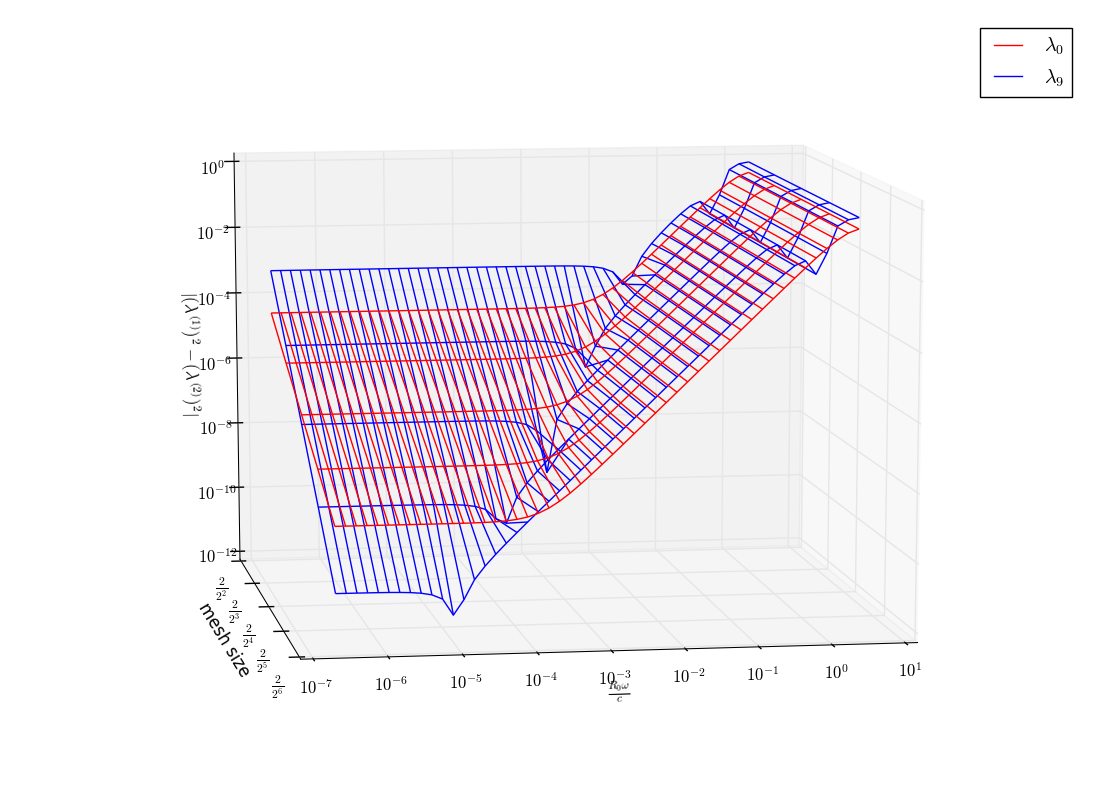}
}
\caption[Anisospectrality vs. mesh finesse and $\tfrac{R_0\omega}{c}$.]{Change of the eigenvalues in dependence on $\tfrac{R_0\omega}{c}$ vs. mesh width for two different indices of refraction (vacuum $n=1$ and
diamond $n=2.42$).
The results were obtained with the linearised PDE \eqref{eq:EV_LinPDE_FE_AlgebraicForm} using $\mathcal{P}_3$ FE. }
\label{fig:MeshRefinementConvergenceStudies}
\end{figure}

The domains of \ref{fig:Intro_IsospectralManifolds} are next in line for the analysis.
The coarse triangulations shown in \ref{fig:Exp_FEMMeshes} form the basis from which sufficiently fine meshes are obtained by uniform refinement, as has been
mentioned in the previous chapter.

As a common starting point the convergence of the chosen numerical method towards a solution for increasingly finer meshes has to be checked. This was done
using $\Cpv_3$ ansatz functions on meshes obtained from those of \ref{subfig:Exp_ID1Meshes} by red-green refinement for refinement orders $2-7$. 
The results obtained using the FE formulation \eqref{eq:EV_LinPDE_FE_AlgebraicForm} of the linearized PDE for vacuum and a medium are shown in \ref{fig:MeshRefinementConvergenceStudies}.
It has to be emphasized that henceforth only the differences between the squares of the eigenvalues are considered due to the way the
eigenvalue search was implemented using ARPACK(++) and the shift \eqref{eq:HamiltonianNullShift} (c.f. \ref{sec:AlgForm_PracticalAspects}). This does not pose a restriction on the
conclusions derived from the data, as has been explained in \ref{sec:Theo_Analysis_IDs}. Hence, whenever a difference in eigenvalues 
is mentioned the difference  of the squares is understood.
Note that as $[\lambda] = \tfrac{1}{L} = [k]$ one may, by merit of $\Omega = 2\pi k$ (here $\Omega$ is the angular frequency of the propagating wave) refer
to $\lambda$ as eigenvalue or eigenfrequency.

In the \ref{fig:MeshRefinementConvergenceStudies} the difference of the eigenvalues computed for both domains of the first pair is shown, for the
ground state and the 9-th eigenstate (corresponding to \eqref{eq:Analysis_IDFirstAnaEigenval}), in dependence on the dimensionless angular velocity
$\tfrac{R_0\omega}{c}$ and the length of a side of the mesh triangles.
Considering only the ground state (red mesh) one can clearly see a flat region corresponding to a range of parameters where the eigenvalues
of the domains remain apparently the same (their difference remains constant). 
According to theory (\ref{sec:Theo_IDsPriorArt}) the difference of the eigenvalues must vanish for $\tfrac{R_0\omega}{c}\rightarrow 0$. 
The domains are, after all, isospectral when at rest. This can be observed with decreasing mesh width.
Therefore it is safe to conclude that the constant difference of the eigenvalues is due to an insufficient spatial resolution of the domain by the mesh and is hence a numerical artefact.
It is interesting to observe, that the difference of the $9^{\text{th}}$ eigenvalues starts off
bigger than the difference of the ground states for a coarse mesh, but drops below the latter for a mesh refinement of $N\geq 4$. This may be explained by the fact
that the domains contain two re-entrant corners which are known \cite[Ch. 4.4]{Deuflhard2013} to limit the order of convergence due to the gradients of solutions
having a logarithmic singularity at the corner. This singularity manifests itself more clearly for functions which vary on shorter length scales (see also \ref{fig:Exp_EigModeGS_ID1_1_Full} 
on p. \pageref{fig:Exp_EigModeGS_ID1_1_Full} and \ref{fig:Exp_EigModeNS_ID1_1_Full} on p. \pageref{fig:Exp_EigModeNS_ID1_1_Full}).
Additionally it is pretty obvious that a certain number of triangles (finite elements) is necessary to reproduce the oscillations of $\psi_9$ (c.f. \ref{fig:Exp_EigModeNS_ID1_1_Full})
sufficiently well.
The fact that the eigenvalues converge for uniform refinement of the mesh can be attributed to an increasing spatial resolution of the problematic parts of the domain.
The eigenfunction is thus increasingly better approximated there.
One can take advantage of this by using adaptive mesh refinement or creating a static mesh which has a finer resolution of the domains in the vicinity of the corners.
This however, would increase the complexity of the implementation and was deemed optional for the current investigation.

\begin{figure}[!t]
\centering
\subfloat[Vacuum]{
\includegraphics[width=0.45\linewidth]{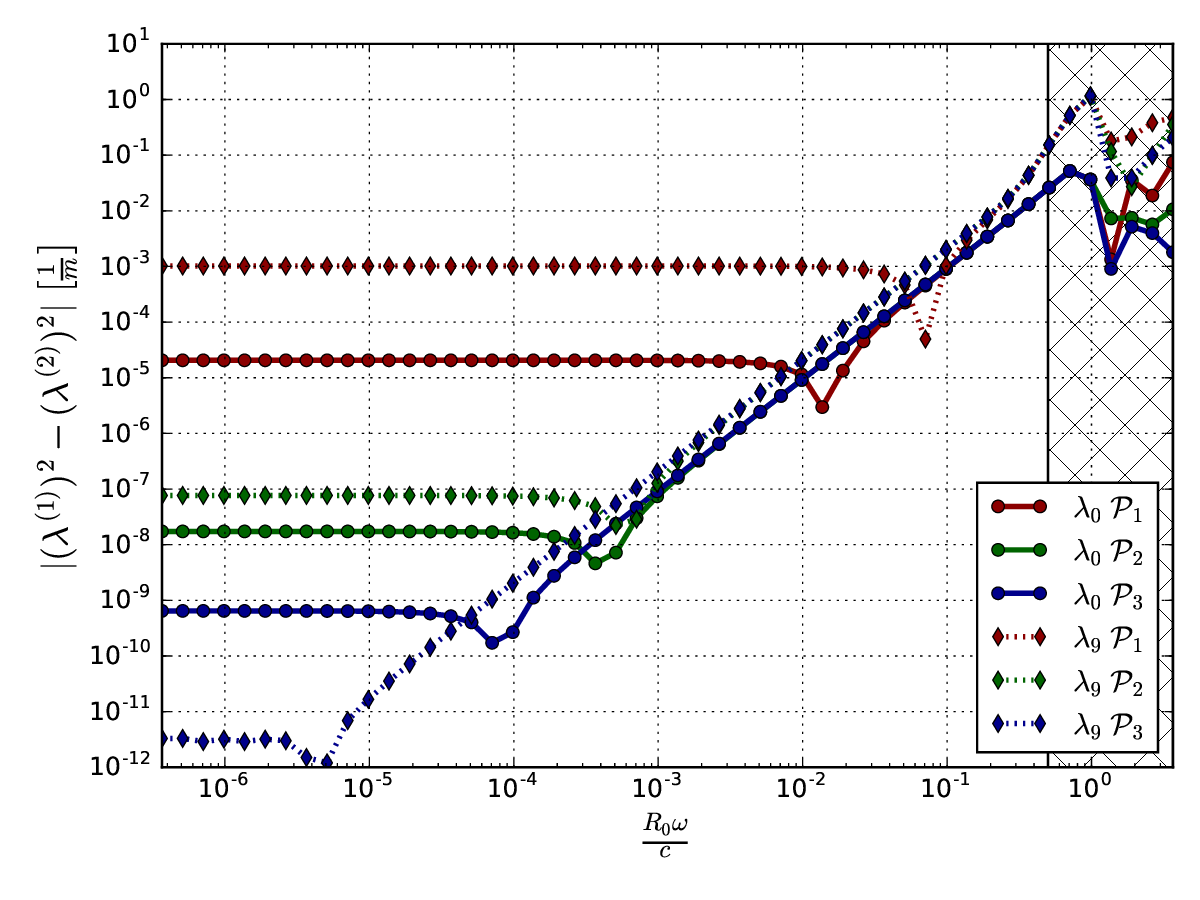}
}\;
\subfloat[Diamond]{
\includegraphics[width=0.45\linewidth]{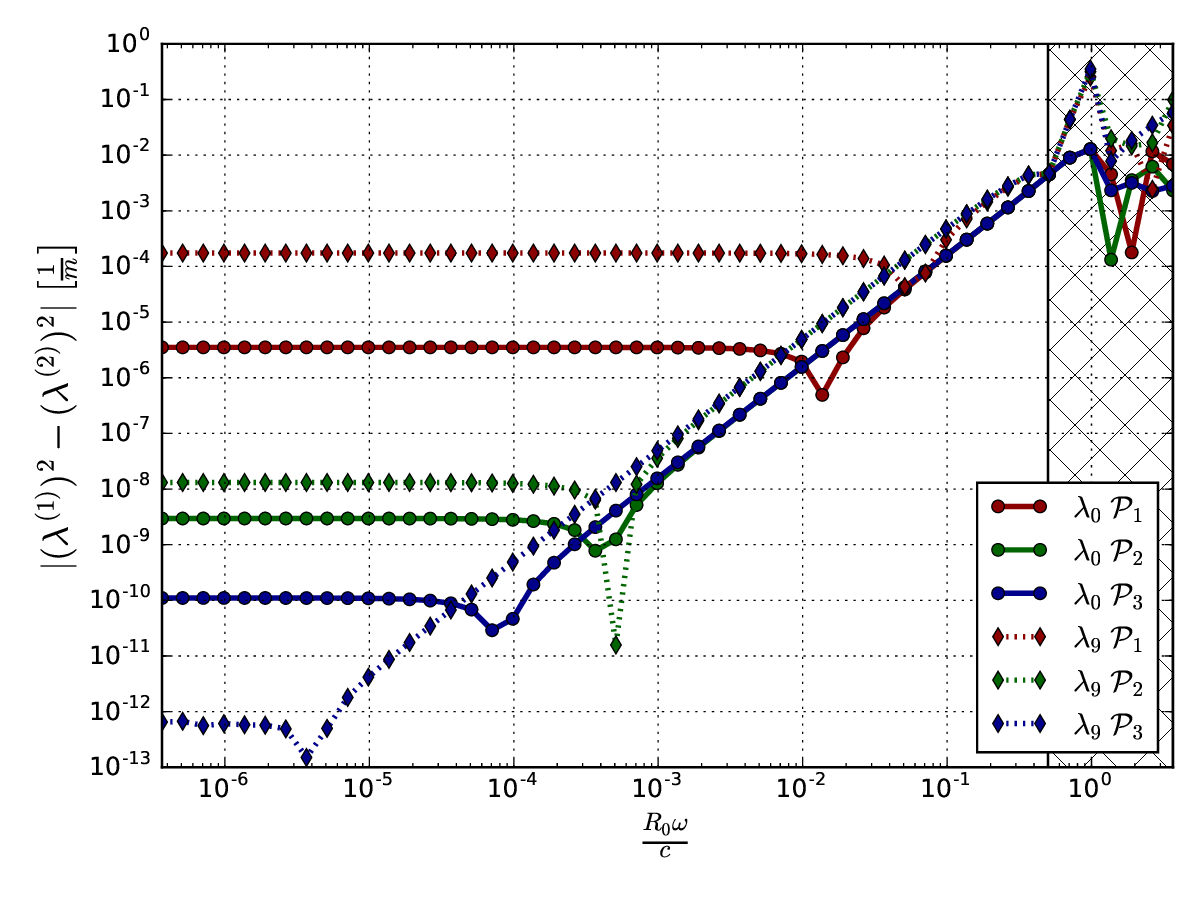}
}
\caption[Ansatz order convergence studies]{The behaviour of the ground state and the ninth eigenstate depending upon $\tfrac{R_0\omega}{c}$ for different
degrees of the ansatz polynomials for two different indices of refraction (vacuum $n=1$ and
diamond $n=2.42$).
The results were obtained with the full PDE \eqref{eq:EV_FullPDE_FE_AlgebraicForm} for a uniform mesh width of $h=\tfrac{1}{2^6}$.}
\label{fig:PolyDegConvergenceStudy}
\end{figure}
The choice of $\Cpv_3$ FE for all further analysis was driven, in part, by the polynomial degree convergence study shown in \ref{fig:PolyDegConvergenceStudy},
which was conducted using the linearized FE form \eqref{eq:EV_LinPDE_FE_AlgebraicForm} of the PDE \eqref{eq:10} for a fixed mesh with width $h=\tfrac{1}{2^5}$.
Note that the mesh width here is the length of a side in the isosceles triangles of the mesh obtained from \ref{fig:Exp_FEMMeshes} by uniform refinement.
It can be seen that the dependence of the anisospectrality on $\tfrac{R_0\omega}{c}$ becomes more prominent with increasing order of the ansatz polynomials.
Again one can see that the deviation of the 9th eigenvalues is initially higher than that of the ground state but for cubic ansatz 
polynomials it is two orders of magnitude smaller than that of the lowest mode of oscillation.
Again this result has been expected due to the fact that with an increasing order of the ansatz polynomials the number of vertices on each element increases ($\Cpv_1: 3,\Cpv_2: 6,\Cpv_3: 10$).
Thus the solution function can be locally better approximated. Looking closer at the results in \ref{fig:PolyDegConvergenceStudy} one can see that there are two dips in the curve for the 9th eigenvalue.
One dip is located in the upper parameter range and is common to all mesh widths and ansatz orders. Upon closer inspection one notices that
this anomaly is situated very close to the upper end of the parameter range where $\tfrac{R_0\omega}{c}\approx 1$.
It is thus at the limit of physically meaningful parameters.
The results in the range $\tfrac{R_0\omega}{c}=10^{-1}\dots 10^0$
may be safely neglected on the grounds that the Galilei-Newton coordinate transformation \eqref{eq:Deriv_GalileiNewtonCoordinateTrafo} used to obtain the PDE is not valid there.
As for the first dip, it happens consistently at the point where the dependence of the eigenvalue difference upon the angular velocity begins to manifest.

The reason for this dip is far less obvious and ties into the implementation of the search for eigenvalues of the QEVP. Apparently
the eigenvalue of domain 2, when determined with the parameters of \ref{tbl:SimParamID1} is consistently higher than the eigenvalue of domain 1,  but the latter 
changes slightly faster due to higher $R_0$ (c.f. \ref{sec:Exp_Validation}), which results in a crossover of the eigenvalues.
The occurrence of a crossover also explains the break in the curve for $\Cpv_3$ elements. At the missing point both eigenvalues coincide, rendering the logarithm of their
difference infinite. Being at the point where the scaling behaviour due to rotation overwhelms numerical imprecisions this dip may be safely neglected.

\subsection{Influence Of The Shape}\label{subsec:Exp_ComparisonID1ToID2}
\begin{figure}[tb]
\centering
\includegraphics[width=0.85\linewidth]{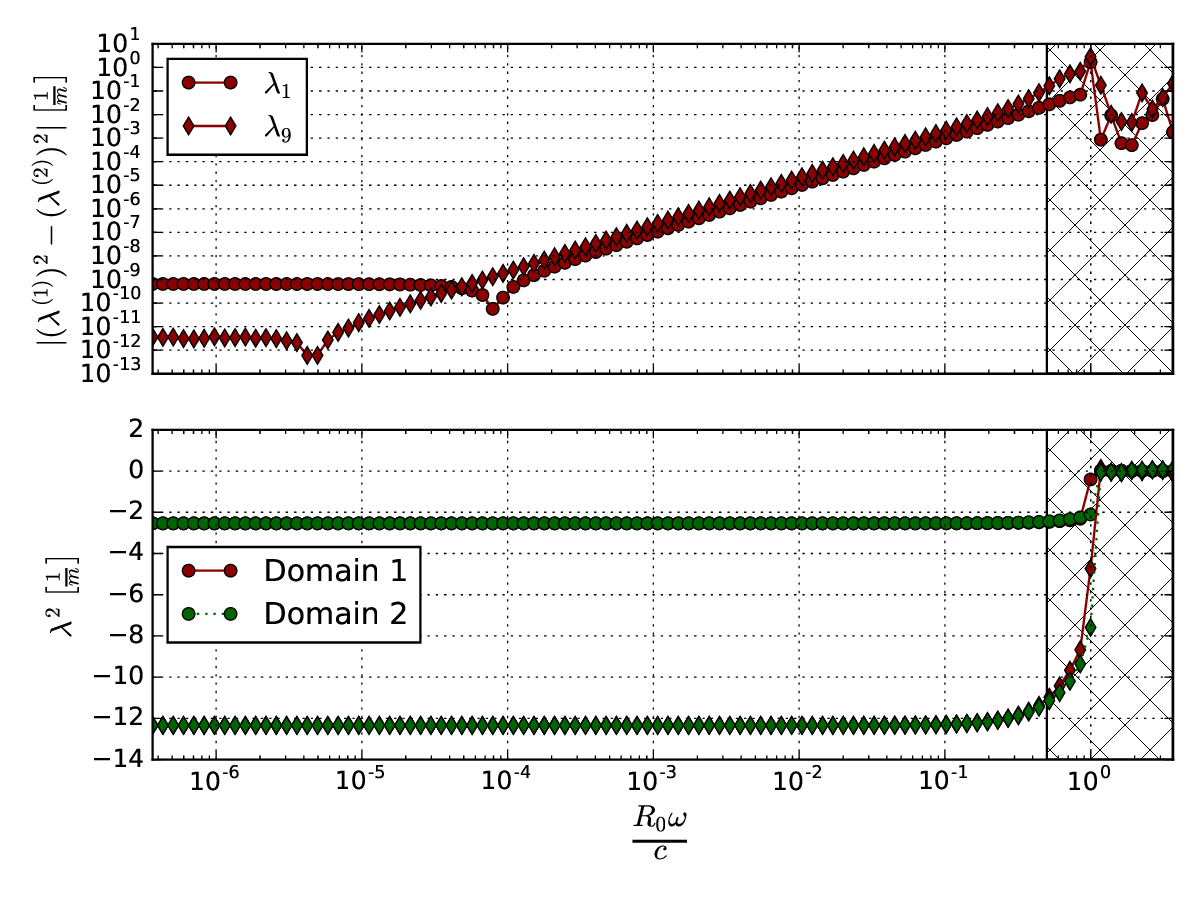}
\caption[Anisospectrality of the domains in fig. \ref{subfig:SimpleIDs}.]{Deviation from the isospectral state for the first pair of isospectral domains of \cite{gordon_isospectral_1992} 
depending on $\tfrac{R_0\omega}{c}$. Computed using \eqref{eq:EV_FullPDE_FE_AlgebraicForm} with $\Cpv_3$ finite elements for vacuum $n=1$. The hatched region marks the domain where
the Galilei-Newton \eqref{eq:Deriv_GalileiNewtonCoordinateTrafo} coordinate transformation is not physically valid.}
\label{fig:Exp_ID1AnisoAndEVFullVacuum}
\end{figure}
\begin{figure}[bt]
\centering
\includegraphics[width=0.85\linewidth]{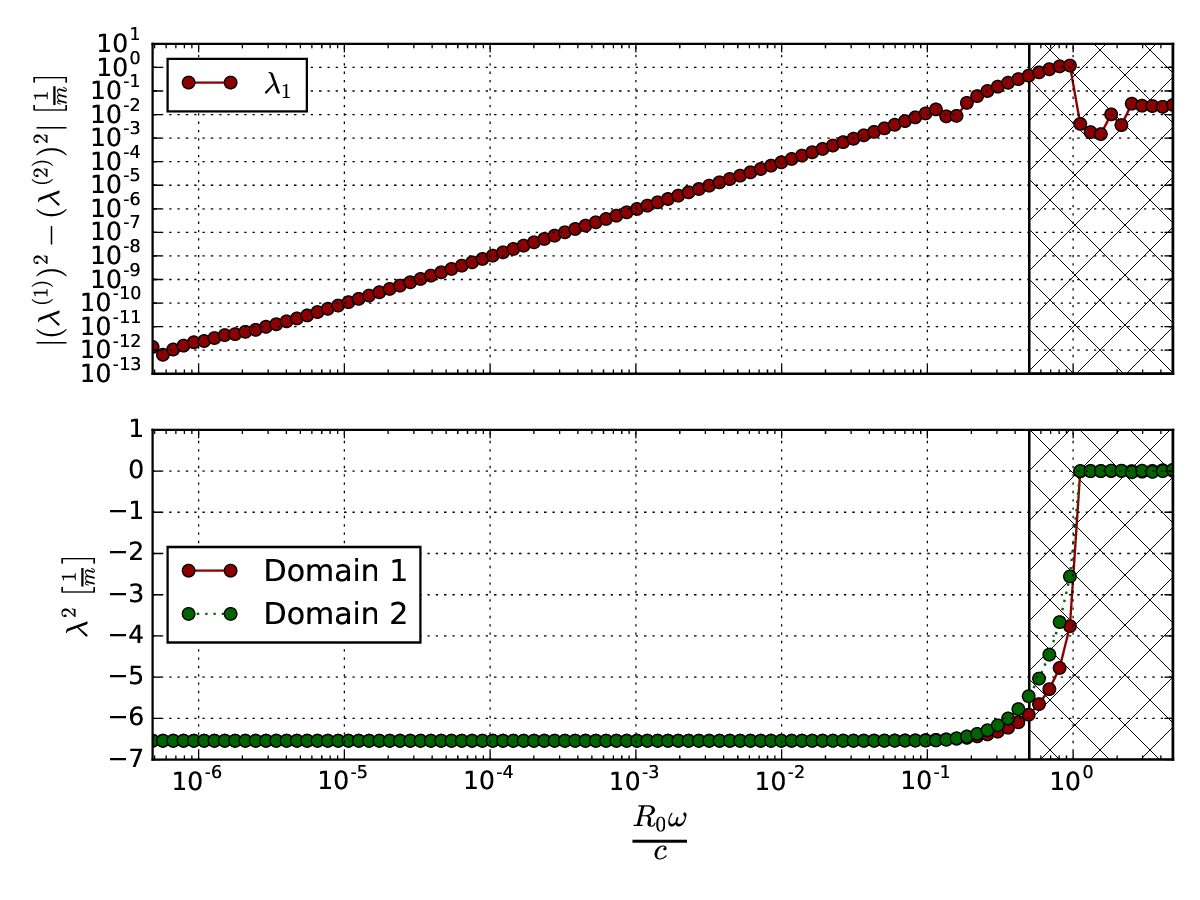}
\caption[Anisospectrality of the domains in fig. \ref{subfig:ElaborateIDs}.]{Deviation from the isospectral state for the second pair of isospectral domains of \cite{gordon_isospectral_1992} 
depending on $\tfrac{R_0\omega}{c}$. Computed using \eqref{eq:EV_FullPDE_FE_AlgebraicForm} with $\Cpv_3$ finite elements for vacuum $n=1$.
The hatched region marks the domain where
the Galilei-Newton \eqref{eq:Deriv_GalileiNewtonCoordinateTrafo} coordinate transformation is not physically valid.}
\label{fig:Exp_ID2AnisoAndEVFullVacuum}
\end{figure}

Next one may consider the influence of the shape on the anisospectrality, which is made possible by using two pairs of isospectral domains (\ref{fig:Intro_IsospectralManifolds}).
The eigenvalues have been computed using the full equation \eqref{eq:EV_FullPDE_FE_AlgebraicForm} for vacuum, subdividing the mesh of \ref{subfig:Exp_ID1Meshes} six times and that of \ref{subfig:Exp_ID2Meshes} five times. Due to the choice of the dimensions of the domains (as shown in \ref{fig:Intro_IsospectralManifolds} and \ref{tbl:SimParamID1})
this choice of refinement results in meshes of the same width $h=\tfrac{1}{2^5}$ for both pairs.
The results are shown in \ref{fig:Exp_ID1AnisoAndEVFullVacuum} (p. \pageref{fig:Exp_ID1AnisoAndEVFullVacuum}) and 
\ref{fig:Exp_ID2AnisoAndEVFullVacuum} (p. \pageref{fig:Exp_ID2AnisoAndEVFullVacuum}).
First it has to be noted, that for the second pair of isospectral domains (the original pair of \cite{gordon_isospectral_1992}) only the evolution
of the anisospectrality of the ground state is shown, for, contrary to the first pair of the domains, the ninth eigenmode is not an eigenfunction whose form is
known analytically for the second pair. It is thus of no particular importance.
Additionally the non-physical parameter range has been made clear by hatching the appropriate part of the plot.

An obvious conclusion which can be drawn from the results presented so far is that domains (or drums) which are isospectral in the sense
given in \ref{sec:Theo_IDsPriorArt} \textit{do not remain isospectral} when subject to uniform rotation.
To be more precise their spectra  diverge quadratically with $\tfrac{\omega}{c}$. Seeing as this happens for both pairs
of isospectral domains one may conclude that the exact shape of the domain has no (obvious) influence on the order of divergence.

Comparison of the results for the first and second pair shows that the eigenvalues of the second pair (fig. \ref{subfig:ElaborateIDs})
diverge immediately upon the onset of rotation, whereas those for the first pair do not. The existence of the plateaus for the first pair has been
explained above, its absence for the second pair may be explained by the fact that the lowest mode of these domains is highly localized in the cross-shaped
regions of the domains and thus experiences the rotation more clearly due to the larger distance from the centre of mass (the axis of rotation). 
Additionally the ground state is localized in a region with four re-entrant corners (in contrast to 2 for the first isospectral pair). It is not hard to imagine that
these corners exert a bigger influence on the state than the two corners of the first isospectral pair.
From the depiction of the eigenvalues themselves (bottom row of the plots \ref{fig:Exp_ID1AnisoAndEVFullVacuum} and \ref{fig:Exp_ID2AnisoAndEVFullVacuum})
one can see that the eigenvalues converge to $0$ for $\frac{R_0\omega}{c}\rightarrow 1$,
where $R_0 = \max\lbrace R_{0}^{(1)},R_{0}^{(1)}\rbrace $ and $R_{0}^{(i)}$ are the radii of circumcircles of the individual domains of a pair.
This behaviour is not exclusive to the full PDE \eqref{eq:10}, but is shared by it's linearized version \eqref{eq:SHDerived}, albeit with some caveats which are to be 
explored in the next section.

A further interesting point is the little dip of the anisospectrality visible in \ref{fig:Exp_ID2AnisoAndEVFullVacuum} for small parameter values. Upon closer inspection
one can see that the dependence of $\left\vert\left(\lambda^{(1)}\right)^2 - \left(\lambda^{(2)}\right)^2\right\vert$ on $\tfrac{R_0\omega}{c}$ becomes unstable
in this parameter range. This is due to the impurities introduced into the eigenvalues due to a missing $\Jmat$-orthogonalization of the Arnoldi basis, as has been described in \ref{sec:AlgForm_PracticalAspects}.
\clearpage

\subsection{The Medium and The Equation}\label{subsec:Exp_MediumAndEquation}
\begin{figure}[t]
\centering
\includegraphics[width=0.7\linewidth]{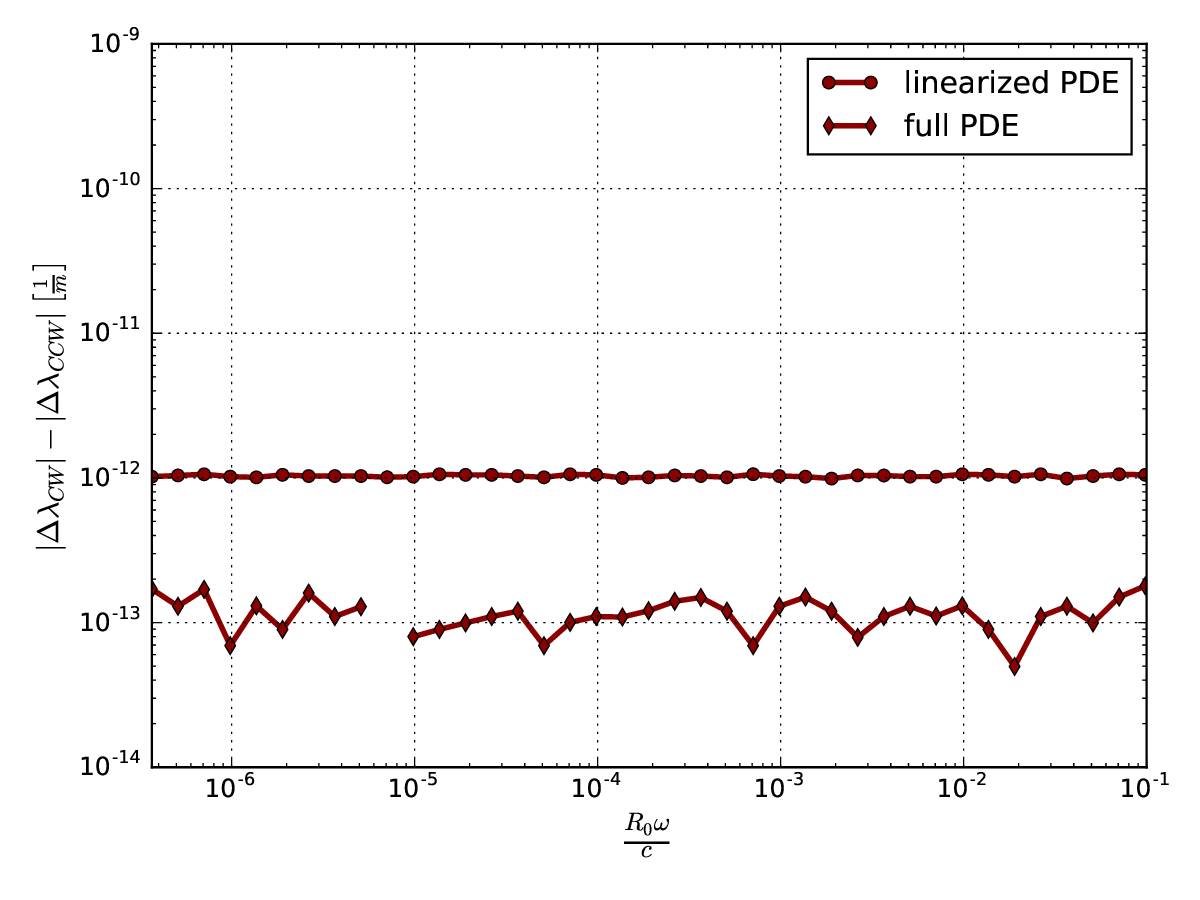}
\caption[Clockwise vs. counter-clockwise rotation.]{The difference of anisospectralities for clock-wise and counter-clock-wise rotation as a function of $\tfrac{R_0\omega}{c}$. Computed for the
pair depicted in fig. \ref{subfig:ElaborateIDs}. The break in the graph for the full PDE is due to a vanishing difference and is purely coincidental.}
\label{fig:Exp_CWvsCCW}
\end{figure}
Prior to proceeding with an investigation of the influence of the medium on anisospectrality we shall have a look at the behaviour of the
equations under a change of the direction of rotation. To this end the anisospectrality was computed for the pair shown in fig. \ref{subfig:ElaborateIDs} with
$\omega \lessgtr 0$ and the difference of both is shown as a function of $\tfrac{R_0\omega}{c}$ in \ref{fig:Exp_CWvsCCW}.
With the relative error for the eigenvalue search set to $10^{-11}$ the difference, which is on the order of $10^{-12}$ or less, can be safely neglected.
This shows that the deviation from isospectrality is invariant under change in the sense of rotation. One may argue that the next leading order contribution to
the anisospectrality will come from terms $\sim\left(\tfrac{\omega}{c}\right)^4$. The break in the curve for the full PDE (\eqref{eq:UnitDisc_PDE_QEVP} is, once again, explainable by a vanishing
difference and the chosen method of graphical representation. The rough nature of said curve is explained by its magnitude and the missing additional orthogonalization of the
Arnoldi-bases (c.f. \ref{sec:AlgForm_PracticalAspects}).
\vspace{1em}
\begin{figure}[p]
\centering
\subfloat[Vacuum]{
\includegraphics[width=0.7\linewidth]{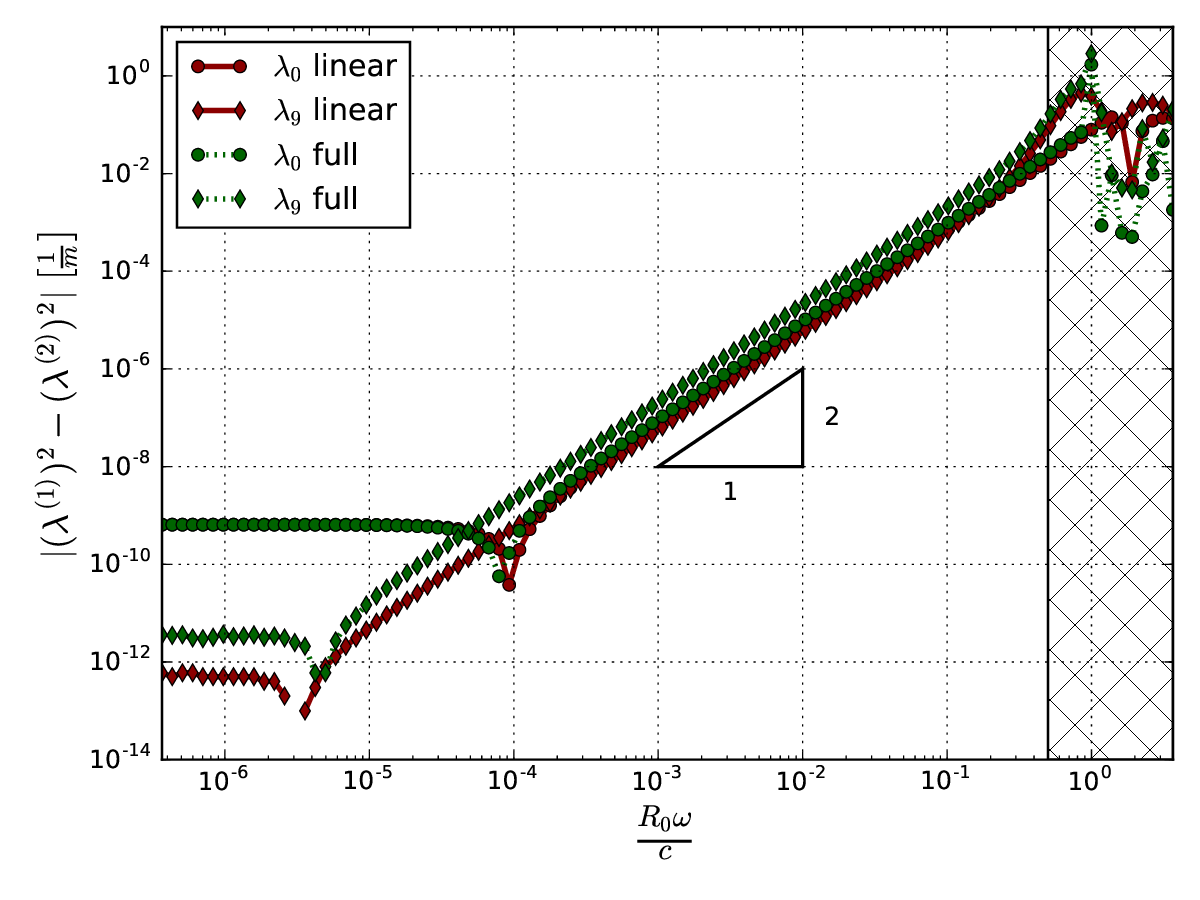}
\label{subfig:Results_AnisoFullVsLinVacuum}
}\;
\subfloat[Diamond]{
\includegraphics[width=0.7\linewidth]{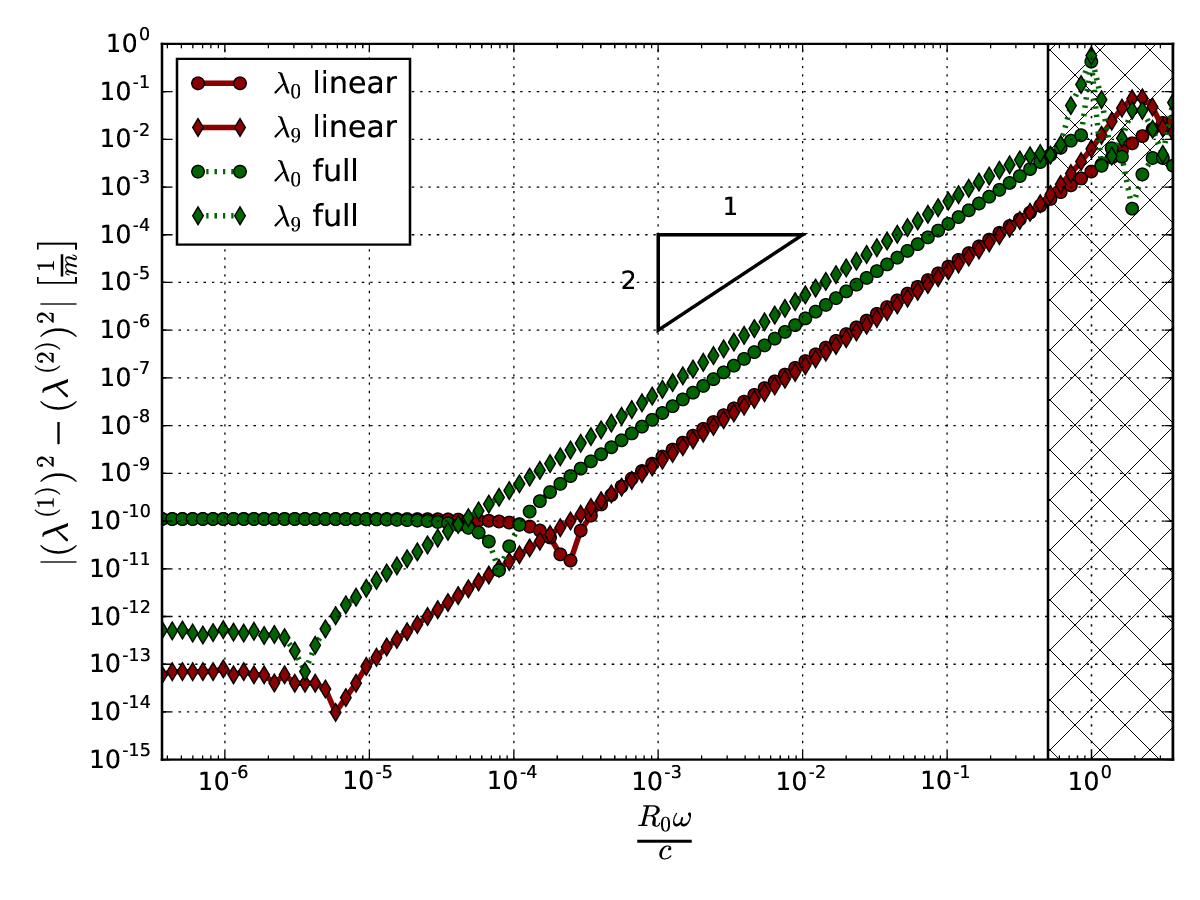}
\label{subfig:Results_AnisoFullVsLinDiamond}
}
\caption[Influence of the material on the anisospectrality.]{Anisospectrality of the first pair of isospectral domains for different materials (vacuum and diamond).
Shown are results for the ground state $\lambda_0$ and the $9^\text{th}$ excited state (c.f. \ref{sec:Theo_Analysis_IDs} ). The presence of a medium attenuates the changes
of eigenvalues/frequencies.}
\label{fig:Results_AnisoFullVsLinVsmaterial}
\end{figure}

Next we turn to the influence of the medium. The results obtained by solving \eqref{eq:EV_FullPDE_FE_AlgebraicForm} and \eqref{eq:EV_LinPDE_FE_AlgebraicForm}
for vacuum and medium for the first pair of isospectral domains are presented in \ref{fig:Results_AnisoFullVsLinVsmaterial} and warrant a close examination.
One notices immediately that the anisospectrality exhibits dips which have already been explained. Both equations result in anisospectralities which exhibit
the same quadratic order of growth with increasing angular frequency.

First we turn our attention to the results obtained for $n=1$. Here the results obtained with \eqref{eq:EV_FullPDE_FE_AlgebraicForm} and \eqref{eq:EV_LinPDE_FE_AlgebraicForm}
are in close proximity, with their difference being less than an order of magnitude.
This is quite easy to explain if one considers both equations.
In case $n=1$ we have in \eqref{eq:AlgForm_EV_PDE_vectorialform} (and thus in \eqref{eq:EV_FullPDE_FE_AlgebraicForm}):
\begin{equation}
\left[n^2-\left(\frac{r\omega}{c}\right)^2\right]\gamma^2 = \left[n^2-\left(\frac{r\omega}{c}\right)^2\right]\cdot \left[ 1-\left(\frac{r\omega}{c}\right)^2\right]^{-1} =
\left[1-\left(\frac{r\omega}{c}\right)^2\right]\cdot \left[ 1-\left(\frac{r\omega}{c}\right)^2\right]^{-1} = 1 \quad.
\end{equation}
Thus the $\lambda^2$-terms in both the linearized and the full PDE are identical. The only remaining difference is the $\left[\tfrac{\omega}{c}\right]^2$
term. Its influence is observed as the slight difference of the anisospectralities. Clearly, the Coriolis term determines the behaviour
of the eigenvalues, closely followed by the $\lambda^2$ term.
This observation justifies - \textit{a posteriori} - the omission of the $\del_\varphi^2$ term during theoretical analysis in \ref{ch:TheoAnalysis}.

Next we look at fig. \ref{subfig:Results_AnisoFullVsLinDiamond}. Here the difference between the two equations manifests itself clearly as a large difference in anisospectrality.
It is therefore not surprising that the crossover
point of the eigenvalues occurs for slower rotations in the case of the full equation than it does in the case of the linearised equation.
Furthermore one notices the displacement of the second dip in the curve to higher angular velocities and thus further into the non-physical realm. Which
conclusively demonstrates that the second dip may be safely ignored.
A change which is more subtle to notice is that the spectral deviations are collectively one order of magnitude smaller for $n=2.42$ than for $n=1$.

The medium therefore clearly separates the linearised and the full equations. Its presence attenuates the influence of rotation on the eigenmodes of the field, i.e.,
it introduces additional "inertia" into the field. Furthermore the numerical
gains obtained by introducing $n\neq 1$ are significant. An index of refraction different from $1$ leads to a sign change in the mass matrix of \eqref{eq:EV_FullPDE_FE_AlgebraicForm},
when $\tfrac{r\omega}{c} > 1$. In this case $\gamma^2 = \tfrac{1}{1-\left(\tfrac{r\omega}{c}\right)^2}< 0$ whereas $n^2 - \left(\tfrac{r\omega}{c}\right)^2 > 0$.

\begin{figure}
\centering
\includegraphics[width=0.7\linewidth,keepaspectratio]{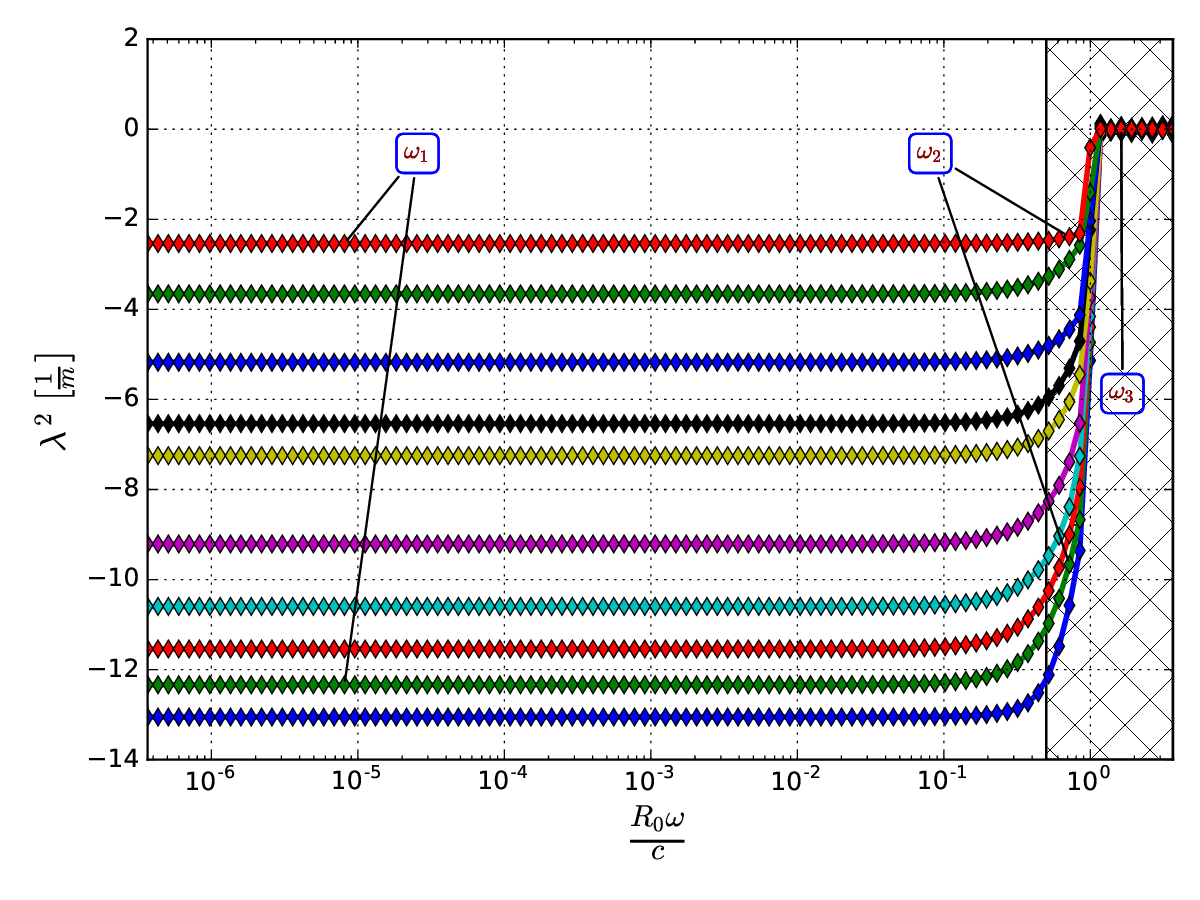}
\caption[Eigenvalues as functions of $\tfrac{R_0\omega}{c}$ for ID1.1.]{Evolution of the eigenvalues of ID 1.1 with increasing rotational speed. The hatched region denotes the non-physical parameter domain.}
\label{fig:Exp_AnnotadedEigenvalues}
\end{figure}

\begin{figure}[tbp]
\centering
\subfloat[The ground state for vacuum.]{
\includegraphics[width=0.75\linewidth,keepaspectratio]{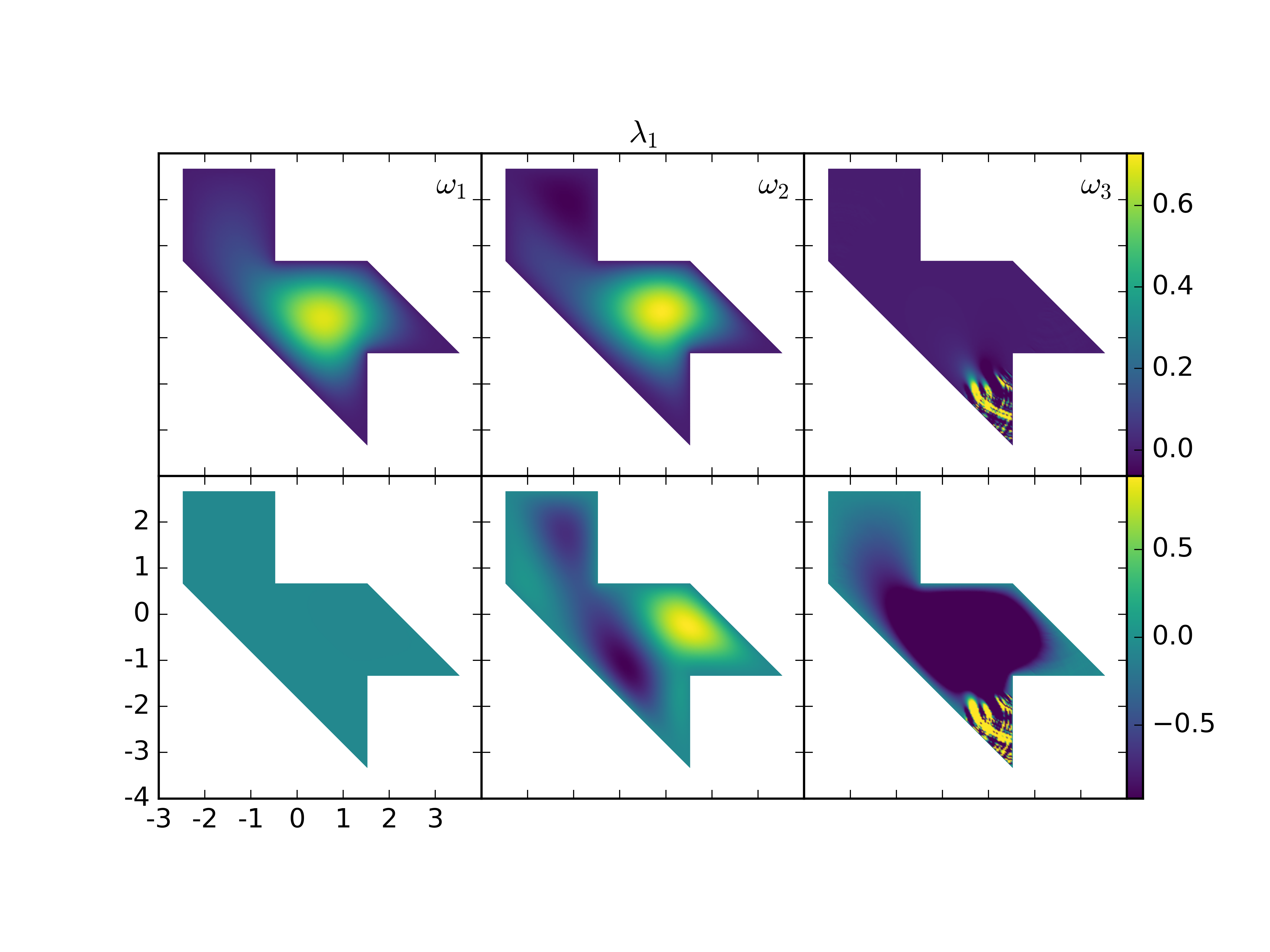}
\label{subfig:Exp_EigModeGS_ID1_1_Full_Vacuum}
}\;
\subfloat[The ground state for diamond $n=2.42$.]{
\includegraphics[width=0.75\linewidth,keepaspectratio]{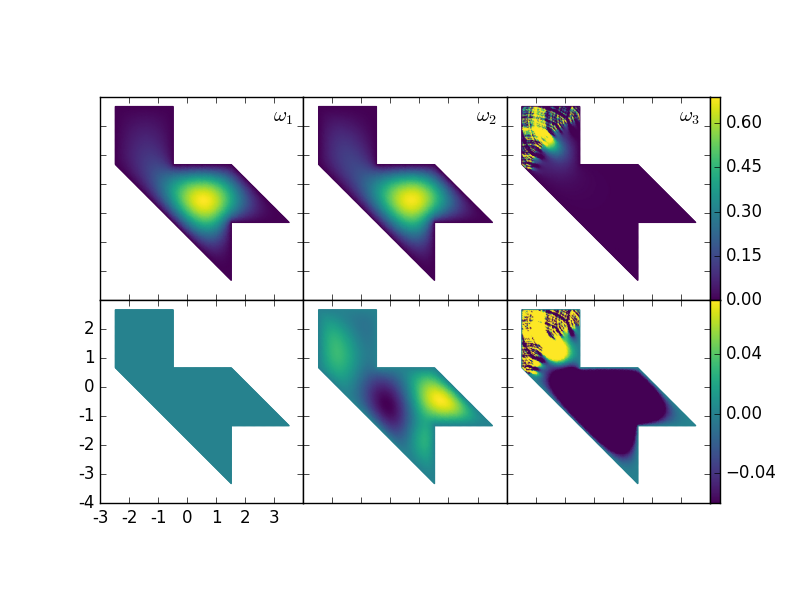}
\label{subfig:Exp_EigModeGS_ID1_1_Full_Diamond}
}
\caption[Ground state of ID1.1 (fig. \ref{subfig:SimpleIDs}) using the full equation, for different angular frequencies]{The ground state (top row) of a domain of the first isospectral pair for different angular frequencies and its deviation (bottom row)
from the lowest eigenmode of the static domain.
Computed using \eqref{eq:EV_FullPDE_FE_AlgebraicForm} and $\Cpv_1$ finite elements, assuming the domain to be made filled with vacuum (fig. \ref{subfig:Exp_EigModeGS_ID1_1_Full_Vacuum}) or
diamond (fig. \ref{subfig:Exp_EigModeGS_ID1_1_Full_Diamond}).
Normalization was performed only for $\omega_1,\omega_2$ because $\omega_3$ lies outside of the 
physical domain.}
\label{fig:Exp_EigModeGS_ID1_1_Full}
\end{figure}
\begin{figure}[tb]
\centering
\subfloat[The case for the full PDE.]{
\includegraphics[width=0.75\linewidth,keepaspectratio]{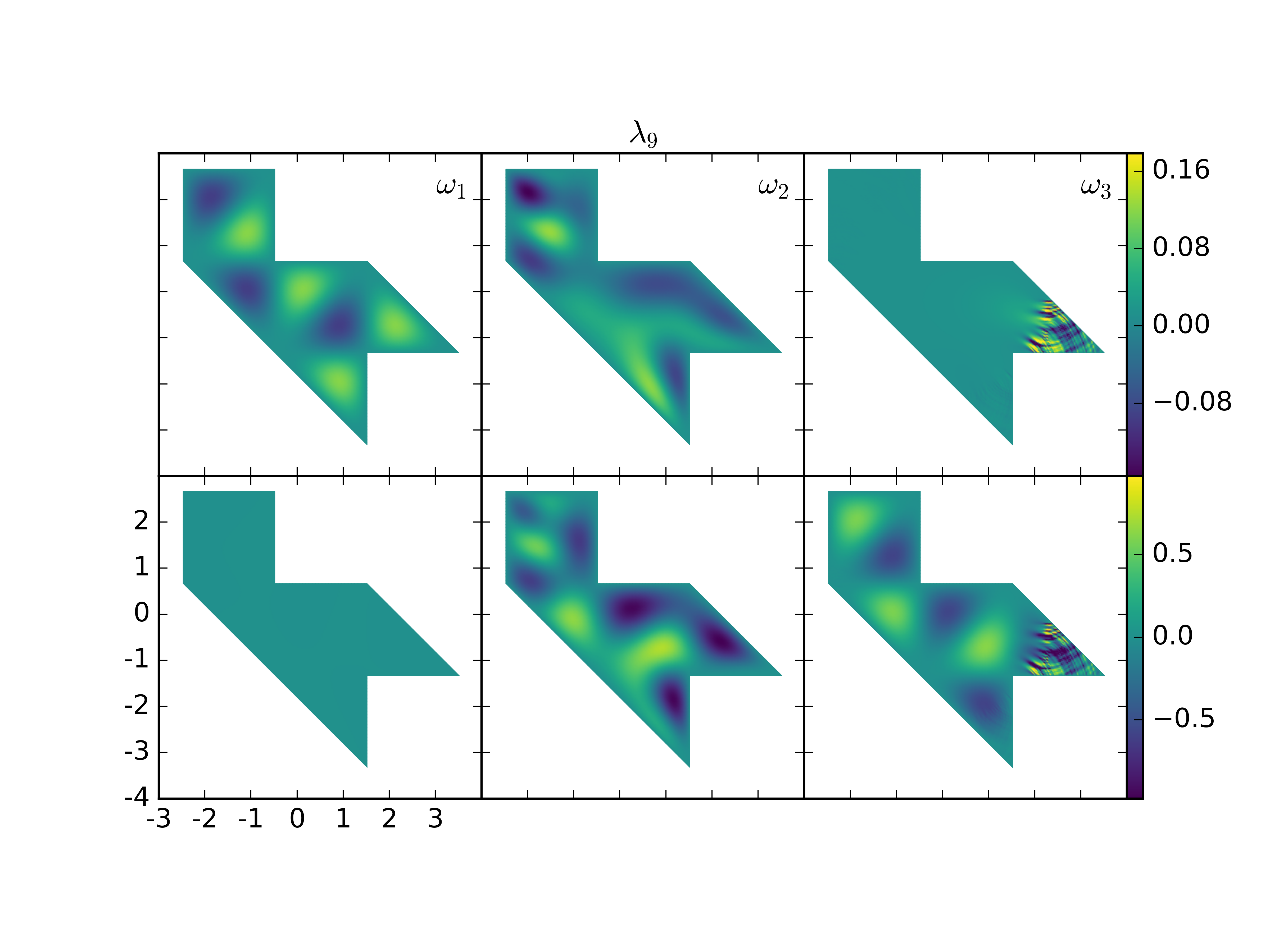}
\label{subfig:Exp_EigModeNS_ID1_1_Full}
}\;
\subfloat[The case for the linearized PDE.]{
\includegraphics[width=0.75\linewidth,keepaspectratio]{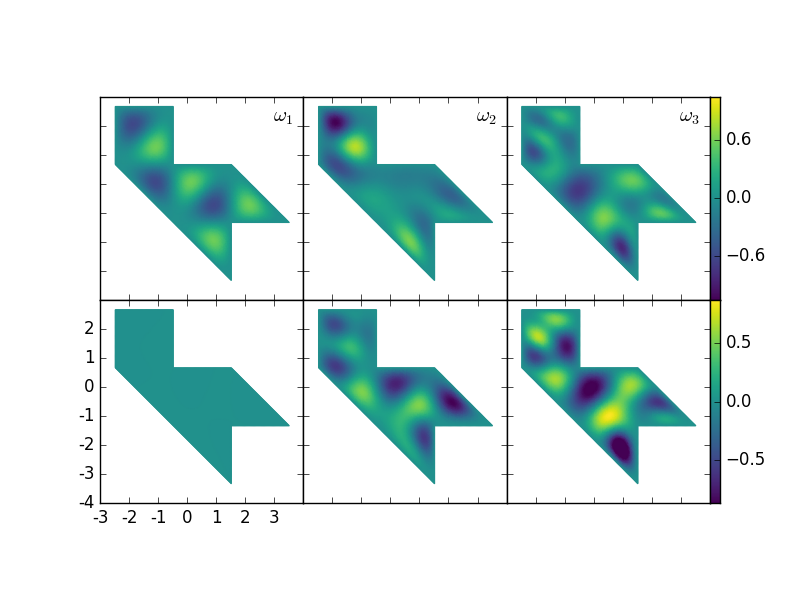}
\label{subfig:Exp_EigModeNS_ID1_1_Lin}
}
\caption[Ninth excited state of ID1.1 using the full and linearized equations, for different angular frequencies.]{The ninth excited state (top row) of a domain of the first isospectral pair for different angular frequencies and its deviation (bottom row) from the eigenmode \eqref{eq:Analysis_EigenfunctionsOfTriangles}.
Computed using \eqref{eq:EV_FullPDE_FE_AlgebraicForm} (fig. \ref{subfig:Exp_EigModeNS_ID1_1_Full}) and \eqref{eq:EV_LinPDE_FE_AlgebraicForm} (fig. \ref{subfig:Exp_EigModeNS_ID1_1_Lin}) and $\Cpv_1$ finite elements, assuming the domain to be made of vacuum.
Normalization was performed only for $\omega_1,\omega_2$ because $\omega_3$ lies outside of the physical domain.}
\label{fig:Exp_EigModeNS_ID1_1_Full}
\end{figure}

We end this section with a look at the eigenmodes of an isospectral drum for different values of the angular velocity and different types of equation.
In \ref{fig:Exp_AnnotadedEigenvalues} the $10$ computed eigenvalues of the first isospectral pair are shown. As has been mentioned above, the eigenvalues
converge towards $0$ as soon as the angular velocity at a point on the largest circumcircle of the domains reaches the speed of light.
These eigenvalues were obtained for $n=1$ with $\Cpv_3$ finite elements using the full equation \eqref{eq:EV_FullPDE_FE_AlgebraicForm}.
Again, the hatched region shows the range of parameters for which the results are physically questionable due to the use of \eqref{eq:Deriv_GalileiNewtonCoordinateTrafo}.
This convergence happens irrespective of whether the full equation of motion \eqref{eq:10} for $E_z$ or its linearized descendent \eqref{eq:SHDerived} are used.

The ground state and the ninth excited mode of the domains along with their deviation from their static shapes are shown in \ref{fig:Exp_EigModeGS_ID1_1_Full} 
(p. \pageref{fig:Exp_EigModeGS_ID1_1_Full}) and in \ref{fig:Exp_EigModeNS_ID1_1_Full} (p. \pageref{fig:Exp_EigModeNS_ID1_1_Full}). 
All eigenmodes were computed using $\Cpv_1$ FE due to practical constraints imposed by the available method of visualizing the results.
The color bars are normalized over the first two columns of their respective rows, for the eigenmode depicted in the third column lies
in the realm of non-physical angular velocities and would hinder the ability to compare the modes and their differences.

In \ref{fig:Exp_EigModeGS_ID1_1_Full} the influence of the medium ($n\neq 1$) on the behaviour of the eigenmodes of \eqref{eq:EV_FullPDE_FE_AlgebraicForm} can be observed.
The presence of the medium makes itself apparent for $\omega_2$, where one can see that it prevents the crest of the oscillation from wandering as far to the boundary as
it does fro vacuum. Next in \ref{fig:Exp_EigModeNS_ID1_1_Full} (p. \pageref{fig:Exp_EigModeNS_ID1_1_Full})  one can see the influence of the type of equation on the eigenmodes. As has been mentioned above, the eigenvalues
(or eigenfrequencies) converge towards $0$ for $\tfrac{R_0\omega}{c}\rightarrow 1$ irrespective of the type of equation. But the influence of terms of higher order than $\tfrac{\omega}{c}$
can be clearly seen in the behaviour of the eigenmodes. In the case of the linearized equation \eqref{eq:SHDerived} (resp. \eqref{eq:EV_LinPDE_FE_AlgebraicForm}) the
eigenmodes retain a sensible shape even when the angular velocity $\omega$ is pushed far into the non-physical domain (the case of $\omega_3$), as can be seen in fig. \ref{subfig:Exp_EigModeNS_ID1_1_Lin} (p. \pageref{subfig:Exp_EigModeNS_ID1_1_Lin}).
This clearly is an unexpected behaviour, for one would expect some drastic changes to take place when one leaves the domain of physics. The absence of such changes is further
evidence to the fact that the linearized equation is a simplification which is certainly not valid for high angular velocities.
In contrast, the full equation \eqref{eq:EV_FullPDE_FE_AlgebraicForm} displays such drastic changes in the eigenmodes of the domains, as can be witnessed in fig. \ref{subfig:Exp_EigModeNS_ID1_1_Full}.

It should perhaps be pointed out that the crests (and troughs) of the eigenmodes are pushed out from the axis of rotation towards the boundaries for increasing velocity of rotation.
When $\tfrac{R_0\omega}{c}=1$ they are concentrated at the boundary only. If $\omega$ is increased even further the radius $R_0$, past which $\tfrac{R_0\omega}{c}>1$ decreases and one
can observe, at least for the correct equation \eqref{eq:10}, that the nonsensical oscillations start to approach the axis of rotation from the boundary.
This can be observed in \ref{fig:Exp_EigModeGS_ID1_1_Full} (p. \pageref{fig:Exp_EigModeGS_ID1_1_Full}) for $\omega_3$, where even the disc with  $r < R_0$ (the still physical domain) can be seen.

\section{Discussion}\label{sec:Exp_Discussion}
In summary it has been demonstrated in this chapter, by numerical experiment, that planar isospectral drums \emph{do not remain isospectral} when
rotated with constant angular velocity. Their spectra deviate proportionally to $\left(\tfrac{\omega}{c}\right)^2$, with the exact shape of the
domains being of secondary importance. By comparing the results obtained for isospectral domains to those obtained for the unit square
one may conclude that both belong to the same class of geometric shapes, whereas the circle belongs to a different class due to its spectrum changing
$\sim\left(\tfrac{\omega}{c}\right)$ and the football-stadium cavity of \cite{sunada_design_2007} lying somewhere at the boundary
between both classes. This indicates that the influence of rotation on the spectrum of a planar domain does not depend primarily on its topology.

Furthermore it has been demonstrated that the direction of rotation does not influence anisospectrality, whereas the material, from
which isospectral domains are manufactured, as well as the type of equation (full or linearized) used to draw the conclusion contribute only constant factors
to the growth of the spectral deviation, with the order of the growth of the latter remaining the same.
It has been demonstrated too, that the contribution of the $\del_\varphi^2$ term in \eqref{eq:10} is negligible, which implies that the orientation of the
domains w.r.t. the axis of rotation is negligible as well, save for very high angular velocities. But in that case the coordinate transformation used
to obtain \eqref{eq:10} does not hold any more.

One can draw a parallel to mechanical systems here. Assuming the domain to be indeed a "drum", a two dimensional piece of a membrane clamped at the boundary such that no displacement out of the plane is possible there, one can argue that the faster this drum rotates the more centrifugal force does the membrane experience and any displacement will thus be pushed outwards to the boundary.
In the present case the drums are rotated with angular frequencies of such magnitudes that their outer edges do reach the speed of light,
where one would expect the problem to turn non-physical and yield nonsensical results.
This circumstance has been observed in the numerical experiments performed, though their emergence is dependent on whether all necessary terms of the 
equation of motion of $E_z$ are accounted for.
If one were to investigate the parallel to mechanical systems further one could argue that the axes of rotation should be chosen such that most of the displacement (i.e., most of the density)
is concentrated around the axis of rotation. Alas, even in this case the spectra of isospectral drums diverge in the way described above.

Note that, with the anisospectrality of the isospectral pairs changing by $8$ orders of magnitude, for $\lambda_1$ and $\lambda_9$ of the first pair of \ref{fig:Intro_IsospectralManifolds},
it is highly unlikely that this is due to a numerical error.

\part{Conclusions and Appendix}
\chapter{Summary and Outlook}

\section{Summary}
The journey through this thesis began with a simple question 
\begin{center}
"Can one hear the shape of a rotating drum?"
\end{center}

An affirmative answer was finally obtained in \ref{ch:Exp_TheExperiment}.
To arrive at this answer we have considered rums of \ref{fig:Intro_IsospectralManifolds}, which are known to be isospectral when at rest.
They have been subjected to rotations around an axis perpendicular to their surfaces. The resulting changes in the spectra have been studied in
\ref{ch:TheoAnalysis} and \ref{ch:Exp_TheExperiment}. To allow for the widest range of angular velocities we assumed the electric field to
be the drumhead and determined its equation of motion in \ref{ch:Derivations}. This feat was achieved using a formulation of 
electrodynamics in the language of differential forms delineated in \ref{ch:PhysicalBasis} for which we had first to delve deeper into
pure mathematics in \ref{ch:MathematicalMinimum}. All of this was performed whilst keeping in mind that, in practice, such
"drums" would have to be manufactured from a material. This motivated the introduction of a simple medium \eqref{eq:Def_LinearConstitutiveRelations}
and led to a re-derivation of the Minkowski relations as they are stated in \cite{minkowski_grundgleichungen_1910}.
The latter was achieved without the use of Lorentz-transformations and was made possible by the introduction of the concept of
a 3+1 decomposition (or foliation) of space-time (\ref{sec:Deriv_SpacetimeFoliations}).

Apart from the main result of \ref{ch:Exp_TheExperiment} the behaviour of the spectra of simple geometric shapes (square and disc)
was investigated analytically in \ref{ch:TheoAnalysis}. The results showed (\ref{ch:TheoAnalysis}, \ref{sec:Exp_Validation}) that
the order of the change of eigenvalues with $\tfrac{\omega}{c}$, where $\omega$ is the angular velocity of the domain, is
different for both. The degenerate eigenfrequencies of the disc drift apart linearly with $\ocf$ and those of the square
and of isospectral domains are doing the same $\propto \left(\ocf\right)^2$.
In \ref{sec:Exp_Validation} we have also encountered a domain which is a hybrid of a disc and a square - the football stadium cavity of \cite{sunada_design_2007}.
The almost-degenerate eigenmodes of this domain have been verified to behave as if the domain were as square or a disc, depending on the
angular velocity of its rotation.

Apart from determining an answer to the main question further investigations of the behaviour of the electric field-turned-drumhead was carried out in \ref{ch:Exp_TheExperiment}.
It was found that the presence of a medium attenuates the effects of rotation on the difference in the spectra of isospectral domains.
The medium has been found to provide additional "inertia" to the transverse magnetic modes of studied here.

Furthermore we have seen that, although the linearized equation \eqref{eq:SHDerived} reproduces the behaviour of the ansisospectrality and of eigenvalues correctly,
it yields, too, sensible eigenmodes for nonsensical angular velocities. This is not the case for the complete PDE \eqref{eq:10}.

All of the conclusions, save for the disc, required the use of numerical methods. The latter were deliberated in \ref{ch:Exp_Introduction},
evolved in \ref{ch:Exp_AlgebraicFormulation} and applied to great effect in \ref{ch:Exp_TheExperiment}.
Along the implementation path we encountered the concept of a gyroscopic quadratic eigenvalue problem and the possibilities of dealing with it.
This led to the introduction of a H/SH matrix pencil and a shift operator for the shift-and-invert Arnoldi method more fitting to the problem. 
The methods for H/SH problems may now be applied to any case of rotating structures, e.g. tyres.

To reiterate: Drums which are isospectral but geometrically different when at rest do not remain isospectral when rotated.

\section{The Way To Go}
With the basic results presented here, multiple paths to tread upon further have been
illuminated.
The most fruitful to me seems to be the theoretical one, for if one were determined
to disprove the findings of this work the easiest way would be by showing a flaw in its
theoretical foundations or the derivations of the equations used. To facilitate this the explicit derivations 
of \ref{ch:Derivations} have been provided.

On a more constructive side one could attempt to prove anisospectrality of
isospectral in the case when one only has a pseudo-metric at hand. A less ambitious
but nevertheless constructive goal is to derive an analytic expression for the dependence
of anisospectrality on the angular frequency $\omega$.

A more in-depth treatment of the derivation of \eqref{eq:10} is warranted, for doubts remain whether it is as meaningful as it is
made out to be (e.g., \cite{sunada_design_2007}). These doubts arise when one considers the 3+1 decomposition used to derive this
equation and the fact that the form of the oscillations of the field could be mapped by a point charge if it is taken around the domain.
But the latter would introduce an additional coframe into the lot.
Additionally the current considerations were limited to two-dimensional domains. In reality these domains would be 3 dimensional. The
presence of the third dimension may exert some influence on the measurable eigenfrequencies of the domains.
Such an investigation could be carried out numerically using i.e. \textsc{Comsol}\footnote{The accompanying code is for 2D FE only.} 
but would require a different evaluation of \eqref{eq:Deriv_LaplaceBeltrami_Coefficients}.

The numerical treatment of the PDE \ref{eq:10} provides a fertile ground for a multitude of numerical
experiments and improvements of the current results. As has been noted in \ref{sec:ExpIntro_QEVP}, the SHIRA algorithm 
has not been used in this work in its full form. An extension of ARPACK++ to permit treatment
of skew-Hamiltonian/Hamiltonian eigenvalue problems and its application to the problem at hand 
is one simple way of improving upon the current results.

Further venues are the use of a fundamentally different numerical method for the solution of \eqref{eq:10}
on the domains of \ref{fig:Intro_IsospectralManifolds}  and an introduction of an 
adaptive mesh, both of which have been remarked upon in \ref{sec:ExpIntro_MethodOverview}. Furthermore an
extensive study of isospectral domains could be made using the software provided with this thesis and a selection of
domains presented in \cite{buser_planar_2010}, for I make no claims of having done such a study.

The choice of the centre of mass as the point around which the domains are rotated in the plane is,
as noted in \ref{sec:Exp_Validation} fairly arbitrary. One may wish to study the influence of the position of the
axis of rotation w.r.t. the domain on the spectrum of the domain. Especially if one were to draw a parallel to 
classical mechanics and Steiner's theorem for the moment of inertia.

\appendix
\chapter{Miscellaneous Calculations}\label{ch:App_CoordTransforms}
\section{The Hodge Dual}\label{sec:App_HodgeDual}
Here the derivation of the relations expressions for the Hodge-duals \eqref{eq:GeneralHodgeDualOfOneForm} and \eqref{eq:GeneralHodgeDualOfTwoForm} are given.
\begin{align*}
*dx^0 &= \iota_{dx^0}\left(\epsilon\right) = \sqrt{|\det g|}\left(\iota_{dx^0}dx^1\wedge dx^2\wedge dx^3 +(-1)dx^0\wedge\iota_{dx^0}(dx^1\wedge dx^2\wedge dx^3)\right)\\
&= \sqrt{|\det g|}\left[\iota_{dx^0}(dx^0)dx^1\wedge dx^2\wedge dx^3 +(-1) dx^0\wedge\iota_{dx^0}(dx^1)dx^2\wedge dx^3\right.\\
&+ \left.(-1)^2 dx^0\wedge dx^1\wedge\iota_{dx^0}(dx^2)dx^3 + (-1)^3 dx^0\wedge dx^1\wedge dx^2\iota_{dx^0}(dx^3)\right]\\
&= \sqrt{|\det g|}\left[ g^{00}dx^1\wedge dx^2\wedge dx^3 - g^{01} dx^0\wedge dx^2 \wedge dx^3 + g^{02} dx^0\wedge dx^1\wedge dx^3
- g^{03} dx^0\wedge dx^1\wedge dx^2\right]\\
\end{align*}
Expressions for $dx^i$ follow in a similar manner. Generalizing the above will yield
\begin{equation}
*dx^i = \sqrt{ \left\vert \det g\right\vert } \left[ g^{i0} dx^1 dx^2 dx^3
- g^{i1}dx^0 dx^2 dx^3 + g^{i2} dx^0 dx^1 dx^3
-g^{i3} dx^0 dx^1 dx^2\right].
\end{equation}
Proceeding to 2-forms one obtains
\begin{align*}
*(dx^0\wedge dx^2) &= \dots = \sqrt{|\det g|}\left[g^{00}\left( g^{21}dx^2dx^3 - g^{22}dx^1dx^3 + g^{23}dx^1dx^2\right)\right.\\
&- g^{01}\left( g^{20}dx^2dx^3 - g^{22}dx^0dx^3 + g^{23} dx^0dx^2\right) + g^{02}\left( g^{20}dx^1dx^3 - g^{21}dx^0dx^3 + g^{23}dx^0dx^1\right)\\
&- \left.g^{03}\left( g^{20}dx^1dx^2 - g^{21}dx^0dx^2 + g^{22}dx^0dx^1\right)\right]
\end{align*}
or in general
\begin{multline}
*(dx^i\wedge dx^j) = \sqrt{\left\vert \det g \right\vert } \left[ g^{i0}\left(
g^{j1}dx^2\wedge dx^3 - g^{j2}dx^1\wedge dx^3 +g^{j3}dx^1\wedge dx^2\right) \right.\\
 - g^{i1}\left( g^{j0}dx^2\wedge dx^3 - g^{j2} dx^0\wedge dx^3 + g^{j3} dx^0\wedge dx^2\right)
 + g^{i2}\left( g^{j0}dx^1\wedge dx^3 - g^{j1} dx^0\wedge dx^3 + g^{j3} dx^0\wedge dx^1\right)\\
\left. - g^{i3}\left( g^{j0} dx^1\wedge dx^2 - g^{j1} dx^0\wedge dx^2 + g^{j2}dx^0\wedge dx^1\right)
\right]\;.
\end{multline}

\section{Uniform Linear Acceleration}
Following \cite{hehl_foundations_2003} and \cite{koks_explorations_2006} one obtains the following expressions for the coordinates
of a point in a coordinate system $K'$, uniformly accelerating along the $x$-axis of the inertial coordinate system $K$ ($\tau$ is the proper time
in the moving system):
\begin{align}
x(\tau) &= \frac{c^2}{g}\left[\cosh\left(\frac{g\tau}{c}\right) -1\right] + x'\cosh\left(\frac{g\tau}{c}\right)\label{eq:UniformLinearAccel_XCoordinateFunction}\\
t(\tau) &= \frac{c}{g}\sinh\left(\frac{g\tau}{c}\right) + \frac{1}{c}x'\sinh\left(\frac{g\tau}{c}\right)\label{eq:UniformLinearAccel_TCoordinateFunction}
\end{align}
These in turn yield
\begin{align*}
dt &= \left(1+\frac{gx'}{c^2}\right)\cosh\left( \frac{g\tau}{c}\right)d\tau + \frac{1}{c}\sinh\left( \frac{g\tau}{c}\right)dx'
& d\tau &= \left[1+\frac{gx'}{c^2}\right]^{-1}\left[\cosh\left(\frac{g\tau}{c}\right) dt - \frac{1}{c}\sinh\left( \frac{g\tau}{c}\right)dx\right]\\
dx &= \left( c+\frac{gx'}{c}\right) \sinh\left(\frac{g\tau}{c}\right)d\tau + \cosh\left( \frac{g\tau}{c}\right)dx' &
dx' &= \cosh\left(\frac{g\tau}{c}\right)dx - c\sinh\left( \frac{g\tau}{c}\right)d\tau\\
dy &= dy' & dy' &= dy\\
dz &= dz' & dz' &= dz
\end{align*}
These transformations of the cobases can then be used in a manner similar to that presented in \ref{ch:Derivations} to obtain the
following relations for the coefficient functions of the electromagnetic 2-forms
\begin{align*}
E_1 &= \left[ 1+\frac{gx'}{c^2}\right]^{-1} E_1' & E_2 &= \left[ 1+\frac{gx'}{c^2}\right]^{-1}\cosh\left(\frac{g\tau}{c}\right) E_2' + B_{12}' c \sinh\left(\frac{g\tau}{c}\right)\\
E_3 &= \left[ 1+\frac{gx'}{c^2}\right]^{-1}\cosh\left(\frac{g\tau}{c}\right) E_3' - B_{31}' c \sinh\left(\frac{g\tau}{c}\right)
& B_{23} &= B_{23}'\\
B_{31} &= B_{31}'\cosh\left(\frac{g\tau}{c}\right) - \frac{1}{c}\left[ 1+\frac{gx'}{c^2}\right]^{-1}E_3'\sinh\left(\frac{g\tau}{c}\right) & & \\
B_{12} &= B_{12}'\cosh\left(\frac{g\tau}{c}\right) + \frac{1}{c}\left[ 1+\frac{gx'}{c^2}\right]^{-1}E_2'\sinh\left(\frac{g\tau}{c}\right) & &
\end{align*}
along with the constitutive relations for a simple medium
\begin{gather*}
\CalD_{23} = \frac{\lambda_0\epsilon}{c}E_1,\quad \CalD_{31} = \frac{\lambda_0\epsilon}{c} E_2\\
\CalD_{12} = \frac{\lambda_0\epsilon}{c}E_3, \quad \CalH_{1} = \frac{\lambda_0 c}{\mu} B_{23}\\
\CalH_2 = \frac{\lambda_0 c}{\mu} B_{31},\quad \CalH_3 = \frac{\lambda_0 c}{\mu}B_{12}\;.
\end{gather*}
After a final transformation to the adapted coframe ($K'$ here) of the uniformly accelerated observer one obtains
\begin{align*}
\CalH_1' &= \left( 1+\frac{gx'}{c^2}\right) \CalH_1 & \CalH_2' &= \left( 1+\frac{gx'}{c^2} \right) \left[\cosh\left( \frac{g\tau}{c} \right)\CalH_2 + c\sinh\left(\frac{g\tau}{c}\right)\CalD_{12}\right]\\
& & \CalH_3' &= \left( 1+\frac{gx'}{c^2} \right) \left[\cosh\left( \frac{g\tau}{c} \right)\CalH_3 - c\sinh\left(\frac{g\tau}{c}\right)\CalD_{31}\right]\\
\CalD_{23}' &= \CalD_{23} & \CalD_{31}' &= \CalD_{31}\cosh\left(\frac{g\tau}{c}\right) - \frac{1}{c}\sinh\left(\frac{g\tau}{c}\right)\CalH_3\\
& & \CalD_{12}' &= \CalD_{12}\cosh\left(\frac{g\tau}{c}\right) + \frac{1}{c}\sinh\left(\frac{g\tau}{c}\right)\CalH_2\;.
\end{align*}
Expanding exemplary the last relation one obtains
\begin{multline}
\CalD_{12}' = \lambda_0\left[\frac{\epsilon}{c}\frac{1}{1+\frac{gx'}{c^2}}\cosh^2\left( \frac{g\tau}{c}\right) E_3' - \epsilon\sinh\cosh \left( \frac{g\tau}{c}\right) B_{31}'\right.\\
\left.+\frac{1}{\mu}\sinh\cosh\left( \frac{g\tau}{c}\right) B_{31}' - \frac{1}{\mu c}\frac{1}{1+\frac{gx'}{c^2}}\sinh^2\left( \frac{g\tau}{c}\right) E_3'\right]\;,
\end{multline}
which does not correspond to the results presented in \cite{anderson_electromagnetic_1969} for an observer accelerating uniformly and an inertial medium. 
With success in deriving the necessary equation in \ref{ch:Derivations} as well as further successful derivations not provided here I am inclined to believe that the expressions
given in \cite[tbl. I case 2]{anderson_electromagnetic_1969} are incorrect.

\section{The Unit Disc}\label{sec:App_UnitDisc}
An analytically tractable case is given by a disc in $\RR^2$. For simplicity the unit disc $\Cd_1=\bigl\{\xx\in\RR^2\; :\; \Vert \xx\Vert_2 \leq 1\bigr\}$
is chosen as the domain. Starting with \eqref{eq:10} and assuming
\begin{align}
E_3(r,\varphi;t) &= \Psi(r,\varphi)e^{c\lambda t}\qquad\lambda\in\CC \notag\\
0 &= \Delta\Psi - \left[\ocf\right]^2\del_\varphi^2\Psi + 2\ocf\lambda\del_\varphi\Psi - \lambda^2 \Psi\quad .\label{eq:TheoAnalysis_PDE_EVP}
\end{align}
Clearly, to remain in the realm of physically meaningful equations one has to require $\tfrac{R_0\omega}{c}\leq 1$, where $R_0=1$ is the upper bound on the extent of
the domain from the origin, through which goes, by derivation, the axis of rotation. Considering the above equation in planar polar coordinates one can see, that a sufficient
condition for $\lambda$ to remain unaffected by rotation is for the associated eigenmode of the disc to be invariant under continuous rotations.
This is clearly the case for the ground state of the disc and shall become apparent in the course of this section.

Next one may assume that $\Psi$ factorizes into a part $R(r)$, dependent only on $r$ and a part $F(\varphi)$, thus:
\begin{equation}
\Psi(r,\varphi) = R(r)F(\varphi)\quad .\label{eq:SeparationAnsatz_Psi_Polar}
\end{equation}
This yields (here the primes $'$ denote differentiation w.r.t. the appropriate variable):
\begin{align*}
0 &= \left[\del_r^2 + \frac{1}{r}\del_r + \frac{1}{r^2}\del_\varphi^2\right] RF  - \left[\ocf\right]^2 R \del_\varphi^2F
+2\ocf\lambda R \del_\varphi F - \lambda^2 RF\\
&= (R''  +\frac{1}{r} R') F + \frac{1}{r^2} R F'' - \left[\ocf\right]^2 R F'' + 2\ocf \lambda R F' -\lambda^2 RF
\end{align*}
and after multiplication with $\tfrac{r^2}{RF}$:
\begin{equation}
0 = r^2\frac{R''}{R} + r\frac{R'}{R} + \frac{F''}{F} - \left[\rocf\right]^2 \frac{F''}{F} + 2\frac{r^2\omega}{c}\lambda\frac{F'}{F} - r^2\lambda^2\quad.
\end{equation}

As a start the behaviour of the eigenmodes in the stationary limit $\omega=0$ should be studied, which amounts to considering
\begin{equation}
0 = r^2\frac{R''}{R} + r\frac{R'}{R} + \frac{F''}{F} - r^2\lambda^2\quad.\label{eq:PDE_OnDisc_With_Separation_Stat}
\end{equation}
Separating the above into $r,\varphi$ dependent parts results in
\begin{equation}
r^2\frac{R''}{R} + r\frac{R'}{R} - r^2\lambda^2 = \sigma = -\frac{F''}{F},\quad \sigma=\text{const.}\quad,
\end{equation}
which in turn yields an ordinary differential equation for $F$:
\begin{equation}
F''(\varphi) + \sigma F(\varphi) = 0\quad .\label{eq:F_ODE}
\end{equation}
Here one can make the classical ansatz
\begin{equation}
F(\varphi) = e^{\kappa\varphi}\;,\label{eq:AzimuthalODE_ClassicalExponentialAnsatz}
\end{equation}
which will result in \[ \kappa^2 + \sigma = 0. \]

The three possible scenarios are:
\begin{enumerate}
\item[$\sigma =0$:] $F''(\varphi) = 0\Longrightarrow F(\varphi) = A_0\varphi + B_0$
\item[$\sigma < 0$:] Then $-\sigma > 0 \Rightarrow \kappa = \pm \sqrt{-\sigma}\in\RR$.
Defining$\mu:=\sqrt{-\sigma}$ yields
\begin{equation*}
F(\varphi) = A_1e^{\mu\varphi} + B_1e^{-\mu\varphi}\;.
\end{equation*}
\item[$\sigma >0$:] Then $F'' + \sigma F = 0$ possesses the general solutions
\begin{equation*}
F(\varphi) = A_2 e^{i\sqrt{\sigma}\varphi} + B_2 e^{-i\sqrt{\sigma}\varphi}\quad .
\end{equation*}
\end{enumerate}
Here $A_i,B_i\in\RR,\; i=0,1\quad A_2,B_2\in\CC$.

$F$ being part of a physically meaningful quantity can not attain different values at the same point in space. This requires
\begin{equation*}
A_0 = A_1 = B_1 = 0\quad.
\end{equation*}
Having chosen a coordinate system adapted to the behaviour of the physical system and having a domain invariant under arbitrary rotations around the origin
one may reasonably expect that $\Psi$ is periodic in $\varphi$, that is
\[
F(\varphi + 2k\pi) = F(\varphi)\quad k\in\ZZ\quad .
\]
For $\sigma >0 $ this leads to the following constraint on $\sigma$:
\begin{equation}
k\sqrt{\sigma} = m\cdot 2\pi\qquad m\in\ZZ\quad .
\end{equation}
Assuming $k=1$ yields $\sqrt{\sigma}=m$.
\begin{equation}
F(\varphi) = A_2 e^{im\varphi} + B_2 e^{-im \varphi}\quad m\in\ZZ .\label{eq:AzimuthalFactor_OnDisc}
\end{equation}
Here the first term corresponds to a wave propagating clock-wise in the plane of the disc and the second term to
a counter-clock-wise propagating wave (or vice-versa, depending on $\sgn(m)$).
Note that if $m=0$ one has $F(\varphi)=A_2+B_2 = \text{const.}$ and the case reduces to $\sigma = 0$.

Having chosen $\sigma  >0$, where from $\sigma = m^2$, and established the form of the azimuthal dependence one can proceed to consider the radial dependence of \eqref{eq:PDE_OnDisc_With_Separation_Stat}:
\begin{equation}
r^2 \frac{R''}{R} + r\frac{R'}{R}  -r^2\lambda^2 = m^2\; \overset{\cdot R}{\Longrightarrow}\;r^2 R'' + rR' - (r^2\lambda^2 + m^2)R = 0\quad .\label{eq:ModifiedBessel_Stat_PreForm}
\end{equation}
Define
\begin{subequations}\label{eq:TheoAnalysis_UnitDiscFirstCoordSubst}
\begin{align}
x &:= \lambda r\\
R(r) &= R(r(x)) =: \tilde{R}(x)\\
R'(r) &= \frac{dR}{dr} = \frac{d\tilde{R}}{r}(x) = \frac{d\tilde{R}}{dx}\frac{dx}{dr}= \tilde{R}' \cdot \lambda
\end{align}
\end{subequations}
thus obtaining
\begin{equation}
\left(\frac{x}{\lambda}\right)^2 R'' + \left(\frac{x}{\lambda}\right) R' -(x^2 + m^2)R
= x^2 \tilde{R}'' + x \tilde{R}' -(x^2+m^2)\tilde{R} = 0\quad .\label{eq:ModifiedBessel_Stat}
\end{equation}
This is a \textit{modified} Bessel's equation, whose general solution is given by
\begin{equation}
\tilde{R}(x) = A_3 I_m(x) +B_3 K_m(x)
\end{equation}
where $I_m,K_m$ are modified Bessel functions of the first and second kind.
Due to $K_m$ having a logarithmic singularity for $r\rightarrow 0\Leftrightarrow x\rightarrow 0$, that is at the centre of the disc, one may neglect its contribution by setting $B_3 = 0$, thus
\[ \tilde{R}(x) = A_3 I_m(x) \] and 
\begin{equation}
\Psi_m(r,\varphi) = A_3 I_m(r\lambda)\left[A_2 e^{im\varphi} + B_2 e^{-im\varphi}\right]\label{eq:Disc_ComplexGeneralSolution}
\end{equation}
Enforcing Dirichlet boundary conditions at $r=1$ is equivalent to requiring:
\begin{equation}
\tilde{R}(\lambda r)\vert_{r=1} = 0\Longleftrightarrow A_3 I_m(\lambda) = 0 \Longrightarrow I_m(\lambda) = 0\quad.
\end{equation}
Here the following identity comes in handy \cite{_dlmf:_????},
\begin{equation}
I_\nu(x) = e^{\pm i\nu\frac{\pi}{2}} J_\nu\left(xe^{\pm i\frac{\pi}{2}}\right)\quad.\label{eq:Bessel_ModBessel_FirstKind_Relation}
\end{equation}
Thus
\[
I_m(\lambda) = 0 \equiv i^{-m} J_m(i\lambda) = 0\;,
\]
which finally yields
\begin{equation}
i\lambda = i\lambda_{m,s} = j_{m,s} \Longleftrightarrow \lambda_{m,s} = -i j_{m,s}\; ,\label{eq:Eigenmodes_Disc_static_Complex}
\end{equation}
where $j_{m,s}$ denotes the $s$-th zero of the ordinary Bessel function $J_m$. One immediately notes that $\lambda_{m,s}$ is purely complex, as
could be anticipated by comparing the above derivation to the solution of the wave equation on a disc with the temporal ansatz $e^{ic\lambda t}$.

\subsection*{Dynamic Analysis}
Next one can consider the PDE \eqref{eq:TheoAnalysis_PDE_EVP} for a rotating disc and assume, guided by the prior derivation,
\begin{equation*}
F(\varphi) = e^{\pm im\varphi}\quad .\label{eq:AzimuthalAnsatz_OnDisc_Rot}
\end{equation*}

Then \eqref{eq:TheoAnalysis_PDE_EVP} becomes
\begin{align*}
 r^2\frac{R''}{R} + r\frac{R'}{R} - (r\lambda)^2 &= -(\pm im)^2 + (\pm im)^2\left(\rocf\right)^2 - 2\rocf r\lambda (\pm im) = m^2 -\left(\rocf\right)^2 m^2 \mp 2\rocf imr\lambda\\
0 &= r^2\frac{R''}{R} + r\frac{R'}{R} - \left[ ( r\lambda)^2 +m^2 \left(1-\left(\rocf\right)^2\right)\mp 2\rocf imr\lambda\right]\\
\overset{\cdot R}{\Rightarrow} 0 &= r^2 R'' + rR' -\left[ ( r\lambda)^2 +m^2 \left(1-\left(\rocf\right)^2\right)\mp 2\rocf imr\lambda\right]R
\end{align*}
One may use \eqref{eq:TheoAnalysis_UnitDiscFirstCoordSubst} to obtain
\begin{align*}
0 &= r^2R'' +rR' -\left[(r\lambda)^2 +m^2\left(1-\left(\rocf\right)^2\right) \mp 2i \rocf m(r\lambda)\right] R\\
&= \left(\frac{x}{\lambda}\right)^2R'' + \left(\frac{x}{\lambda}\right) R' -\left[ x^2 +m^2\left( 1-\left(\frac{x\omega}{\lambda c}\right)^2\right) \mp 2i\frac{x\omega}{\lambda c} mx\right] R\\
&= x^2 \tilde{R}'' +x\tilde{R}' - \left[x^2\left(1\mp\frac{2i\omega}{c\lambda} m -\left(\frac{\omega}{c\lambda}\right)^2m^2\right) +m^2\right] \tilde{R}\;.
\end{align*}
With $m\in\ZZ$ fixed define
\begin{subequations}\label{eq:App_UnitDiscSecondSubstitution}
\begin{align}
u(x) &:= \sqrt{x^2\left( 1\mp 2\frac{i\omega}{c\lambda} m -\left(\frac{\omega}{c\lambda}\right)^2 m^2\right)} = x\sqrt{\left( 1\mp 2\frac{i\omega}{c\lambda} m -\left(\frac{\omega}{c\lambda}\right)^2 m^2\right)}\\
\frac{du}{dx} &= \frac{x}{u}\left( 1\mp 2\frac{i\omega}{c\lambda} m -\left(\frac{\omega}{c\lambda}\right)^2 m^2\right)\\
x &= \frac{u}{\sqrt{ 1\mp 2\frac{i\omega}{c\lambda} m -\left(\frac{\omega}{c\lambda}\right)^2 m^2}}\\
\hat{R}(u) &:= R(x(u))\;.
\end{align}
\end{subequations}
With these relations the above equation may be rewritten as
\begin{equation}
0 = u^2 \hat{R}''(u) + u\hat{R}'(u) -\left[u^2+m^2\right]\hat{R}(u)\;.\label{eq:TheoAnalysis_SecondModifiedBesselEquation}
\end{equation}
This is again a modified Bessel equation. Using the same reasoning as above to eliminate $K_m$ and enforcing homogeneous Dirichlet boundary conditions
on $S^1=\del \CalD$ one obtains
\begin{gather*}
iu = j_{m,s} \Leftrightarrow u = -ij_{m,s}\\
\lambda \sqrt{1\mp2\frac{i\omega}{c\lambda}m -\left(\frac{\omega}{c\lambda}\right)^2 m^2 } = -ij_{m,s}\;.
\end{gather*}
Seeing as one is looking only for values of $\lambda$ for which the equation holds true one can solve for $\lambda$, obtaining
\begin{align}
-j_{m,s}^2 &= \lambda^2 \left( 1 \mp 2\frac{i\omega}{c\lambda} m -\left(\frac{\omega}{c\lambda}\right)^2 m^2\right) \notag\\
0 &= \lambda^2  \mp 2\frac{i\omega}{c} m\lambda -\left(\frac{\omega}{c}\right)^2 m^2 + j_{m,s}^2 \notag\\
\lambda_{1\vert 2} &= \frac{1}{2}\left[-\left( \mp\frac{2i\omega m}{c}\right )\pm \sqrt{ \left(\mp\frac{2i\omega m}{c}\right)^2 -4\left(j_{m,s}^2 -\left(\frac{m\omega}{c}\right)^2\right) }\right]\notag\\
&= \frac{1}{2}\left[ \pm \frac{2i\omega m}{c}\pm \sqrt{-4\left(\frac{m\omega}{c}\right)^2 -4 j_{m,s}^2 +4\left(\frac{m\omega}{c}\right)^2 }\right]\notag\\
&= \frac{1}{2}\left[\pm\frac{2i\omega m}{c} \pm \sqrt{-4j_{m,s}^2}\right] = \pm i \left[ j_{m,s} +\frac{m\omega}{c}\right]\label{eq:App_UnitDiscEVEvolution}
\end{align}

One can see that the eigenvalues (or eigenfrequencies, if one were to use $\Omega = -i c\lambda$) are real and change linearly with $\omega$, the angular velocity
of the disc. Choosing the positive sign in the above equation, i.e., choosing $\lambda_1(\omega)$ and noting that $m\in \ZZ$ it follows that the co- and counterpropagating waves
diverge $\propto \frac{2|m|\omega}{c}$, indeed
\begin{equation}
\lambda(\omega;m) - \lambda(\omega;-m) = i \left[ j_{m,s} + \frac{m\omega}{c} - j_{-m,s} -\frac{-m\omega}{c}\right] = i \frac{2m \omega}{c}\;.\label{eq:App_UnitDiscEVDivergence}
\end{equation}
Which is, save for the $i$, in accordance with the result presented by \citeauthor{sunada_sagnac_2006} in \cite{sunada_sagnac_2006}.
It has been shown here, that there are, in fact, no terms of higher than linear order in $\tfrac{\omega}{c}$ provided we consider the case where $n=1$.

\chapter{The Software}\label{ch:App_Software}
\em{This section differs from the original version as a reference to an enclosed storage medium is not suitable for an online publication!}
The source code and data can be found in a Git repository  at https://github.com/Anton-Le/RotatingIsospectralDrums .

The source code, its documentation and the data are provided with the intent to facilitate the reproduction or refutation of the results of this thesis.
The source code is hereby placed under the GNU General Public License v. 3.

\chapter{Acknowledgments}

I am thankful to my advisor, Prof. Nils Schopohl, who suggested the topic of this thesis and has been supportive of my quest
to use differential forms for the theoretical analysis. I have enjoyed many an hour of invigorating discussions with him about the formal structure of
electrodynamics as well as  different types of numerical methods which could have been used for this work.

I owe a great debt of gratitude to Arno Petri for his help in finding the current repository of ARPACK++ \cite{_reuter_arpackpp}. 
In doing so he spared me time-consuming implementations of interfaces to the FORTRAN routines of ARPACK. He was kind enough, too, to
have a look at this thesis prior to its completion, though the subject is not his area of expertise.

Next on the list is Michael Benner, who helped me clarify the mathematical notions of \ref{ch:MathematicalMinimum} as well as the subject of the $3+1$ decomposition.
These discussions have proven useful in showing me that I am not a mathematician at heart, though parts of this thesis may suggest otherwise.

He is closely followed by  Marius Dommermuth and Maximilan Becker, the members of Prof. Schopohl's group during the time I have been researching the topic of this thesis.
Discussions with them helped to show me the concepts of the theory I had not understood sufficiently well. I also enjoyed many talks with them on topics not related to this work.

Last but not least I thank my parents for the support they have given me towards the end of this mental marathon.
\backmatter

\printbibliography[heading=bibintoc]
\listoffigures

\end{document}